\documentclass[preprint,12pt]{article}

\usepackage[margin=1in]{geometry}
\usepackage{fancyhdr}
\usepackage{amsmath}
\usepackage{amssymb}
\usepackage{easymat}
\usepackage{graphicx}
\usepackage{listings}
\usepackage{cases}
\usepackage{xcolor}
\colorlet{revisionblue}{black}
\usepackage{pdflscape}
\usepackage[english]{babel}
\usepackage{amsthm}
\newtheorem{lemma}{Lemma}
\newtheorem{thm}{Theorem}

\newtheorem{assum}{Assumption}

\newtheorem{eg}{Example}
\newtheorem{prop}{Proposition}

\newtheorem*{conditionst}{Condition ST}
\newtheorem*{conditionns}{Condition CO}

\usepackage[flushleft]{threeparttable}
\usepackage{bbm}

\usepackage{tikz}
\usetikzlibrary{patterns}
\usepackage{pgfplots}
\pgfplotsset{compat=1.15}

\usepackage{natbib}
\RequirePackage[hidelinks,pdfstartview=FitB]{hyperref}

\usepackage{subcaption}
\usepackage{float}
\usepackage{setspace}
\usepackage{multirow}
\usepackage{array}
\usepackage{longtable}
\usepackage{threeparttablex}

\newcommand{\suppref}[2]{\ref{#1}}
\newcommand{\mainref}[2]{\ref{#1}}
\newcommand{\mainequationref}[2]{\eqref{#1}}

\newcommand{\startsupplement}{%
  \clearpage
  \setcounter{section}{0}%
  \setcounter{equation}{0}%
  \setcounter{figure}{0}%
  \setcounter{table}{0}%
  \setcounter{prop}{0}%
  \setcounter{assum}{0}%
  \setcounter{eg}{0}%
  \renewcommand{\thesection}{S.\arabic{section}}%
  \renewcommand{\theequation}{S.\arabic{equation}}%
  \renewcommand{\theprop}{S.\arabic{prop}}%
  \renewcommand{\theassum}{S.\arabic{assum}}%
  \renewcommand{\thefigure}{S.\arabic{figure}}%
  \renewcommand{\thetable}{S.\arabic{table}}%
  \renewcommand{\theeg}{S.\arabic{eg}}%
  \renewcommand{\theHsection}{S.\arabic{section}}%
  \renewcommand{\theHequation}{S.\arabic{equation}}%
  \renewcommand{\theHfigure}{S.\arabic{figure}}%
  \renewcommand{\theHtable}{S.\arabic{table}}%
}

\title{Estimation and  Inference for Synthetic Control Methods with Spillover Effects}
\author{Jianfei Cao
	\footnote{Department of Economics, Northeastern University, j.cao@northeastern.edu.}
\and Connor Dowd
\footnote{Joint Committee on Taxation, connor.dowd@jct.gov.}
}
\date{}

\begin{document}

\maketitle

\abstract{
Estimation and inference procedures for synthetic control methods often do not allow for the existence of spillover effects, which are plausible in many applications. In this paper, we consider estimation and inference for synthetic control methods, allowing for spillover effects. We propose estimators for both direct treatment effects and spillover effects and show that they are asymptotically unbiased. In addition, we propose an asymptotically valid inference procedure. Our estimation and inference procedure applies to cases with multiple treated units and/or multiple post-treatment periods, and to ones where the underlying factor model is either stationary or cointegrated.
We discuss the bias from misspecified spillover structures and propose a test for correct specification.
Simulation shows that our method beats existing methods when the remaining pure donor is substantially different from the treated unit.
We apply our method to a classic empirical example that investigates the effect of California's tobacco control program as in \cite{Abadie2010} and find evidence of spillovers.
}

\section{Introduction}

The synthetic control method (SCM), introduced by \cite{Abadie2003}, has gained popularity for estimating treatment effects in settings with few treated units and post-treatment periods, such as studying state-level policies. By leveraging pre-treatment data, SCM often provides better counterfactual estimates than methods like difference-in-differences in comparative case studies \citep{Abadie2018,Ferman2016}. However, like many other methods, SCM relies on the Stable Unit Treatment Value Assumption (SUTVA), which assumes untreated units are unaffected by the treatment. This assumption is often violated, and the structure of SCM makes it particularly vulnerable to bias when SUTVA does not hold. For instance, when California increases cigarette taxes, SUTVA implies that consumers won't shift purchases to neighboring states like Nevada, an assumption we show is likely to be violated in Section \ref{empirical example}. Similar SUTVA violations are common in many applications using geographically aggregated data.

The presence of spillover effects can severely bias SCM in estimating the direct treatment effect (formally defined later by $\alpha_1=y_{1,T+1}(1,0,\dots,0)-y_{1,T+1}(0,\dots,0)$). Post-treatment control contamination leads to a biased counterfactual estimate and, consequently, a biased treatment effect estimate. 
Compared to difference-in-differences, this contamination-induced bias can be substantially larger, if spillovers are concentrated in heavily-weighted control units.
Moreover, if spillovers propagate along the same channels as the underlying factor model, SCM may actively select for bias-inducing units. We illustrate these bias effects in our simulation section (Section \suppref{section_sim_spatial}{S.3.5}), showing a scenario where SCM disproportionately assigns weights to control units with higher spatial correlation, amplifying the bias when spillover effects also increase with spatial correlation.

The problem of spillover effects can be partially solved by eliminating contaminated units in estimation, known as the \emph{pure-donor method}, where a pure donor refers to a control unit not experiencing spillover effects. 
Similar methods include those developed for multiple treated units, which are often equivalent to the pure-donor method \citep[see][]{Cavallo2013, Firpo2018, Kreif2016, Robbins2017, Xu2017}.
However, these methods may be concerning for several reasons. 
First, there are cases where most, if not all, control units are affected by the spillover, making it impractical to exclude all such units. 
We aim to develop a practical solution for these scenarios that allows effective analysis even when spillovers are widespread.
{\color{revisionblue}
Second, contaminated units are often crucial in forming the synthetic control.
Their exclusion can leave a pure-donor pool that differs substantially from the treated unit and cause a loss of efficiency \citep{DiStefano2020}.
Our simulations show that in this case our method has markedly lower root mean square error (RMSE) and greater power than the alternatives.}
In addition, by using fewer control units, the pure-donor method exhibits a large worst-case scenario bias when the spillover structure is misspecified. 
Specifically, this method presents a larger identified set of potential biases. 
We argue that this bias can be significantly mitigated by using a full-sample method. 
We illustrate this through a sensitivity analysis conducted in our empirical application. 

The goal of this paper is to relax the SUTVA condition and to perform estimation and testing. Particularly, we look at the cases where there are spillover effects, which are defined by a Rubin model as the difference between the actual outcomes and the counterfactual ones. To facilitate estimation, we assume some knowledge about the spillover effects is known. Specifically, the treatment effect and the spillover effects are linear in some unknown parameters.
We give examples where this assumption is plausible. 
Thanks to the known spillover structure, we can propose an asymptotically unbiased estimator for the treatment and spillover effects. We also characterize the asymptotic distribution of the estimator.
Compared with existing methods, our method uses information from all known units in estimation.
Our method can also often deal with cases where all units are contaminated by spillovers, at the cost of assuming more structure on the spillovers.  

We follow the setup in \cite{Ferman2016}, where we focus on cases with an imperfect pre-treatment fit. 
We will proceed under the assumption that treatment assignment is not correlated with time-varying unobservables, under which \cite{Ferman2016} show that the demeaned SCM ensures asymptotically unbiased estimation of treatment effects.
We also rely on the imperfect pre-treatment fit to identify the null distribution of the proposed statistic. 
In terms of the asymptotic framework, we consider cases with many pre-treatment periods and a fixed number of control units. 
As suggested by \cite{Ferman2016}, even in cases where large-$T$ asymptotics is not justified, our results can be interpreted as the SCM weights not converging to weights that reconstruct the factor loadings of the treated unit when the number of pre-treatment periods is large. 
{\color{revisionblue}While our formal results use fixed-$N$, large-$T$ asymptotics, Section \suppref{section_largeN}{S.2.1} discusses the additional requirements for a possible large-$(N,T)$ extension.}
Monte-Carlo simulation shows that our methods produce reasonable estimation and testing results as long as there is a moderate number of pre-treatment periods. 

Additionally, we propose an inferential procedure based on \cite{Andrews2003}'s end-of-sample instability test, or $P$-test. 
We generalize the $P$-test to the synthetic control method with and without spillover effects.
Similar to the $P$-test, our testing procedures use the idea of approximating the null distribution of the statistic using pre-treatment data.  
By using pretreatment time variation, our inference procedure avoids requiring comparable null distributions across treated and control units.

We give high-level conditions under which our methods are valid. In addition, our conditions adapt to factor models with either stationary or cointegrated common factors, which are often used to justify the usage of synthetic control methods. 
Furthermore, we consider extensions where treatment applies to multiple units or periods, and where there are extra covariates.

Throughout, we assume the spillover structure is known.
We illustrate through our empirical application in Section \ref{empirical example} how contextual knowledge can be used to inform the structure of the {\color{revisionblue}spillovers}.
Moreover, we propose a test for correct specification ($\kappa_A$-statistic) to alleviate concern regarding structure misspecification.
{\color{revisionblue}Section \suppref{sec_simulation_kappaA}{S.3.3} studies its finite-sample behavior under range and functional-form misspecification.}
In Section \suppref{sec_multi_post_period}{S.2.3}, we extend this test to settings with multiple post-treatment periods, and give conditions under which consistent structure estimation is possible.

Another assumption we adhere to in the paper is that there is no selection on time-varying unobservables, a condition similarly employed by \citet{Ferman2016}. 	
For potential extension to cases without this assumption, readers may explore settings with a perfect pre-treatment fit  \citep{Abadie2010,DiStefano2020}, or settings where the number of controls approaches infinity \citep{arkhangelsky_synthetic_2021, ferman_properties_2021}.
	
We revisit the empirical example from \cite{Abadie2010} on California's 1989 cigarette tax. \cite{Abadie2010} used data from 1970 onward, excluding 12 states potentially affected by spillovers or interventions, leaving 38 control states. Despite this, we find evidence of SUTVA violations in most post-tax years, even within the 38 control states, and in the 12 excluded states.
Our estimates of the tax's impact are consistently smaller in magnitude than \cite{Abadie2010}'s for the first four post-treatment years, potentially due to adjustments for significant early spillovers in Nevada, which was heavily weighted in standard SCM.
We also conduct a sensitivity analysis of worst-case bias from misspecified spillover structures, examining how the identified bias set expands with increasing spillover magnitude. The pure-donor method exhibits a larger identified bias set than our method.

This paper contributes to three developing literatures. First, it complements the literature on statistical inference in SCM by providing formal results without assuming SUTVA. Many works have developed inference methods for settings with few treated units and short pre- and post-treatment periods, such as difference-in-differences \citep{Conley2011}, which can be viewed as a special case of SCM with equal weights, and placebo tests that permute across units \citep{Hahn2017}. Most related work to ours is \cite{Chernozhukov2021}, who also use variation across periods rather than units to perform testing. \cite{Li2019} proposes a testing procedure based on projection onto convex sets and results from \cite{Fang2018}. 
\cite{cao_synthetic_2025} consider synthetic control methods in the context of staggered adoption. 
However, few works consider spillovers. To our knowledge, the only inferential methods allowing for spillovers are \cite{DiStefano2020} and \cite{Grossi2020}. \cite{DiStefano2020} require perfect pre-treatment fit, while our method allows for imperfect fit. \cite{Grossi2020} use a penalized SCM similar to \cite{Abadie2021} and assume exchangeable groups, which we do not require. Furthermore, our method applies to cointegrated factor models, of interest even without spillovers.

We also contribute to the growing literature on estimating treatment and spillover effects. This fast-growing literature looks into both estimation of treatment effects in the presence of spillover effects, as well as estimation of spillover effects themselves. For example, \cite{Vazquez-Bare2017} considers a framework with clusters, allowing spillovers within but not across clusters, and estimates heterogeneous treatment and spillover effects. \cite{Basse2019} and \cite{Rosenbaum2007} use randomization tests for inference with spillovers. See \cite{Basse2019} and \cite{Vazquez-Bare2017} for literature reviews.
However, this literature rarely considers panel data settings with few treated units and short post-treatment periods, partly due to insufficient information about spillovers. We address this by assuming pre-specified spillover structures linear in underlying parameters, enabling estimation and testing of spillover effects. We also introduce a test for correct specification of spillover structures and provide conditions for consistent spillover structure estimation with multiple post-treatment periods, contributing to existing work on spillover structure estimation in panel data \citep{de_paula_identifying_2023,Manresa2013-cr}.

Third, our results extend the literature on \cite{Andrews2003}'s end-of-sample instability tests. \cite{Andrews2003} uses data across time periods to approximate the null distribution of the test statistic and applies this idea to OLS, IV, and GMM.
{\color{revisionblue}\cite{Hahn2017} advocated applying Andrews' test in the context of synthetic control. We develop the formal asymptotic theory for this approach, both with and without spillover effects.}
\cite{Chernozhukov2021} propose a permutation test that is more general, but similar in cases where serial correlation matters. As \cite{Andrews2006} extend \cite{Andrews2003}'s results to the cointegrated cases, we also show that our method is still valid for a cointegrated factor model.

The rest of this paper is organized as follows. Section \ref{section_model} introduces a potential outcome framework with spillover effects and defines a known spillover structure that will be useful later.
Section \ref{section_estimation} proposes an estimator and derives its asymptotic distribution. It also discusses an example of a factor model and an invertibility assumption of the proposed estimator. Section \ref{section_inference} considers the $P$-test introduced by \cite{Andrews2003} and \cite{Andrews2006} and explains how it can be applied in our setting. 
In Section \ref{section_discussion}, we discuss spillover misspecification and comparison with the alternative methods. 
An empirical application of our method is presented in Section \ref{empirical example}. 
{\color{revisionblue}In the Supplement, Section \suppref{section_spillover_examples}{S.1} presents empirical examples of spillover structures. Section \suppref{section_extension}{S.2} discusses large-panel asymptotics, multiple treated units and post-treatment periods, the connection to Powell (2022), additional covariates, and an example without pure donors.}
Section \suppref{section_simulation}{S.3} presents all the Monte Carlo simulation results, and Section \suppref{section_proof}{S.4} contains all the proofs.
{\color{revisionblue}Section \suppref{section_app_details}{S.5} reports details of the empirical application including robustness checks, numerical results, and donor-pool sensitivity analysis.}

\section{Model Specification}
\label{section_model}
\subsection{A Rubin model with spillover effects}

We start our discussion with a Rubin potential outcome framework. 
We consider a standard synthetic control setting where only one unit is treated and only one period is available after the treatment is implemented. 
We consider cases with multiple treated units and multiple post-treatment periods in Section \suppref{section_extension}{S.2}.

In Rubin's model with a violation of SUTVA, the potential outcomes are functions of treatment assignments on all units. 
Assume the outcome of unit $i$ at time $t$ is 
$y_{i,t}=y_{i,t}(d_t),$
where $d_t=(d_{1,t},\dots,d_{N,t})'$ and $d_{i,t}=1$ if unit $i$ has been treated at time $t$.  
Assume unit 1 is treated between time $T$ and $T+1$, and there are another $N-1$ units that are not directly treated by the policy. Thus, we observe an $N\times (T+1)$ panel as shown in Figure \ref{data_matrix}. 

\begin{figure}
		\centering
		\renewcommand{\arraystretch}{1.5}
		\begin{tabular}{ccc|c}
			$y_{1,1}(0,\dots,0)$ & \dots & $y_{1,T}(0,\dots,0)$ &$y_{1,T+1}(1,0,\dots,0)$  \\
			\hline
			$y_{2,1}(0,\dots,0)$ & \dots & $y_{2,T}(0,\dots,0)$ & $y_{2,T+1}(1,0,\dots,0)$  \\
			\vdots &$\ddots$ & \vdots & \vdots \\ 
			$y_{N,1}(0,\dots,0)$ & \dots & $y_{N,T}(0,\dots,0)$ & $y_{N,T+1}(1,0,\dots,0)$
		\end{tabular}
		\begin{tabular}{l}
			\} \text{\footnotesize \ \  treated unit}\\
			\multirow{3}{*}{$\left\}
				\begin{array}{ll}
					\\
					\text{\footnotesize control units}	\\
					\\
				\end{array}
				\right.$}\\
			\\
			\\
		\end{tabular}
		
		\ 
		
		\scriptsize  \quad \ \ \ \ \ \ \ \ \ $\uparrow$ treatment
		
	\caption{Observed panel with one treated unit and one post-treatment period}
	\label{data_matrix}
\end{figure}

Note that we only observe outcomes with $d_{t}=(0,\dots,0)'$ for $t=1,\dots,T$ and $d_{T+1}=(1,0,\dots,0)'$. This is the fundamental limitation of the dataset we are currently studying. Unless other homogeneity conditions are assumed, we cannot say anything about $y_{i,T+1}(d_{T+1})$ for $d_{T+1}\not\in \{(0,\dots,0)',(1,0,\dots,0)' \}$, because only one unit is treated and only one post-treatment period is available. For notation simplicity, let
\begin{equation*}
	\begin{cases}
		y_{i,t}(0) = y_{i,t}(0,\dots,0)\\
		y_{i,t}(1) = y_{i,t}(1,0,\dots,0)
	\end{cases}
\end{equation*}
for each $(i,t)$. Let $\alpha_i=y_{i,T+1}(1)-y_{i,T+1}(0)$ be the potential deviation from unit $i$'s counterfactual outcome $y_{i,T+1}(0)$ where no unit is treated at time $T+1$. That is, $\alpha_1$ is the direct treatment effect on unit 1, while $\alpha_i$ with $i\ne 1$ is the indirect effect or spillover effect. 
The whole effect vector $\alpha=(\alpha_1,\dots,\alpha_N)'$ can be of interest in our setting. 
Throughout, we consider the case where $N$ is fixed and $T$ goes to infinity. 

Our analysis will start with the following linear model, which will be formally defined later: 
\begin{equation}
	\label{eq_stacked_linear_model}
	Y_{t}(0)=a+B Y_{t}(0)+u_{t},
\end{equation}
where $Y_{t}(0)=(y_{1,t}(0),\dots, y_{N,t}(0))'$ is the stacked untreated outcomes, $a\in \mathbb R^{N\times 1}$, $B\in\mathbb R^{N\times N}$ with diagonal entries being zero, and $u_{t}\in \mathbb R^{N\times 1}$ is a mean-zero stationary process. 
This equation characterizes the relationship between any unit, including both treated and controls, and the rest.  
The plan is to use the pre-treatment data to estimate $a$ and $B$, and hope this relationship carries over to the post-treatment period, from which we can learn an asymptotically unbiased estimator of both treatment and spillover effects. 
Importantly, although $a$ and $B$ will be learned using SCM in this paper, all high-level conditions discussed later do not rely on a specific restriction, allowing for alternative estimation strategies. 

\subsection{Known spillover structures}

Throughout the paper, we assume that some knowledge about the spillover effects is known.

Namely, assume that the full effect vector $\alpha$ is a linear transformation of some unknown parameter $\gamma\in \mathbb{R}^k$, i.e., $\alpha = A\gamma$, where $\alpha\in \mathbb R^{N\times 1}$, $A\in \mathbb R^{N\times k}$, and $\gamma\in \mathbb R^{k\times 1}$. 
Here, $A$ contains the information about the spillover structure, and $\gamma$ contains sufficient information about the magnitude of spillover effects. 
Typically, $\gamma$ has fewer dimensions than $\alpha$ does ($k<N$). 
Consider the extreme example where we impose no restrictions on the possible spillovers. 
This is a special case of $\alpha=A\gamma$ with $A$ being the identity matrix and $\gamma=\alpha$. 
{\color{revisionblue}However, under the SCM weights used below, this unrestricted specification does not satisfy the identification condition in Section \ref{section_invertibility}.}

{\color{revisionblue}
Our leading example specifies the range of units that may experience spillovers without restricting their magnitudes.
\begin{eg}
	\label{nonparametric spillover}
	(Limited range)
	Assume that spillover effects may occur at known locations but are zero elsewhere. For example, suppose spillovers may occur only at units whose distance from unit $1$ is less than a known cutoff $\bar{d}$. Let $\mathcal E$ contain unit $1$ and the $p$ units that may experience spillovers. Let $\mathcal P$ contain the remaining units, which are pure donors. Without loss of generality, order the units as $(\mathcal E,\mathcal P)$, with unit $1$ first. Let $\alpha_{\mathcal E}=(\alpha_1,\ldots,\alpha_{p+1})'\in\mathbb R^{p+1}$ and $\alpha_{\mathcal P}=(\alpha_{p+2},\ldots,\alpha_N)'\in\mathbb R^{N-p-1}$ denote the corresponding subvectors of $\alpha$. The treatment and spillover effect vector then satisfies
	$$
	\alpha=A\gamma,
	\qquad
	A=\begin{bmatrix}
	I_{p+1}\\
	0_{(N-p-1)\times(p+1)}
	\end{bmatrix},
	\qquad
	\gamma=\alpha_{\mathcal E},
	\qquad
	\alpha_{\mathcal P}=0.
	$$
	Here, $k=p+1$ and $I_{p+1}$ is the $(p+1)\times(p+1)$ identity matrix. Thus, each unit in $\mathcal E$ has its own unrestricted effect parameter. The first component is the treatment effect on unit $1$, and the other $p$ components are the potential spillover effects.
\end{eg}
}

The assumptions in Example \ref{nonparametric spillover} are often plausible.
Later we will illustrate how contextual knowledge can be used to inform the structure of the {\color{revisionblue}spillovers} through our empirical application in Section \ref{empirical example}.
%
If misspecification of the spillover structure is a concern, one can choose an $A$ matrix that incorporates more potential spillovers, i.e., a larger $p$. 
The consequences of misspecification are discussed in Section \ref{section_misspecification}. 

Besides Example \ref{nonparametric spillover}, we will explore other possibilities, including the case where some units experience equal spillover effects, as discussed in Example \ref{equally hit} in Section \ref{section_invertibility}, and the case where spillover effects decrease exponentially as the distance from the main treated unit increases, as in Example \suppref{eg_spatial}{S.1} in Section \suppref{section_continuous_treatment}{S.2.6}. 
{\color{revisionblue}Although our theoretical results allow more general choices of $A$, Example \ref{nonparametric spillover} is our baseline spillover specification throughout the paper and supplement, except in the discussions and illustrations based on Examples \ref{equally hit} and \suppref{eg_spatial}{S.1}.}

To alleviate concern regarding structure misspecification, we propose a test for correct specification ($\kappa_A$-statistic) in Section \ref{section_misspecification}.
In Section \suppref{sec_multi_post_period}{S.2.3}, we extend this test to settings with multiple post-treatment periods, and give conditions under which consistent structure estimation is possible.

\section{An Asymptotically Unbiased Estimator}
\label{section_estimation}
\subsection{SCM without spillover effects}

We consider a version of SCM proposed in \cite{Ferman2016}, which starts with obtaining the synthetic control weights by solving the optimization problem
\begin{equation}
(\widehat a_1,\widehat b_1')'
=\underset{\tilde{a}\in \mathbb{R}, \tilde{b}\in {\color{revisionblue}\mathcal{W}^{(1)}}}{\arg\min}\sum_{t=1}^T(y_{1,t}-\tilde{a}-Y_t'\tilde{b})^2,
\end{equation} 
where $Y_t=(y_{1,t},\dots,y_{N,t})'$ and ${\color{revisionblue}\mathcal{W}^{(1)}}=\{(w_1,\dots,w_N)'\in\mathbb{R}^{N}_+: w_1=0, \sum_{j=2}^Nw_j=1 \}$. This restricts the estimation weights such that they sum to 1, are all non-negative, and the same-unit weight is 0, as well as including an intercept term.
An estimator of the treatment effect $\alpha_{1}$ is given by 
$	\widehat{\alpha}_{1}=y_{1,T+1}-(\widehat{a}_1+Y_{T+1}'\widehat{b}_1),$
i.e., the counterfactual value $y_{1,T+1}(0)$ is approximated by $\widehat{a}_1+Y_{T+1}'\widehat{b}_1$. 
Here, we do not restrict the intercept but require the other coefficients to be non-negative and sum up to one.
See \cite{Doudchenko2016} for a discussion of other choices of restriction sets.
The intercept term $a$ is important in our setting because it takes out the bias by recentering the estimator. 
\cite{Ferman2016} show that when the pre-treatment {\color{revisionblue}fit} is imperfect and there is no selection on time-varying unobservables, this estimator is asymptotically unbiased, while the original SCM as in \cite{Abadie2010} has bias. 

\subsection{The proposed estimator under spillovers}
\label{section_estimator}

In order to back out the spillover effects, we first define the individual synthetic control weights and their limits. Namely, for each $i$, let the individual-specific synthetic control weights (and the intercept) be 
\begin{equation}
	\label{optimization}
		(\widehat{a}_i,\widehat{b}_{i}')'
=\underset{\tilde{a}\in \mathbb{R}, \tilde{b}\in {\color{revisionblue}\mathcal{W}^{(i)}}}{\arg\min}\sum_{t=1}^T(y_{i,t}-\tilde{a}-Y_t'\tilde{b})^2,
\end{equation} 
where ${\color{revisionblue}\mathcal{W}^{(i)}}=\{(w_1,\dots,w_N)'\in \mathbb{R}^{N}_+: w_i=0, \sum_{j=1}^Nw_j=1 \}$. These are the same restrictions as above (sum-to-one, non-negativity, same-weight is 0). Then, let the probability limit of the intercept and weights be
\begin{equation*}
	a_i=\text{plim}_{T\rightarrow \infty}\ \widehat{a}_i,\ b_i=\text{plim}_{T\rightarrow \infty }\ \widehat{b}_i,
\end{equation*} 
and we only consider cases where they are well-defined. 
We show later by Lemma \ref{estimation_lemma} in Section \ref{section_factor_model} that  $a_i$ and $b_i$ exist for each $i$ in factor models with stationary or cointegrated common factors. In general, $a_i$ and $b_i$ do not coincide with the weights that reconstruct the factor loadings \citep{Ferman2016}.

For each $(i,t)$, define the specification error by 
\begin{equation}
	\label{linear combination form}
	u_{i,t}=y_{i,t}(0)-(a_i+Y_t(0)'b_i).
\end{equation}
Define $a=(a_1,\dots,a_N)'$ and $B=(b_1,\dots,b_N)'$. 
Stacking Equation \eqref{linear combination form} for all $i$'s gives 
$u_t=Y_t(0)-(a+BY_t(0)),$
where  $u_t=(u_{1,t},\dots,u_{N,t})'$. 
This gives us Equation \eqref{eq_stacked_linear_model}. 
Since $\alpha = Y_{T+1}(1)-Y_{T+1}(0)$,
we have at period $T+1$
\begin{equation}
	\label{error equation}
	u_{T+1}=(I-B)(Y_{T+1}-\alpha)-a,
\end{equation}
where $Y_{T+1}=(y_{1,T+1},\dots,y_{N,T+1})'$. We will use this equation to estimate the whole effect vector $\alpha$.  
%
%
%

We form estimators for $(a,B)$ using synthetic control methods as in \eqref{optimization}. We do that for each $i=1,\dots, N$, as if each $i$ is the treated unit and other units are controls. Then, the estimators for $a$ and $B$ are $\widehat{a}=(\widehat{a}_1,\dots,\widehat{a}_N)'$ and $\widehat{B}=(\widehat{b}_1,\dots,\widehat{b}_N)'$, respectively. 
Define $M=(I-B)'(I-B)$ and let $\widehat{M}=(I-\widehat{B})'(I-\widehat{B})$ be an estimator for $M$.  

Recall that the effect vector is $\alpha=A\gamma$. 
Let an estimator of $\gamma$ be such that 
\begin{equation}
	\label{gamma estimation}
	\widehat{\gamma} = \underset{g\in \mathbb{R}^k}{\arg\min}\Vert (I-\widehat{B})(Y_{T+1}-Ag)-\widehat{a} \Vert
	=(A'\widehat{M}A)^{-1}A'(I-\widehat{B})'((I-\widehat{B})Y_{T+1}-\widehat{a}).
\end{equation}
Note that the first-order condition implies 
$	A'(I-\widehat{B})'\widehat u_{T+1}=0,$
where $\widehat u_{T+1} = (I-\widehat B) (Y_{T+1}-\widehat\alpha)-\widehat a$,
i.e., it requires that some weighted sum of the residuals be zero. Under that condition, the treatment and spillover effect vector $\alpha$ can be estimated by $\widehat{\alpha}=A\widehat{\gamma}$. 
One way to interpret this estimator is that the proposed $\widehat \alpha$ is the most consistent with the (constrained) linear model learned using pre-treatment data.

\begin{assum}
	\label{unbiased assumption}
	\emph{(a)} $\{ u_t\}_{t\ge 1}$ is stationary, and has mean zero;
	
	\emph{(b)} $\Vert \widehat{a}-a\Vert=o_p(1)$, $\Vert \widehat{B}-B\Vert =o_p(1)$;
	
	\emph{(c)} $\Vert (\widehat{B}-B)Y_{T+1}(0)\Vert =o_p(1)$;
	
	\emph{(d)} $A'MA$ is non-singular. 
\end{assum}

Part (a) generally requires that there is no regime shift or structural break. 
It also corresponds to the assumption of no selection on time-varying unobservables as in \cite{Ferman2016}. 
Parts (b) and (c) require that there are at least a moderate number of pre-treatment periods so that the synthetic control weights are well-estimated.
We show later that Assumption \ref{unbiased assumption}(a)-(c) are satisfied in factor models with either stationary or cointegrated common factors. 
We will discuss Part (d) later in Section \ref{section_invertibility}.

\begin{thm}
	\label{unbiasedness}
	Suppose Assumption \ref{unbiased assumption} holds. Then, $\widehat{\alpha}-(\alpha+Gu_{T+1})\rightarrow_p0$ as $T\rightarrow \infty$, where $G=A(A'MA)^{-1}A'(I-B)'$. Moreover, $E[Gu_{T+1}]=0$. 
\end{thm}

The structure of the limiting distribution is similar to the case in \cite{Ferman2016}, as it is inconsistent but asymptotically unbiased.
Note that consistent estimators are impossible because only one post-treatment period with one treated unit is available, so the error term for that one unit and period does not shrink in any limit we consider.

\subsection{The factor model as an example}
\label{section_factor_model}

Factor models are often used to justify the usage of synthetic control methods. 
See \cite{cao_principal_2020} for a review. 
Here we show that our assumptions are satisfied by factor models with stationary and cointegrated common factors. We follow \cite{Ferman2016} and consider a factor model such that for $i=1,\dots,N$ and $t=1,\dots,T+1$,  
\begin{equation}
	\label{factor model form}
	y_{i,t}(0)=\eta_t+\lambda_t'\mu_i+\varepsilon_{i,t},
\end{equation}
where $\lambda_t$ is an $F$-dimensional vector of common factors, with $F$ fixed, and $\varepsilon_{i,t}$ is the noise that is uncorrelated with $\lambda_t$. For notation simplicity, we write $Y_t(0) = (y_{1,t}(0),\dots,y_{N,t}(0))'$, $Y_t = (y_{1,t},\dots,y_{N,t})'$, and $\varepsilon_t=(\varepsilon_{1,t},\dots,\varepsilon_{N,t})'$.

We focus on two sets of conditions in our discussion. 

\begin{conditionst}[model with stationary common factors]
	Assume $\{(\lambda_t,\varepsilon_t)\}_{t\ge 1}$ is stationary, ergodic for the first and second moments, and has finite $(2+\delta)$-moment for some $\delta>0$. 
	The time fixed effect $\eta_t$ satisfies $\Vert (\widehat B-B)\eta_{T+1}\Vert=o_p(1)$.  
	Assume $\Omega_y=Cov[(\lambda_t'\mu_1+\varepsilon_{1,t},\dots,\lambda_t'\mu_N+\varepsilon_{N,t})']$ is  positive definite.
\end{conditionst}

\noindent\textbf{Remarks:} 1. Stationarity implies that there is no selection on time-varying unobservables. 
The time fixed effect $\eta_t$ is allowed to be non-stationary or to explode, but not faster than the rate of $\widehat B$. 
We can do this because the time fixed effect always gets canceled out, given the simplex restriction of the synthetic control methods. 

2. We show in the proof of Lemma 1 that in this case
$a_i=E[y_{i,1}(0)-Y_{1}(0)'b_i]$ 
and
$b_i=\underset{w\in {\color{revisionblue}\mathcal{W}^{(i)}}}{\arg\min}\ (w-e_i)'\Omega_y (w-e_i)$,
where $e_i$ is a unit vector with one at the $i$-th entry and zeros everywhere else, and ${\color{revisionblue}\mathcal{W}^{(i)}}=\{(w_1,\dots,w_N)\in \mathbb{R}_+^N: w_i=0,\sum_{j\ne i }w_j=1 \}$. Note that in general $b_i$ does not recover the factor structure, because $\mu_i\ne (\mu_1,\dots,\mu_N)b_i$ in general.

3. We do not impose any restriction on the factor loadings $\{\mu_i \}_{i=1}^N$ except for $\Omega_y$ being positive definite. In the stationary case, the key  for the treatment estimator to be asymptotically unbiased and the test proposed below to be valid is to include an intercept in the optimization problem \eqref{optimization}.  

\begin{conditionns}	[model with cointegrated $\mathcal{I}(1)$ common factors] Rewrite Equation \eqref{factor model form} as 
$
		y_{i,t}(0)=	(\lambda_t^1)'\mu_i^1+	(\lambda_t^0)'\mu_i^0+\varepsilon_{i,t},
$
	and $\eta_t$ can be either in $\lambda_t^1$ or $\lambda_t^0$. Assume $\{(\lambda_t^0,\varepsilon_t) \}_{t\ge 1}$ is stationary, ergodic for the first and second moments, and has a finite $4$-th moment. Without loss of generality, $E[\varepsilon_{i,t}]=0$. Assume $\{\lambda_t^1\}_{t\ge 1}$ is $\mathcal{I}(1)$. Further assume for each $i$, $y_{i,t}(0)$ is such that weak convergence holds for $T^{-1/2}y_{i,[rT]}(0)\Rightarrow \nu_i(r)$, where $\Rightarrow$ is weak convergence and process $\nu_i(r)$ is defined on $[0,1]$ and has a bounded continuous sample path almost surely.  For each $i$, let ${\color{revisionblue}\mathcal{W}^{(i)}}=\{(w_1,\dots,w_N)\in \mathbb{R}_+^N:w_i=0,\sum_{j\ne i}w_j=1 \}$. Assume for each $i$, there exists $w^{(i)}\in {\color{revisionblue}\mathcal{W}^{(i)}}$ such that $\mu_i^1=\sum_{j=1}^Nw^{(i)}_j\mu_j^1$. That is, $(w^{(i)}-e_i)$ is a cointegrating vector for $Y_t(0)$, where $e_i$ is a unit vector with $i$-th entry being one and zeros everywhere else.
\end{conditionns}

Note that Condition CO puts restrictions on the factor loadings. The restrictions are similar to those in \cite{Ferman2016}. 

The relevance of the factor model is given by the following lemma:
\begin{lemma}
	\label{estimation_lemma}
	Suppose $A'MA$ is non-singular. 
	Then, either Condition ST or Condition CO implies Assumption \ref{unbiased assumption}.
\end{lemma}
Thus, results derived in Theorem 1 apply to factor models with Condition ST or Condition CO.

\subsection{Invertibility of $A'MA$}
\label{section_invertibility}

In Assumption \ref{unbiased assumption}(d), we require that the matrix \(A'MA\) must be invertible.
{\color{revisionblue}An advantage of this condition is that it can be checked empirically. The matrix $A$ is chosen by the researcher and $B$ can be consistently estimated, so $A'\widehat{M}A$ is directly computable. We therefore recommend that users of our method calculate and report its smallest eigenvalue for every spillover structure they use. A smallest eigenvalue close to zero indicates that the chosen structure is close to a non-identified configuration, in which case estimates of $\gamma$ are unstable and confidence intervals are wide.}

This section explores the implications of this key assumption, providing both examples and counterexamples, so that the reader can develop intuition for what the condition requires and for what a small eigenvalue reflects. In addition, we discuss the connection to a network framework by giving conditions under which \(I-B\) has rank \(N-1\), where a single pure donor unit is sufficient to ensure the invertibility of \(A'MA\).
Extending beyond this section, we discuss how the invertibility of 
$A'MA$ parallels the full rank assumption of the design matrix in a linear model in Section \suppref{section_continuous_treatment}{S.2.6}.

\subsubsection{Intuition and a toy example}

First, note the invertibility of $A'MA$ is testable in principle. Recall that $M=(I-B)'(I-B)$, so that $A'MA = A'(I-B)'(I-B)A$. $A$ is defined by the econometrician ahead of time. We can consistently estimate $B$ so the data informs us of the validity of this assumption. 

To understand this assumption better, we replace $\alpha$ by $A\gamma$ in Equation \eqref{error equation} and have 
\begin{equation}
	\label{id equation}
	(I-B)A\gamma=(I-B)Y_{T+1}-a-u_{T+1}.
\end{equation}
Equation \eqref{id equation} is the key to learning $\alpha$. Under mild regularity conditions, $a$ and $B$ are identified from the model and learned by the synthetic control method. We do not observe $u_{T+1}$, but the distribution of $u_{T+1}$ can be learned using pre-treatment data under stationarity of $\{u_t \}_{t\ge 1}$. Therefore, if $A'MA$ is non-singular, or equivalently, $(I-B)A$ has full rank, we can form an estimator of $\gamma$ whose limiting distribution is identified by multiplying both sides of Equation \eqref{id equation} by $(A'MA)^{-1}A'(I-B)'$. Note that we do not point-identify $\gamma$ or $\alpha$, because we have only one observation of the outcome in the post-treatment period. 

The following example is useful in illustrating the invertibility assumption:
\begin{eg}
	\label{equally hit}
	(Homogeneous spillovers)
	Assume that  a subset of control units, but not all of them, are equally affected by the spillover effects, i.e. $\alpha=A\gamma = (\alpha_1,b,\dots, b,0_{1\times l})'$, where 
	$$
	A=\begin{bmatrix}
		1&0\\0_{k\times 1}&1_{k\times 1}\\ 0_{l\times 1} & 0_{l\times 1}
	\end{bmatrix}, \ \
	\gamma=\begin{bmatrix}
		\alpha_1\\b
	\end{bmatrix},
	$$
	$\alpha_1$ is the treatment effect, and $b$ is the homogeneous spillover effect. 
\end{eg}

For illustration, consider a three-unit case, where unit 1 is treated. WLOG, let the synthetic control weight matrix $B$ be
	$$B=\begin{bmatrix}
		0 & w_1 &1-w_1\\
		w_2 & 0 & 1-w_2 \\
		w_3 & 1-w_3 & 0
	\end{bmatrix}.$$
	Suppose the researcher first assumes unit 2 and 3 are equally exposed to the spillover effects. 
	That is, we have $\alpha=(\gamma_1,\gamma_2,\gamma_2)'$ with 
	$$A_1=\begin{bmatrix}
		1 & 0 \\
		0 & 1\\
		0 & 1
	\end{bmatrix}\text{ and }\gamma=\begin{bmatrix}
		\gamma_1\\
		\gamma_2
	\end{bmatrix},
\text{ and thus }
(I-B)A_1=\begin{bmatrix}
	1 & -1\\
	-w_2 & w_2\\
	-w_3 & w_3
\end{bmatrix},
$$
leading to a non-invertible $A'MA$.
	Intuitively, the problem here is there are two control observations we want to take a difference from, to determine the treatment effect. 
	
	If they instead assume only one of the controls is exposed to the spillover effects, $A'MA$ is non-singular in general. 
	In this case,  we have  $\alpha=(\gamma_1,\gamma_2,0)'$ with 
	$$A_2=\begin{bmatrix}
		1 & 0 \\
		0 & 1\\
		0 & 0
	\end{bmatrix}\text{, }\gamma=\begin{bmatrix}
		\gamma_1\\
		\gamma_2
	\end{bmatrix}, 
	\text{ and thus }
	(I-B)A_2=\begin{bmatrix}
		1 & -w_1\\
		-w_2 & 1\\
		-w_3 & w_3-1
	\end{bmatrix}.$$
	It can be shown that $(I-B)A_2$ always has full rank for $(w_1,w_2,w_3)\in [0,1]^3$. 

This applies to more general settings. That is, if all controls are equally hit by the spillover effects, then $(I-B)A$ does not have full rank and $A'MA$ is non-invertible. Allowing a few units to be exempt from the spillover effects makes $(I-B)A$ have full rank in general. 

\subsubsection{Results on rank of $I-B$}
\label{sec_IB_rank_Nminus1}

We extend the main idea of the previous section to the more interesting case of Example \ref{nonparametric spillover}, where the range of spillover effects is bounded but their magnitudes can vary. In this case, the matrix $(I-B)A$ is obtained by eliminating columns corresponding to units that are neither treated nor exposed to spillovers. 
The invertibility assumption is more plausible when a moderate number of columns are eliminated from $(I-B)$, suggesting that only a limited range of units are exposed to spillovers.

We explore the properties of $I-B$ here. 
Note that $I-B$ has a maximum rank of $N-1$, implying a necessary condition for invertibility in Example \ref{nonparametric spillover} is having at least one pure donor. 
For the sufficient condition, we show its connections to network frameworks, which is formalized below.

Let $\mathcal{X} = \{1, \dots, N\}$ be the set of nodes corresponding to units. The matrix $B$ can then define an adjacency matrix, with the directed edge $e: \mathcal{X}^2 \rightarrow \{0,1\}$ defined by $e(i,j) = \mathbbm{1}\{b_{i,j} > 0\}$. As a result, $(\mathcal{X}, e)$ defines a directed network. Let a path from $i_1 \in \mathcal{X}$ to $i_k \in \mathcal{X}$ be a sequence of nodes $p(i_1, i_k) = (i_1, \dots, i_k)$ where $e(i_1, i_2) = e(i_2, i_3) = \dots = e(i_{k-1}, i_k) = 1$. 
\begin{prop}
	\label{prop_IminusB_rank_network}
	If for each ordered pair $(i, j) \in \mathcal{X}^2$ there exists a path $p(i,j)$, then $I-B$ has rank $N-1$.
\end{prop}

This result helps us understand when the invertibility of $A'MA$ is {\color{revisionblue}satisfied}.
The following two examples illustrate networks excluded by the connectedness assumption.
First, it rules out disconnected networks, which imply that $I-B$ is a block-diagonal matrix. This results in at least two groups of nodes, each of which is capable of predicting outcomes within the group without outside data, e.g.,
\[
I-B=\begin{bmatrix}
	1 & -1 & 0 & 0\\
	-1 & 1 & 0 & 0\\
	0 & 0 & 1 & -1\\ 
	0 & 0 & -1 & 1
\end{bmatrix}.
\]
Here, the blocks without the parameter of interest are redundant. Normally, the rank of $I-B$ is $N$ minus the number of blocks (so $N-1$ in a connected network with only one block). 
Second, it rules out networks with sinks, where a few units can predict all other units, e.g.,
\[
I-B=\begin{bmatrix}
	1 & -1 & 0 & 0\\
	-1 & 1 & 0 & 0\\
	-1 & 0 & 1 & 0\\ 
	-1 & 0 & 0 & 1
\end{bmatrix}.
\]
Here, units 3 and 4 assign all their donor weight to unit 1, but do not contribute to unit 1's synthetic control.

When the invertibility condition fails, one possible remedy is to restrict the analysis to a smaller set of units while preserving the effects of interest. Whether these units are needed depends on the spillover structure and the effects of interest.
The invertibility condition should then be verified for the reduced system after re-estimating the synthetic-control weights.

\section{Statistical Inference}
\label{section_inference}

In this section, we discuss formal results on inference. At a high level, our test uses pre-treatment data to form the null distribution of a pre-specified post-treatment quantity.
We only consider cases with imperfect pre-treatment fit to facilitate the identification of the null distribution. 
In Section \ref{no spillover effect test}, we consider the case without spillover effects, and state the assumptions under which Andrews' $P$-test \citep{Andrews2003} is valid.
This result is of independent interest.
In Section \ref{spillover effect test}, we generalize the $P$-test to cases where spillover effects cannot be ignored, and allow for a more general set of hypotheses.

\subsection{Cases without spillover effects}
\label{no spillover effect test}

Suppose for now that there are no spillover effects ($\alpha_2=\dots=\alpha_N=0$). We want to test for the existence of treatment effect on unit 1. The null and alternative hypotheses of interest are $H_0:\alpha_1 = 0$ and $H_1:\alpha_1\ne 0$, respectively.
The test procedure we consider here is the end-of-sample instability test ($P$-test) in \cite{Andrews2003}. The usage of Andrews' test in the context of synthetic control methods is mentioned in \cite{Ferman2018} and \cite{Hahn2017}. We formalize this idea and derive conditions under which Andrews' test delivers valid inference results.

We assume that  $\alpha_1$ is not a function of $T$ under $H_1$. That is, we consider fixed, not local, alternatives, as in \cite{Andrews2003} and \cite{Andrews2006}. Specifically, $\alpha_1$ does not change as $T$ grows, which facilitates our analysis of the test statistic under $H_1$. 

Now we translate our hypothesis into the linear formulation considered in \cite{Andrews2003}. Namely, we have  
$	y_{1,t}=a_1+(a_1^*-a_1)\mathbbm 1\{ t=T+1\}+Y_t'b_1+u_{1,t}.$
A non-zero treatment effect is equivalent to a shift in the intercept $a_1$ (or equivalently, change of the distribution of $u_{1,t}$, at $t=T+1$). The null and alternative hypotheses become 
$H_0:a_1^*=a_1$ and $		H_1:a_1^*\ne a_1$, respectively. 
Let the synthetic control regression residuals be $\widehat{u}_{1,t}=y_{1,t}-\widehat{a}_1-Y_t'\widehat{b}_1$.
If there is no treatment effect, the distribution of $\widehat{u}_{1,T+1}$ should be asymptotically equivalent to that of $\widehat{u}_{1,t}$ for $t\le T$. 
Using this idea, define the test statistic by $	P=\widehat{u}_{1,T+1}^2.$
For notational simplicity, let $\beta_1=(a_1,b_1')'$,
$\widehat{\beta}_1=(\widehat{a}_1,\widehat{b}_1')'$, and $x_t=(1,Y_t')'$.
For any $\beta\in\mathbb{R}^{N+1}$, define
$P_t(\beta)=(y_{1,t}-x_t'\beta)^2$.
Then $P=P_{T+1}(\widehat{\beta}_1)$, where $\widehat{\beta}_1$ is estimated
using all $T$ pre-treatment periods.
For each $t\le T$, let $\widehat{\beta}_1^{(t)}$ denote the coefficient
estimator used to construct the $t$th pre-treatment statistic, and define
$P_t=P_t(\widehat{\beta}_1^{(t)})$.
The estimators $\widehat{\beta}_1$ and $\widehat{\beta}_1^{(t)}$ must be
consistent for $\beta_1$ in the sense of Assumption
\ref{no spillover assumption}(d).
A simple choice is the original full-sample synthetic-control estimator, which
sets $\widehat{\beta}_1^{(t)}=\widehat{\beta}_1$ for every $t$.
Another choice is the leave-one-pre-treatment-period-out construction, which
re-estimates $\widehat{\beta}_1^{(t)}$ after omitting period $t$.
\cite{Andrews2003} and \cite{Andrews2006} suggest using the leave-one-out estimator for better finite sample performance.
For a significance level of $\tau$, we reject $H_0$ if $P$ is larger than the $(1-\tau)$-quantile of $\{P_t \}_{t=1}^T$. 

To establish the validity of the proposed test, let $P_\infty$ be a random variable with the same distribution as $P_{T+1}(\beta_1)$.
Define the empirical CDF of $\{P_t \}_{t=1}^T$ by
$	\widehat{F}_{P,T}(x)=T^{-1}\sum_{t=1}^{T}\mathbbm{1}\{P_t\le x\},$
and let $F_P(x)$ be the distribution function of $P_1(\beta_1)$. 
We reject $H_0$ if $P > \widehat{q}_{P,1-\tau}$, where $\widehat{q}_{P,1-\tau}=\inf \{x\in \mathbb{R}:\widehat{F}_{P,T}(x)\ge 1-\tau \}$. 
Finally, let $q_{P,1-\tau}$ be the $(1-\tau)$-quantile of $P_1(\beta_1)$. 

\begin{assum}
	\label{no spillover assumption}
	\emph{(a)} $\{ u_{1,t}\}_{t\ge 1}$ is stationary, ergodic, and has mean zero; \\
	\emph{(b)} $E[|u_{1,t}|]<\infty$; \\
	\emph{(c)} $\exists$ positive definite $\{C_T \}_{T\ge 1}$ such that $\max_{t\le T+1}\Vert C_T^{-1} x_t\Vert=O_p(1) $;\\
	\emph{(d)}  $\Vert C_T(\widehat{\beta}_1-\beta_1)\Vert=o_p(1)$, and $\max_{t=1,\dots,T}\Vert C_T(\widehat{\beta}_1^{(t)}-\beta_1)\Vert=o_p(1)$; \\
	\emph{(e)} The distribution function of $P_1(\beta_1)$ is continuous and increasing at its $(1-\tau)$-quantile. 
\end{assum}

Assumption \ref{no spillover assumption} is similar to those in  \cite{Andrews2003}. 
Part (a) does not allow for a structural break. 
Part (b) and (c) are moment conditions. 
Part (d) requires scaled consistency of $\widehat{\beta}_1$ and uniform scaled consistency of the estimators $\widehat{\beta}_1^{(t)}$ used in the pre-treatment statistics. 
Part (e) generally requires that $u_{1,t}$ follows a continuous distribution.

\begin{thm}
	\label{thm_no_spillover}
	Suppose Assumption \ref{no spillover assumption} holds. Then, as $T\rightarrow \infty$, 	\\
	\emph{(a)} $P\rightarrow_d P_\infty$ under $H_0$ and $H_1$;\\
	\emph{(b)} $\widehat{F}_{P,T}(x)\rightarrow_p F_P(x)$ for all $x$ in a neighborhood of $q_{P,1-\tau}$ under $H_0$ and $H_1$;  \\
	\emph{(c)} $\widehat{q}_{P,1-\tau}\rightarrow_p q_{P,1-\tau} $ under $H_0$ and $H_1$; \\
	\emph{(d)} $\Pr(P>\widehat{q}_{P,1-\tau})\rightarrow \tau$ under $H_0$. 
\end{thm}

Theorem \ref{thm_no_spillover} states that the distribution of our test statistic $P$ can be approximated by the empirical distribution of $\{ P_t\}_{t=1}^T$.
Specifically, Part (d) shows that the proposed test is asymptotically valid in the sense that the rejection probability under the null is convergent to the nominal level. 

Combining this result with the Dominated Convergence Theorem, we have that under the alternative hypothesis \(H_1\), the rejection probability \(\Pr(P > \widehat{q}_{P, 1-\tau})\) converges to \(\Pr((u_{1,1} + \alpha_1)^2 > q_{P, 1-\tau})\), where \(q_{P, 1-\tau}\) is the \((1-\tau)\)-quantile of \(u_{1,1}^2\). Since the distribution of \(u_{1,1}\) can be estimated from the pre-treatment data, this result provides a practical method to approximate the power curves of the test.

We also show the relevance of the factor model in this context by the following lemma:
\begin{lemma}
	\label{lemma_inference_no_spillover}
	Suppose the distribution function of $P_1(\beta_1)$ is continuous and increasing at its $(1-\tau)$-quantile. For the full-sample construction $\widehat{\beta}_1^{(t)}=\widehat{\beta}_1$, either Condition ST or CO implies Assumption \ref{no spillover assumption}.
\end{lemma}

{\color{revisionblue}In addition, the same idea provides a practical diagnostic for the stationarity condition in Assumption \ref{no spillover assumption}(a). Treat period $T$ as a pseudo-treatment period and apply the $P$-test using periods $1,\ldots,T-1$ for estimation and the reference distribution. 
}

\subsection{Cases with spillover effects}
\label{spillover effect test}

In this section, we generalize Section \ref{no spillover effect test} to cases allowing for non-zero spillover effects. We propose a testing procedure that is based on Andrews' $P$-test and accounts for the spillover effect. The null and alternative hypotheses we consider are $H_0:C\alpha=d$ and $H_1:C\alpha\ne d$, with known $C$ and $d$. For example, we want to test for the hypothesis that there is no treatment effect at the treated unit (unit 1), then we let $C=(1,0,0,\dots,0)\in \mathbb{R}^{1\times N}$ and $d=0$. This effectively makes Section \ref{no spillover effect test} a special case of our test, although Theorem 2 has slightly stronger results than Theorem 3 does. If we want to test that there is a spillover, then we can let $C=[0_{(N-1)\times 1} \ I_{N-1}]\in \mathbb{R}^{(N-1)\times N}$ and $d=(0,\dots,0)'\in \mathbb{R}^{(N-1)\times 1}$. 

The test statistic we consider here is $P=(C\widehat{\alpha}-d)'{\color{revisionblue}\Xi_T}(C\widehat{\alpha}-d)$ for some weighting matrix ${\color{revisionblue}\Xi_T}\rightarrow_p {\color{revisionblue}\Xi}$.
For notational simplicity, let $\theta_0=[a\ B]$, $\widehat{\theta}=[\widehat{a}\ \widehat{B}]$, and $x_t=(1,Y_t')'$.
The estimator $\widehat{\theta}$ is obtained using all $T$ pre-treatment periods.
For any $\theta=[a_\theta\ B_\theta]\in\mathbb{R}^{N\times(N+1)}$, define
$M(\theta)=(I-B_\theta)'(I-B_\theta)$ and
\[
G(\theta)=A\{A'M(\theta)A\}^{-1}A'(I-B_\theta)'.
\]
Thus, $G(\theta_0)=G$ and $G(\widehat{\theta})=\widehat{G}$.
By Theorem \ref{unbiasedness}, $P$ is asymptotically equivalent to $u_{T+1}'G'C'{\color{revisionblue}\Xi} CGu_{T+1}$.
To construct critical values, define
\begin{equation*}
	\widehat{P}_t(\theta)=(Y_t-\theta x_t)'G(\theta)'C'{\color{revisionblue}\Xi_T}CG(\theta)(Y_t-\theta x_t)
\end{equation*}
and its population version $P_t(\theta)=(Y_t-\theta x_t)'G'C'{\color{revisionblue}\Xi}CG(Y_t-\theta x_t)$.
For each $t\le T$, let $\widehat{\theta}^{(t)}=[\widehat{a}^{(t)}\ \widehat{B}^{(t)}]$ denote the estimator used to construct the $t$-th pre-treatment statistic, and define $\widehat{P}_t=\widehat{P}_t(\widehat{\theta}^{(t)})$.
The estimators $\widehat{\theta}$ and $\widehat{\theta}^{(t)}$ need to be consistent for $\theta_0$ in the sense of Assumption \ref{spillover assumption}(d).
A simple choice is the original full-sample system of synthetic-control estimators, which sets $\widehat{\theta}^{(t)}=\widehat{\theta}$ for every $t$.
Another choice is the leave-one-pre-treatment-period-out construction, which re-estimates $\widehat{\theta}^{(t)}$ after omitting period $t$.
This construction also recomputes $G(\widehat{\theta}^{(t)})$ from the corresponding $\widehat{B}^{(t)}$ after omitting period $t$.
\cite{Andrews2003} and \cite{Andrews2006} suggest using the leave-one-out estimator for better finite sample performance.
For a significance level of $\tau$, we reject $H_0$ if $P$ is larger than the $(1-\tau)$-quantile of $\{\widehat{P}_t \}_{t=1}^T$. 

To establish the validity of the proposed test, let $P_\infty=P_1(\theta_0)$. Define
	$\widehat{F}_{P,T}(x)=T^{-1}\sum_{t=1}^{T}\mathbbm{1}\{\widehat{P}_t\le x\},$
and let $F_P(x)$ be the distribution function of $P_\infty$. Finally, let $\widehat{q}_{P,1-\tau}=\inf \{x\in \mathbb{R}:\widehat{F}_{P,T}(x)\ge 1-\tau \}$, and $q_{P,1-\tau}$ be the $(1-\tau)$-quantile of $P_\infty$. 
The assumptions and validity of the proposed testing procedure are given as follows.

\begin{assum}
	\label{spillover assumption}
	\emph{(a)} Assumption 1 holds;
	
	\emph{(b)} $\{ u_t\}_{t\ge 1}$ is ergodic and $E[\Vert u_t\Vert^2]<\infty$;
	
	\emph{(c)} There exists a non-random sequence of positive definite matrices $\{D_T \}_{T\ge1}$ such that $\max_{t\le T+1}\Vert D_T^{-1}x_t\Vert =O_p(1)$;
	
	\emph{(d)} The estimators satisfy $\Vert (\widehat{\theta}-\theta_0)D_T\Vert_F=o_p(1)$, $\max_{t=1,\dots,T}\Vert (\widehat{\theta}^{(t)}-\theta_0)D_T\Vert_F=o_p(1)$, and $\max_{t=1,\dots,T}\Vert \widehat{B}^{(t)}-B\Vert_F=o_p(1)$, where $\Vert\cdot\Vert_F$ is the Frobenius norm;
	
	\emph{(e)} The distribution function of $P_1(\theta_0)$ is continuous and increasing at its $(1-\tau)$-quantile;
	
	\emph{(f)} ${\color{revisionblue}\Xi_T}\rightarrow_p {\color{revisionblue}\Xi}$ as $T\rightarrow \infty$.
\end{assum}

Assumption \ref{spillover assumption}(b)-(f) are similar to Assumption \ref{no spillover assumption} as well as those in \cite{Andrews2003}.
Part (d) requires scaled consistency of $\widehat{\theta}$ and uniform scaled consistency of the estimators $\widehat{\theta}^{(t)}$ used in the pre-treatment statistics. It also requires uniform consistency of $\widehat{B}^{(t)}$, which ensures uniform consistency of $G(\widehat{\theta}^{(t)})$.

\begin{thm}
	\label{thm_spillover}
	Suppose Assumption \ref{spillover assumption} holds. Then, under $H_0$, as $T\rightarrow \infty$, 	\\
	\emph{(a)} $P\rightarrow_d P_\infty$,\\
	\emph{(b)} $\widehat{F}_{P,T}(x)\rightarrow_p F_P(x)$ for all $x$ in a neighborhood of $q_{P,1-\tau}$,   \\
	\emph{(c)} $\widehat{q}_{P,1-\tau}\rightarrow_p q_{P,1-\tau}$, \\
	\emph{(d)} $\Pr(P>\widehat{q}_{P,1-\tau})\rightarrow \tau$. 
\end{thm}

Just like Theorem \ref{thm_no_spillover}, Theorem \ref{thm_spillover} shows that we can approximate the null distribution of $P$ using its pre-treatment counterparts. 
Part (d) shows the asymptotic validity of the test proposed in this section. 

Similar to Theorem \ref{thm_no_spillover}, we can extend the results to derive power curves. Under the alternative hypothesis \(H_1\),
$$\Pr(P > \widehat{q}_{P,1-\tau}) \rightarrow \Pr\left(((C\alpha-d)+CGu_1)' {\color{revisionblue}\Xi} ((C\alpha-d)+CGu_1)> q_{P,1-\tau}\right),$$
where \(q_{P,1-\tau}\) is the \((1-\tau)\)-quantile of \(u_1' G' C' {\color{revisionblue}\Xi} C G u_1\). The right hand side is estimable from the pre-treatment data, allowing us to approximate the power curves for the test.

We show the relevance of the factor model in this context by the following lemma:

\begin{lemma}
	\label{lemma inference spillover}
	For the full-sample construction $\widehat{\theta}^{(t)}=\widehat{\theta}$ and $G(\widehat{\theta}^{(t)})=\widehat{G}$, suppose that $A'MA$ is non-singular and the distribution function of $P_1(\theta_0)$ is continuous and increasing at its $(1-\tau)$-quantile. {\color{revisionblue}Then, under either Condition ST or Condition CO, Assumption \ref{spillover assumption} is satisfied with $\Xi_T=I$. It is also satisfied with the inverse-covariance weight
	\[
		\Xi_T=(C\widehat{G}(T^{-1}\sum_{t=1}^T\widehat{u}_t\widehat{u}_t')\widehat{G}'C')^{-1},
	\]
	provided that $CGE[u_tu_t']G'C'$ is positive definite.}
\end{lemma}

\section{Discussion}
\label{section_discussion}
\subsection{Structure misspecification}
\label{section_misspecification}

In this section, we first characterize the bias resulting from spillover structure misspecification and demonstrate that the misspecification bias in our proposed method is linear in  the overlooked spillover effects. In contrast, the bias in the usual synthetic control method depends on all spillover effects. 
Subsequently, we introduce a statistic, $\kappa_A$, to assess spillover specification and propose an associated test.

\subsubsection{Misspecification bias}
\label{section_misSpeBiasCharacterization}

We follow Example \ref{nonparametric spillover} and assume that the spillover structure specifies the range of exposed units. Assume the effect vector is $\alpha=(\alpha_1,\dots,\alpha_{k},\dots,\alpha_{k+p},0,\dots,0)'$. 
The (asymptotic) bias for the usual synthetic control method (SCM) is
\begin{equation}
	\label{eq_SCM_misspecification_bias}
	\delta_{1,SCM}=-\sum_{i=2}^{k+p}b_{1,i}\alpha_i.
\end{equation}

When applying the proposed method in the paper, suppose we include only units 2 through $k$ in the spillover structure. The bias of our estimator for the entire vector $\alpha$ then becomes
\begin{equation}
	\delta_{SP} = (A(A'MA)^{-1}A'(I-B)'(I-B)-I) (0,\dots,0,\alpha_{k+1},\dots,\alpha_{k+p},0,\dots,0)'. \label{eq_bias} 
\end{equation}
Consequently, the bias for the treatment effect estimator is a linear combination of the omitted spillovers
$$\delta_{1,SP}=\sum_{i=k+1}^{k+p} c_i \alpha_i,$$
where each $c_i$ is determined by the first entry of Equation \eqref{eq_bias}.

For concreteness, consider the case where units 2 and 3 are affected by spillover effects, but unit 3 is omitted in our method. The biases are then given by
$$\delta_{1,SCM}=-b_{1,2}\alpha_2 - b_{1,3}\alpha_3, \quad \text{and} \quad \delta_{1,SP}=-\frac{\mathrm{det}\left( [\tilde{b}_1,\tilde{b}_2]'[\tilde{b}_2,\tilde{b}_3] \right) }{\mathrm{det}(A'MA)}\alpha_3, $$
where $\tilde{b}_i=e_i-(b_{1,i},b_{2,i},\dots,b_{N,i})'$ and $e_i$ is the unit vector with one in the $i$-th entry and zeros elsewhere.
There is generally no guarantee that either $\delta_{1,SCM}$ or $\delta_{1,SP}$ will be smaller. 
We suggest that the researcher be conservative about choosing the structure.
That is, if a certain unit is suspected to be affected by spillover effects, it should be included in the spillover structure.

\subsubsection{$\kappa_A$-statistic}
\label{section_kappaA_statistic}

In addition to results from the previous section, we note that the goodness of fit is informative about the accuracy of the spillover structure.
To see this, define
$$\kappa_A = \Vert (I-\widehat{B})(Y_{T+1} - \widehat{\alpha}) - \widehat{a} \Vert$$
as a function of $A$, where $\widehat{\alpha}$ is the estimated effect vector under the assumption that $A$ correctly specifies the spillover effects. 
This statistic is useful in evaluating the correctness of spillover specifications, which is motivated by the observation that $\kappa_A$ is small when $A$ is correctly specified.
We formalize this idea by the proposition below.

Given $A$, define $\Gamma_A = (I-B)A(A'(I-B)'(I-B)A)^{-1}A'(I-B)'$, the projection onto the span of columns of $(I-B)A$. The sample analog is defined as $\widehat{\Gamma}_A = (I-\widehat{B})A(A'(I-\widehat{B})'(I-\widehat{B})A)^{-1}A'(I-\widehat{B})'$. Andrews' test can be applied to test for correct spillover structure specification. Similar to Section \ref{spillover effect test}, we approximate the null distribution of $\kappa_A$ using $\{\Vert (I-\widehat \Gamma_A)\widehat u_t\Vert\}_{t=1}^T$. Define
$\widehat{q}_{\kappa, 1-\tau}^A = {\color{revisionblue}\inf}\{x \in \mathbb{R} : \widehat F_{\kappa, T}^A(x) \geq 1-\tau\},$
where $\widehat F_{\kappa, T}^A(x) = T^{-1} \sum_{t=1}^T \mathbbm{1}\{\Vert (I-\widehat \Gamma_A)\widehat u_t\Vert \leq x\}$. We reject the null hypothesis that $A$ correctly specifies the spillover effects if $\kappa_A > \widehat q_{\kappa, 1-\tau}^A$. Let $q_{\kappa, 1-\tau}^A$ be the $(1-\tau)$-quantile of $\Vert (I-\Gamma_A)u_1\Vert$.

\begin{prop} 
	\label{prop_test_for_A}
	Suppose Assumption \ref{spillover assumption} holds with part (e) replaced by the requirement that the distribution function of $\Vert(I-\Gamma_A)u_1\Vert$ is continuous and strictly increasing at its $(1-\tau)$-quantile. Then, when $A$ is correctly specified, $\Pr(\kappa_A > \widehat q_{\kappa, 1-\tau}^A) \rightarrow \tau$.
	More generally, for a fixed effect vector $\alpha$, if $\Vert(I-\Gamma_A)u_1 + (I-\Gamma_A)(I-B)\alpha\Vert$ has no point mass at $q_{\kappa, 1-\tau}^A$, then
	\[
	\Pr(\kappa_A > \widehat q_{\kappa, 1-\tau}^A) \rightarrow \Pr\bigl(\Vert(I-\Gamma_A)u_1 + (I-\Gamma_A)(I-B)\alpha\Vert > q_{\kappa, 1-\tau}^A\bigr).
	\]
\end{prop}

When $A$ is correctly specified, $(I-\Gamma_A)(I-B)\alpha = 0$, making $q_{\kappa, 1-\tau}^A$ precisely the $(1-\tau)$-quantile of $\Vert(I-\Gamma_A)u_{T+1}\Vert$. Conversely, $(I-\Gamma_A)(I-B)\alpha \neq 0$ generally when $A$ is not a correct specification, which enables the test to have power.{\color{revisionblue}\interfootnotelinepenalty=10000\footnote{\color{revisionblue}This test can be modified to form
 a pre-treatment diagnostic for the assumption in Condition CO that weights exist to reconstruct the loadings on the nonstationary factors. 
 Estimate $\widehat a$ and $\widehat B$ using periods $1,\ldots,T-1$, and let $\kappa=\Vert (I-\widehat B)Y_T-\widehat a\Vert$. Compare the resulting statistic with the empirical $(1-\tau)$-quantile of $\{\Vert(I-\widehat B)Y_t-\widehat a\Vert\}_{t=1}^{T-1}$. Since period $T$ is untreated, this checks a pre-treatment implication of the reconstruction assumption under the other maintained conditions. 
 We thank an anonymous referee for suggesting it. 
 }}

{\color{revisionblue}Rejection therefore provides evidence against the prespecified structure, while nonrejection does not validate it. This distinction is especially important with one or few post-treatment periods, when the diagnostic may have limited power. The researcher should choose $A$ using institutional, geographic, or other domain knowledge and use $\kappa_A$ as a specification diagnostic and robustness aid rather than as a model-selection rule. If several structures are substantively plausible, we recommend prespecifying a small set and reporting the fixed-$A$ estimates and inference separately for each. When multiple post-treatment periods are available and the spillover structure is stable, consistent discrimination among a finite set of well-separated candidate structures can be possible. Section \suppref{sec_multi_post_period}{S.2.3} provides the details.}

{\color{revisionblue}
\subsection{Comparison with alternative estimators}
\label{section pure donor}
}
{\color{revisionblue}
In this section, we consider four estimators of treatment effects in the presence of spillovers, including our method (SP), the pure-donor method (PD), the inclusive synthetic control method (iSCM) of \citet{DiStefano2020}, and the penalized synthetic control method (PSCM) used by \citet{Grossi2020}.

\subsubsection{A common pure donor representation}
\label{section_pd_representation}

We focus on the limited-range structure in Example \ref{nonparametric spillover}.
Let \(\mathcal E\) contain the treated unit and all potentially exposed units, ordered with the treated unit first, and let \(\mathcal P\) contain the pure donors.
For a vector \(v\), \(v_{\mathcal S}\) denotes its subvector indexed by \(\mathcal S\).
For a matrix \(Q\), \(Q_{\mathcal S\mathcal T}\) selects rows in \(\mathcal S\) and columns in \(\mathcal T\), while \(Q_{\mathord\cdot\mathcal T}\) selects all rows and the columns in \(\mathcal T\).
In particular, \(Y_{\mathcal S,t}\) collects the outcomes of units in \(\mathcal S\) at time \(t\), while \(y_{1,t}\) is the treated unit's outcome.

The conventional pure-donor (PD) estimator excludes the other units in \(\mathcal E\) and estimates the treated unit's direct effect as
\[
\widehat\alpha_{1}^{PD}
=y_{1,T+1}-\widehat a^{PD}-(\widehat b^{PD})'Y_{\mathcal P,T+1}.
\]
Here, \(\widehat a^{PD}\) is the fitted intercept and \(\widehat b^{PD}\) is the \(\lvert\mathcal P\rvert\)-vector of pure-donor weights.
The donor weights are nonnegative and sum to one.
By using only pure donors, PD avoids contamination from spillover-affected controls and, under the usual stationarity, centering, and first-stage convergence conditions, yields an asymptotically unbiased estimator of the direct effect.
After removing the fitted intercept, its counterfactual prediction is therefore a convex combination of the pure-donor outcome paths.
In a factor model, this restricts the matched time-varying component to the convex hull of the pure-donor loadings.
PD can consequently have poor pretreatment fit and a large counterfactual prediction error when the treated unit is not well approximated by that convex hull.

Our spillover-adjusted estimator (SP) also has a pure-donor representation under this specification.
Let \(A\) be the \(N\times\lvert\mathcal E\rvert\) matrix that maps one unrestricted effect for each unit in \(\mathcal E\) into the \(N\)-vector of unit-level effects.
We specialize the limited-range structure to
\[
A=
\begin{bmatrix}
I_{\lvert\mathcal E\rvert}\\
0_{\lvert\mathcal P\rvert\times\lvert\mathcal E\rvert}
\end{bmatrix}.
\]
Let \(\widehat a\) be the \(N\)-vector of first-stage fitted intercepts and \(\widehat B\) the \(N\times N\) matrix of first-stage SCM weights, as defined by equation \eqref{optimization}.
Define \(\widehat R=I_N-\widehat B\), and let \(\widehat R_{\mathord\cdot\mathcal E}\) and \(\widehat R_{\mathord\cdot\mathcal P}\) collect the columns indexed by \(\mathcal E\) and \(\mathcal P\), respectively.
Assume \(\widehat R_{\mathord\cdot\mathcal E}\) has full column rank.
Its left inverse is
\[
\widehat R_{\mathord\cdot\mathcal E}^{*}
=\left(\widehat R_{\mathord\cdot\mathcal E}'\widehat R_{\mathord\cdot\mathcal E}\right)^{-1}
\widehat R_{\mathord\cdot\mathcal E}'.
\]
Equation \eqref{gamma estimation} then gives the \(\lvert\mathcal E\rvert\)-vector of SP effect estimates
\[
\widehat\alpha_{\mathcal E}^{SP}
=Y_{\mathcal E,T+1}
-\widehat R_{\mathord\cdot\mathcal E}^{*}\widehat a
+\widehat R_{\mathord\cdot\mathcal E}^{*}\widehat R_{\mathord\cdot\mathcal P}Y_{\mathcal P,T+1}.
\]
The direct-effect component can be written as
\[
\widehat\alpha_1^{SP}
=y_{1,T+1}-\widehat a^{SP}-(\widehat b^{SP})'Y_{\mathcal P,T+1},
\]
where \(\widehat a^{SP}\) is the first entry of \(\widehat R_{\mathord\cdot\mathcal E}^{*}\widehat a\), and \((\widehat b^{SP})'\) is the first row of \(-\widehat R_{\mathord\cdot\mathcal E}^{*}\widehat R_{\mathord\cdot\mathcal P}\).
Thus, \(\widehat a^{SP}\) is the effective intercept and \(\widehat b^{SP}\) is the \(\lvert\mathcal P\rvert\)-vector of induced pure-donor weights.
These induced weights sum to one when the first-stage SCM weights do, but they may be negative.
SP can therefore extrapolate beyond the pure-donor convex hull and into its affine hull.
This extra flexibility can reduce counterfactual prediction error when PD has poor support.
SP is not simply an unconstrained regression on the pure donors because its induced weights are constructed from the full system of first-stage SCM weights in \eqref{optimization}, which are regularized through nonnegativity and sum-to-one constraints.

Following \citet{DiStefano2020}, our inclusive synthetic control method (iSCM) benchmark shares the same affine first-stage SCM fits as SP but uses a different second stage.
The matrices \(\widehat B_{\mathcal E\mathcal E}\) and \(\widehat B_{\mathcal E\mathcal P}\) contain the rows of \(\widehat B\) indexed by \(\mathcal E\) and the columns indexed by \(\mathcal E\) and \(\mathcal P\), respectively, while \(\widehat a_{\mathcal E}\) is the corresponding intercept subvector.
Let \(\widehat H_{\mathcal E}=I_{\lvert\mathcal E\rvert}-\widehat B_{\mathcal E\mathcal E}\).
When \(\widehat H_{\mathcal E}\) is nonsingular, iSCM uses only the equations for units in \(\mathcal E\) and obtains
\[
\widehat\alpha_{\mathcal E}^{iSCM}
=Y_{\mathcal E,T+1}
-\widehat H_{\mathcal E}^{-1}\widehat a_{\mathcal E}
-\widehat H_{\mathcal E}^{-1}\widehat B_{\mathcal E\mathcal P}Y_{\mathcal P,T+1}.
\]
Its direct-effect estimator is again pure-donor, with induced weights given by the first row of \(\widehat H_{\mathcal E}^{-1}\widehat B_{\mathcal E\mathcal P}\).
Under the usual simplex first stage, these weights are nonnegative and sum to one.
Thus, holding the first stage fixed, iSCM uses \(\lvert\mathcal E\rvert\) affected-unit equations to recover \(\lvert\mathcal E\rvert\) unrestricted effects, whereas SP uses all \(N\) equations.
Using only the affected-unit equations can therefore be less precise.
This shared affine first stage differs from the no-intercept simplex specification displayed by \citet{DiStefano2020} and isolates the methods' second-stage distinction.
In the simulation below, iSCM has a higher RMSE than SP once the pure donors are sufficiently far from the treated unit.

Finally, \citet{Grossi2020} apply the penalized synthetic control method (PSCM) of \citet{Abadie2021} under prespecified partial-interference clusters.
Units outside the treated cluster are pure donors under that structure, so the direct-effect estimator remains PD-like.
PSCM chooses donor weights using unit- and cluster-level predictors. It seeks a weighted donor average that fits the treated unit while penalizing differences between the treated unit and individual donors.
This regularization is particularly attractive with many disaggregated units, where ordinary SCM weights may be nonunique, although its performance depends on the cluster structure and penalty choice.

Consequently, SP, PD, iSCM, and PSCM all have the common form
\[
\widehat\alpha_1^m
=y_{1,T+1}-\widehat a^m-(\widehat b^m)'Y_{\mathcal P,T+1},
\]
where \(m\in\{SP,PD,iSCM,PSCM\}\) indexes the estimator, \(\widehat a^m\) is its effective intercept, and \(\widehat b^m\) is its \(\lvert\mathcal P\rvert\)-vector of pure-donor weights.
This common representation also makes the relevant risk comparison transparent.
Consider stationary untreated outcomes with finite second moments.
For an estimator whose fitted intercept and donor weights converge to fixed limits under the usual stationarity and centering conditions, asymptotic unbiasedness reduces its limiting RMSE comparison to a variance comparison.
For such an estimator \(m\), let \(b_m\in\mathbb R^N\) denote its limiting donor-weight vector after setting its entries outside \(\mathcal P\) to zero.
Set \(h_m=(1,0,\dots,0)'-b_m\), and let \(\Omega_y=\operatorname{Var}[Y_t(0)]\).
The limiting variance is
\[
V_m=h_m'\Omega_yh_m.
\]
SP, PD, and our iSCM share this formula but use different limiting pure-donor weights.
It also applies to PSCM when its tuning is fixed, asymptotically deterministic, or based on an independent validation sample.
There is no uniform variance ranking.
Thus, we will compare the estimators' performance through simulation.

%

\subsubsection{Monte Carlo RMSE comparison}
\label{section_estimator_comparison_simulation}

This section studies the finite-sample performance of SP and the alternative estimators.
We report RMSE for the stationary design here.
The supplement reports bias, standard deviation, and RMSE for both the stationary and \(\mathcal{I}(1)\) designs, which yield qualitatively similar conclusions.

\paragraph{Monte Carlo design.}
We consider a three-factor model
\[
Y_{it}(0)=\lambda_i(r)'f_t+\varepsilon_{it}.
\]
The common factors follow the stationary processes
\begin{align*}
f_{1t}&=0.5f_{1,t-1}+\nu_{1t},\\
f_{2t}&=1+\nu_{2t}+0.5\nu_{2,t-1},\\
f_{3t}&=0.5f_{3,t-1}+\nu_{3t}+0.5\nu_{3,t-1}.
\end{align*}
The innovations \(\nu_{kt}\) are independent standard normal draws across factors and time.
We initialize the processes from their exact stationary distributions.
The implementation generates the 50 pretreatment periods and the single post-treatment period used in the experiment.
The idiosyncratic errors \(\varepsilon_{it}\) are independent standard normal draws across units and time and are independent of the factor innovations.

The design has \(N=51\) units, indexed from 1 to 51 and ordered as one treated unit, 13 potentially exposed units, and 37 pure donors.
We generate 50 pretreatment periods and one post-treatment period, in which we evaluate the treatment-effect estimate.
The loading vectors follow
\[
\lambda_i(r)=(-6,0,0)'+\mu_i(r)d_i.
\]
Let \(c_i=(i-1)\bmod 3\). The three unit-length directions satisfy
\[
d_i=
\left(
\sqrt{1-0.2^2},
0.2\cos(2\pi c_i/3),
0.2\sin(2\pi c_i/3)
\right)'.
\]
We set \(\mu_1(r)=0\), assign the exposed-unit values in unit order as 13 evenly spaced points over \([-0.15,0.15]\), and assign the pure-donor values in unit order as \(\mu_j(r)=rq_j\), where the 37 values \(q_j\) are evenly spaced over \([3,6]\).
These loading vectors are fixed across Monte Carlo replications.
We use
\[
r\in\{0,0.10,0.20,0.30,0.35,0.40,0.45,0.50,0.55,0.60,0.70,0.80,1.00\}.
\]
At \(r=0\), the pure donors have the same loading as the treated unit.
For small positive values of \(r\), the pure donors remain close matches and provide favorable support for PD.
Because each \(d_i\) has unit length, at every positive value of \(r\), the distances between the pure-donor and treated loading vectors range from \(3r\) to \(6r\).
Increasing \(r\) therefore moves every pure donor farther from the treated unit, making a convex pure-donor match more difficult and increasing the value of SP's flexibility.
This dimension is important because the controls most similar to the treated unit may also be the ones exposed to spillovers, leaving a less comparable pure-donor pool.
We therefore vary \(r\) to examine how each estimator's RMSE responds to this deterioration in support.

In the post-treatment period, the treated unit receives an effect of 5 and every exposed unit receives a spillover effect of 3.

\paragraph{Estimators and evaluation.}
We compare five estimators over 5,000 independently generated replications.
SP follows the limited-range specification in Example \ref{nonparametric spillover}.
PD applies conventional SCM to the treated unit using only the 37 pure donors.
Following \citet{DiStefano2020}, the iSCM benchmark uses the same affine first-stage fits as SP but applies its affected-unit correction to 14 unrestricted effects.
PSCM is a practical adaptation of \citet{Grossi2020}. The treated and exposed units form one cluster, the pure donors form three fixed clusters, and own-unit and leave-one-out cluster-average pretreatment outcomes enter a no-intercept penalized SCM.
Its penalty is selected by leave-one-pure-control-out prediction using pure-control post-treatment outcomes.
In addition, we include PD-OLS for reference.
It fits an intercept and unrestricted pure-donor coefficients, which may be negative and need not sum to one.
For estimator \(m\), we report
\[
\operatorname{RMSE}_m(r)
=\left\{\frac{1}{5{,}000}\sum_{s=1}^{5{,}000}
\left(\widehat\alpha_{1}^{m,(s)}(r)-5\right)^2\right\}^{1/2}.
\]
Here, \(s\) indexes replications and \(m\in\{SP,PD,iSCM,PSCM,PD\text{-}OLS\}\).

\begin{figure}[H]
\centering
\captionsetup{font={small,color=revisionblue},labelfont={color=revisionblue}}
\includegraphics[width=.685\textwidth]{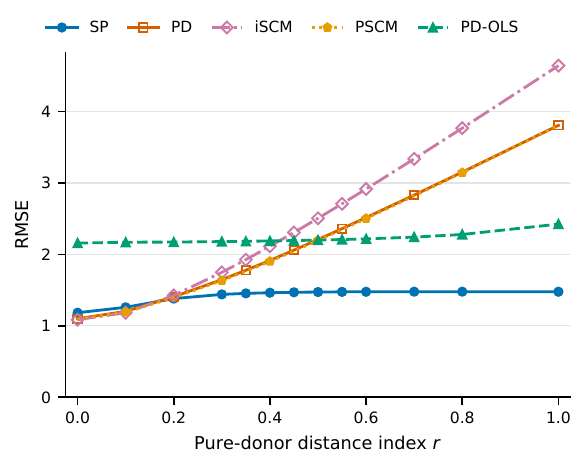}
\caption{Treatment-effect RMSE. SP denotes the proposed spillover-adjusted estimator. PD denotes conventional SCM restricted to pure donors. iSCM denotes our adapted inclusive-SCM benchmark following \citet{DiStefano2020}. PSCM denotes the penalized synthetic-control benchmark adapted from \citet{Grossi2020}. PD-OLS denotes the pure donor method with weights estimated by OLS instead of simplex restrictions. Each curve reports post-treatment-period RMSE over 5,000 Monte Carlo replications under the stationary dependent-factor design with 50 pretreatment periods, one post-treatment period, one treated unit, 13 exposed units, and 37 pure donors.}
\label{fig:estimator_comparison_rmse}
\end{figure}


\paragraph{Results.}
Figure \ref{fig:estimator_comparison_rmse} and Table \ref{tab:estimator_comparison_rmse} show how performance changes as pure-donor support deteriorates.
Near \(r=0\), the four regularized synthetic-control estimators have similar RMSEs, with SP slightly less precise, while PD-OLS is notably worse.
As \(r\) increases, SP's RMSE changes little.
The convex-weight alternatives deteriorate substantially, while PD-OLS increases gradually.
SP has the lowest RMSE over most of the displayed range.
At large \(r\), iSCM performs worst, while PD and PSCM are nearly indistinguishable.

\begin{table}[H]
\centering
\color{revisionblue}
\captionsetup{font={small,color=revisionblue},labelfont={color=revisionblue}}
\caption{Treatment-effect RMSE at selected values of the pure-donor distance index.}
\label{tab:estimator_comparison_rmse}
\renewcommand{\arraystretch}{1.15}
\small
\begin{threeparttable}
\begin{tabular}{lcccccc}
\hline
\hline
& \multicolumn{6}{c}{Pure-donor distance index ($r$)} \\
\cline{2-7}
Method & 0 & 0.2 & 0.4 & 0.6 & 0.8 & 1 \\
\hline
SP     & 1.184 & \textbf{1.382} & \textbf{1.464} & \textbf{1.477} & \textbf{1.479} & \textbf{1.478} \\
PD     & 1.099 & 1.408 & 1.916 & 2.515 & 3.152 & 3.807 \\
iSCM   & \textbf{1.087} & 1.427 & 2.114 & 2.916 & 3.768 & 4.642 \\
PSCM   & 1.094 & 1.396 & 1.903 & 2.503 & 3.145 & 3.807 \\
PD-OLS & 2.158 & 2.173 & 2.189 & 2.218 & 2.279 & 2.424 \\
\hline
\end{tabular}
\begin{tablenotes}
\small
\item Notes: Each entry reports post-treatment-period RMSE over 5,000 Monte Carlo replications. The lowest RMSE in each column is in bold.
\end{tablenotes}
\end{threeparttable}
\end{table}

These patterns are consistent with how the estimators trade off regularization and extrapolation.
iSCM uses only the affected-unit equations in its second stage, and its induced pure-donor weights remain a convex combination.
This combination helps explain its loss of precision when the treated unit lies far from the pure-donor convex hull.
PD and PSCM are similar in this relatively low-dimensional aggregate design, where \(T\) and \(N\) are approximately equal and PSCM's additional penalization and cluster predictors provide little improvement over PD.
SP combines two features that matter in this setting.
Its first-stage simplex fits regularize estimation in the small-to-moderate data regime, while its induced pure-donor weights may be negative and can therefore extrapolate outside the convex hull.

The comparison with PD-OLS helps illustrate the roles of regularization and extrapolation.
PD-OLS can extrapolate, but it does not regularize the pure-donor coefficients.
It is therefore the least precise method when the pure donors closely resemble the treated unit.
As \(r\) grows, the benefit of extrapolation increasingly outweighs the cost of unregularized estimation, and PD-OLS eventually outperforms PD, PSCM, and iSCM.
It nevertheless remains less precise than SP, which combines first-stage regularization with extrapolation.

}

\subsection{Comparison with alternative inferential methods}
\label{discussion on other methods}
\suppressfloats[t]


{\color{revisionblue}
\subsubsection{Inferential methods}
\label{section_inference_theory_comparison}

In this section, we analyze the rationale of the inference method associated
with each of the four estimators compared in Section
\ref{section_estimator_comparison_simulation}.
Our paper proposes the SP estimator combined with Andrews' end-of-sample
instability test, developed in Section \ref{spillover effect test}, and we
write SP-A for that pairing.
We consider the conventional pure-donor estimator combined with the standard
in-space placebo test, where the placebo pool holds only the pure donors, since
those are the units the spillover does not reach (PD-P).
Andrews' test also applies to PD, as described in Section
\ref{no spillover effect test}, so we consider that pairing as well (PD-A).
\citet{DiStefano2020} propose the iSCM estimator together with a placebo test
adjusted for its affected-unit correction (iSCM-P).
\citet{Grossi2020} propose the PSCM estimator together with a residual
resampling test (PSCM-O).
We also discuss a corrected version of that test (PSCM-C), which we describe
below.

SP-A combines SP with Andrews' end-of-sample instability test.
The test approximates the null distribution using pretreatment time-series
variation, and its validity therefore relies on stationarity of the error
process.
As discussed in Section \ref{section_pd_representation}, SP uses the full
system of unit equations and can retain precision when the pure donors provide
weak support for the treated unit.
The Monte Carlo results below show that this precision translates into greater
power.

PD-P is the standard placebo test applied to the pure-donor estimator, and both
its statistic and its placebo pool are built from the units the spillover does
not reach.
Andrews' test applies to PD as well, as Section
\ref{no spillover effect test} describes, which gives PD-A.
Both procedures are based on PD and therefore inherit its estimation error.
When PD has a larger mean squared error than SP, it is less precise, and a
given direct effect is correspondingly more difficult to detect.
Section \ref{section_estimator_comparison_simulation} shows that this mean
squared error gap emerges when the pure donors provide weaker support for the
treated unit.
The Monte Carlo results below confirm the resulting power implication, as both
PD-based procedures become less powerful than SP-A when the pure donors become
less relevant to the treated unit.

iSCM-P backs out the post-treatment counterfactual outcomes of the treated and
spillover units by subtracting the estimated direct and spillover effects from
their observed outcomes.
It then pools these estimated counterfactuals with the observed post-treatment
outcomes of the pure donors to form the placebo null distribution.
\citet{DiStefano2020} provide no theoretical or simulation justification
for this procedure.
The two components of its null distribution pool differ in a fundamental way.
For the treated and spillover units, the untreated outcomes are counterfactuals
and must be estimated, so they contain estimation error.
For the pure donors, the untreated outcomes are observed directly because
these units receive neither the treatment nor a spillover.
Their placebo statistics may still contain synthetic-control fitting error,
but they do not contain the additional error from estimating their untreated
post-treatment outcomes.
Consequently, the two groups have different sampling distributions under the
null, and pooling them can provide a poor approximation to the null
distribution of the treated-unit statistic.
This mismatch is small when the counterfactuals are estimated precisely.
As the pure donors become less similar to the treated unit, however, the RMSE
of the iSCM effect estimates rises and the mismatch becomes more pronounced.
Consistent with this mechanism, the Monte Carlo results below show that iSCM-P
is approximately size-correct when the pure donors closely resemble the
treated unit but increasingly over-rejects as their distance from the treated
unit grows.

\citet{Grossi2020} introduce an inference method that resamples pure-control
residual paths and refits PSCM, which we call PSCM-O.
Namely, for each pure control \(i\in\mathcal P\), let \(\widehat Y_{i,t}\) denote its
penalized synthetic-control fit using pure controls in other clusters at the
selected penalty, and let
\(\widehat e_{i,t}=Y_{i,t}-\widehat Y_{i,t}\) denote the corresponding residual.
In draw \(b\), PSCM-O samples one residual path with replacement for each pure
control, independently across pure controls.
Writing \(J_i(b)\) for the sampled source unit, define
\(\widehat e_{i,t}^{*(b)}=\widehat e_{J_i(b),t}\).
PSCM-O constructs the pseudo donor panel as
\[
Y_{i,t}^{*(b)}=\widehat Y_{i,t}
    +\widehat e_{i,t}^{*(b)},
\qquad i\in\mathcal P,\quad t=1,\ldots,T+1.
\]
The same source unit \(J_i(b)\) is used over all periods, so the procedure
resamples entire residual paths rather than individual residuals.
Refitting the PSCM weights on the pretreatment portion of this pseudo panel
while holding the selected penalty fixed, and then applying those weights to
the pseudo post-treatment outcomes, gives \(\widehat Y_{1,T+1}^{*(b)}\).
The corresponding pseudo effect is
\[
\widehat\tau_O^{*(b)}
=Y_{1,T+1}-\widehat Y_{1,T+1}^{*(b)}.
\]
The empirical distribution of \(\widehat\tau_O^{*(b)}\) is used as the
reference distribution for inference on the original PSCM effect estimate.
\citet{Grossi2020} provide no theoretical or simulation-based justification for
this procedure.
The procedure also performs poorly in our simulations because the treated
outcome \(Y_{1,T+1}\) remains fixed across draws and its post-treatment
innovation is therefore omitted from the reference distribution.
PSCM-O over-rejects even when the pure-donor loadings match the treated loading
exactly.

We apply a simple correction to PSCM-O that improves its size control, at
least under this exact loading match, and call the resulting procedure PSCM-C.
PSCM-C leaves the pseudo donor panel and its PSCM refit unchanged.
It independently samples one additional residual from the pure controls to
represent the treated unit's untreated post-treatment innovation.
If \(\ell(b)\) denotes the source unit for this additional draw, the correction
can equivalently be written as constructing the pseudo treated outcome
\[
Y_{1,T+1}^{*(b),C}
=Y_{1,T+1}+\widehat e_{\ell(b),T+1}
\]
and the pseudo effect
\[
\widehat\tau_C^{*(b)}
=Y_{1,T+1}^{*(b),C}-\widehat Y_{1,T+1}^{*(b)}
=\widehat\tau_O^{*(b)}+\widehat e_{\ell(b),T+1}.
\]
The correction therefore adds a separate residual to the treated side rather
than altering the resampled residual paths used to construct the donor panel.
The PSCM point estimate is unchanged, and the interval is formed from the
corrected pseudo effects using the same pivotal method as PSCM-O.
This correction makes PSCM-C less prone to over-rejection than PSCM-O, and it
achieves approximately correct size when the pure donors provide good support
for the treated unit.
For the same reason as iSCM-P, however, its reference distribution becomes a
poor approximation when the pure donors differ substantially from the treated
unit, and PSCM-C then over-rejects.
}

{\color{revisionblue}
\subsubsection{Monte Carlo comparison of inferential methods}
\label{section_inference_comparison_simulation}

This section studies the finite-sample performance of our procedure and the
alternatives.
We report the stationary design and nominal rejection probabilities here.
The supplement reports the complete results for both the stationary and
\(\mathcal{I}(1)\) designs.

\paragraph{Monte Carlo design.}
We reuse the stationary untreated-outcome process and loading geometry of
Section \ref{section_estimator_comparison_simulation} and restrict attention to
\(r\in\{0,0.5,1\}\), which spans the range from pure donors that match the
treated loading exactly to pure donors that are far from it.
Each exposed unit again receives a spillover effect of 3.
The direct effect assigned to the treated unit now varies over
\(\{0,0.5,1,\dots,5\}\), so that the value 0 measures size and the remaining
values trace out power.
We use 5,000 Monte Carlo replications and a nominal level of five percent
throughout.

\paragraph{Procedures.}
We compare the six procedures defined in Section
\ref{section_inference_theory_comparison}, namely SP-A, PD-P, PD-A, iSCM-P,
PSCM-O, and PSCM-C. 
For both Andrews procedures, SP-A and PD-A, the post-treatment statistic uses the fit from all
pre-treatment periods, while each pre-treatment reference statistic is formed
after omitting that period. PD-A re-estimates $\widehat{\beta}_1^{(t)}$, and
SP-A refits the entire first-stage system. Thus, both use a leave-one-out reference
construction, which has been shown to improve finite-sample performance.
In both PSCM-O and PSCM-C, we use 999 resampling draws.

\paragraph{Results.}
Figure \ref{fig:inference_comparison_power} reports nominal rejection
probabilities under the stationary design. We focus on four patterns.

SP-A is the only procedure that holds size and retains power across the whole
range.
Its size stays close to the nominal level at every value of \(r\), while its
rejection probability against a direct effect of five falls only from 0.989 at
\(r=0\) to 0.917 at \(r=1\). This decline is modest relative to PD-P and PD-A,
whose rejection probabilities fall much more sharply.

PD-P behaves as the discussion above predicts.
It does not over-reject at any value of \(r\), but its rejection probability
against a direct effect of five falls from 0.988 to 0.203 as the pure donors move
away from the treated unit. PD-A remains close to the nominal size in this
stationary design, while its rejection probability falls from 0.996 to 0.298.
When weaker pure-donor support makes
the fit worse in both periods, the larger pretreatment error raises the
benchmark for rejection. Poorer fit therefore reduces power without by itself
causing over-rejection.

iSCM-P does distort size.
Its rejection frequency under the null rises from 0.066 at \(r=0\) to 0.584 at
\(r=1\).
Its second stage uses only the affected-unit equations, so its induced pure
donor weights stay inside the convex hull and the resulting prediction error
grows with \(r\). The placebo distribution pools estimated counterfactuals for
the affected units with directly observed pure-donor outcomes. The pure-donor
outcomes do not contain counterfactual-estimation error. As \(r\) rises, the
pooled distribution therefore fails to reproduce the growing estimation error
in the treated-unit statistic, which leads to over-rejection.

PSCM-O over-rejects everywhere, including at \(r=0\), where the pure donors
match the treated unit's loading exactly, with a rejection frequency of 0.593 against a nominal 0.05.
PSCM-C brings that frequency to 0.059, which supports the explanation that the
omitted treated shock causes the distortion under an exact loading match.
The correction only partially repairs the test at \(r=0.5\) and \(r=1\), where
the size falls to 0.258 and 0.475 rather than to 0.05.
Adding a pure-control residual cannot by itself correct the remaining
discrepancy associated with weaker pure-donor support.

\begin{figure}[H]
\centering
\captionsetup{font={small,color=revisionblue},labelfont={color=revisionblue}}
\includegraphics[width=\textwidth]{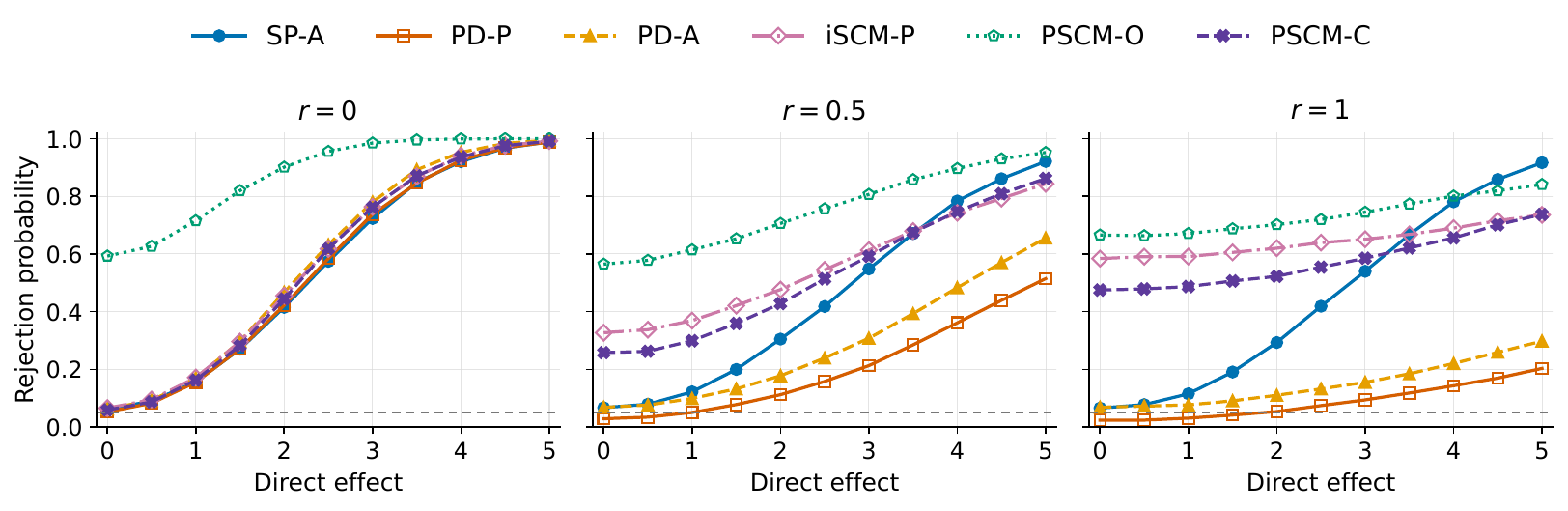}
\caption{Nominal rejection probability against the direct effect at three
values of the pure-donor distance index, for the six procedures of Section
\ref{section_inference_theory_comparison}. Each procedure uses its nominal
five-percent critical value. The dashed line marks the nominal level. Curves are
based on 5,000 Monte Carlo replications under the
stationary design of Section \ref{section_estimator_comparison_simulation}.}
\label{fig:inference_comparison_power}
\end{figure}
}

\section{Estimating the Effects of California's Proposition 99}
\label{empirical example}

\begin{figure}
	\centering
	\begin{subfigure}{.5\textwidth}
		\centering
		\includegraphics[width=\linewidth]{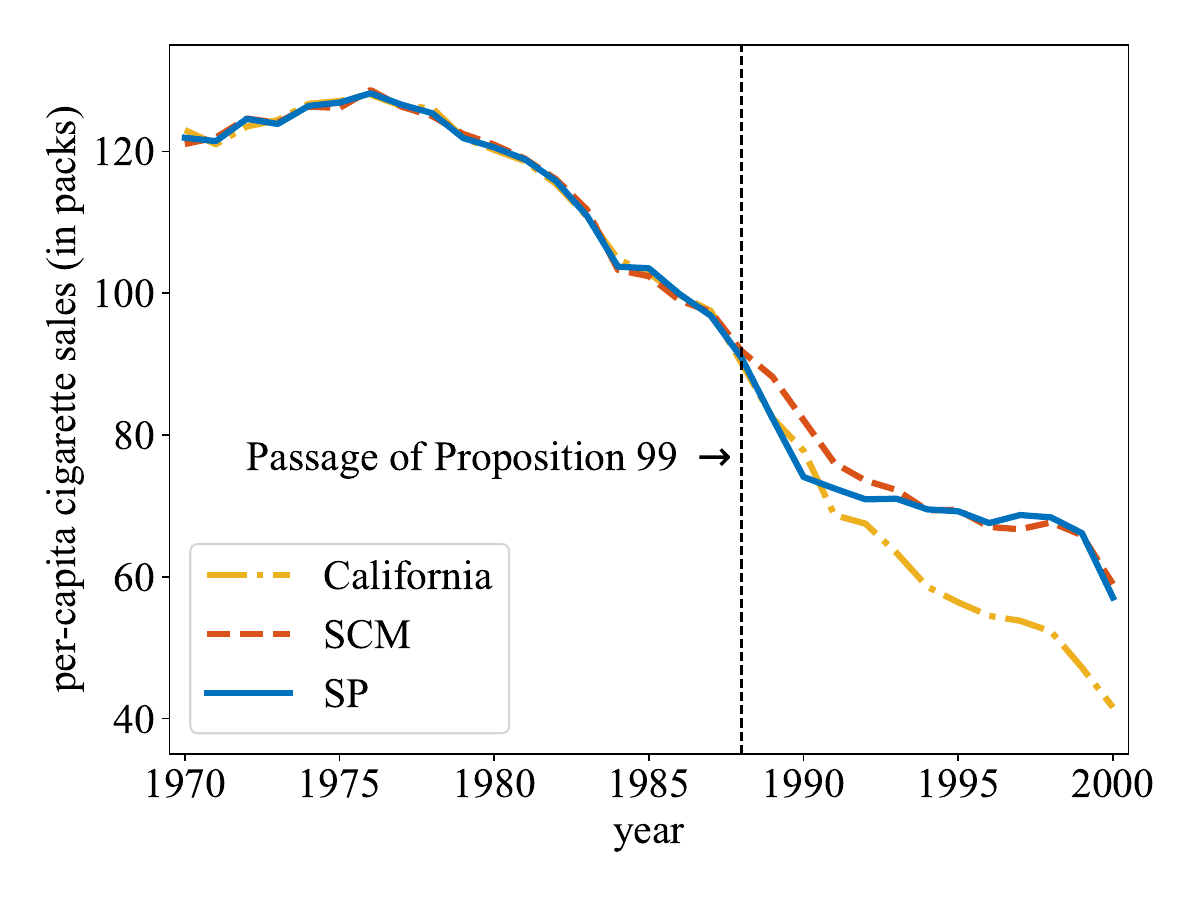}
		\caption{trends in per-capita cigarette sales}
		\label{fig_trend}
	\end{subfigure}%
	\begin{subfigure}{.5\textwidth}
		\centering
		\includegraphics[width=\linewidth]{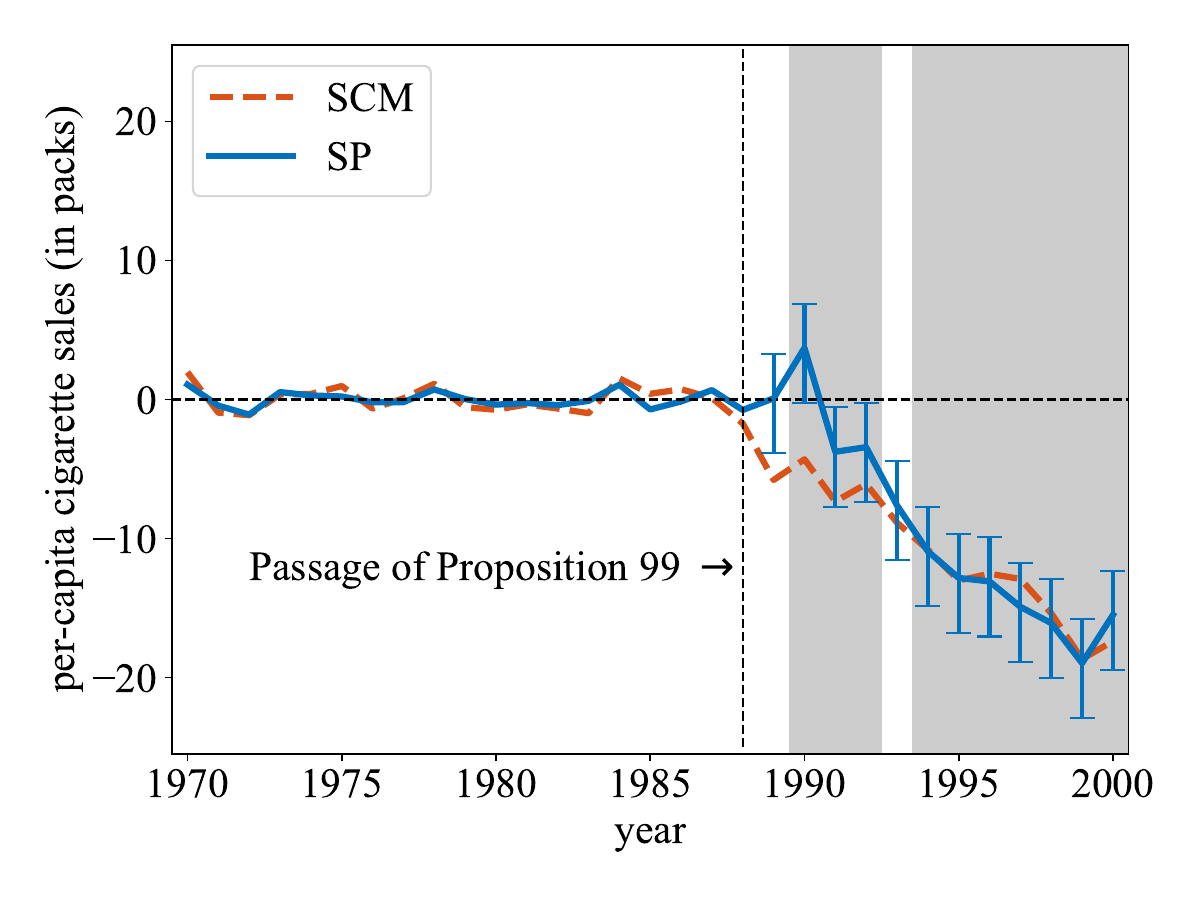}
		\caption{treatment effect estimates}
		\label{fig_gap}
	\end{subfigure}
	\begin{subfigure}{.5\textwidth}
	\centering
	\includegraphics[width=\linewidth]{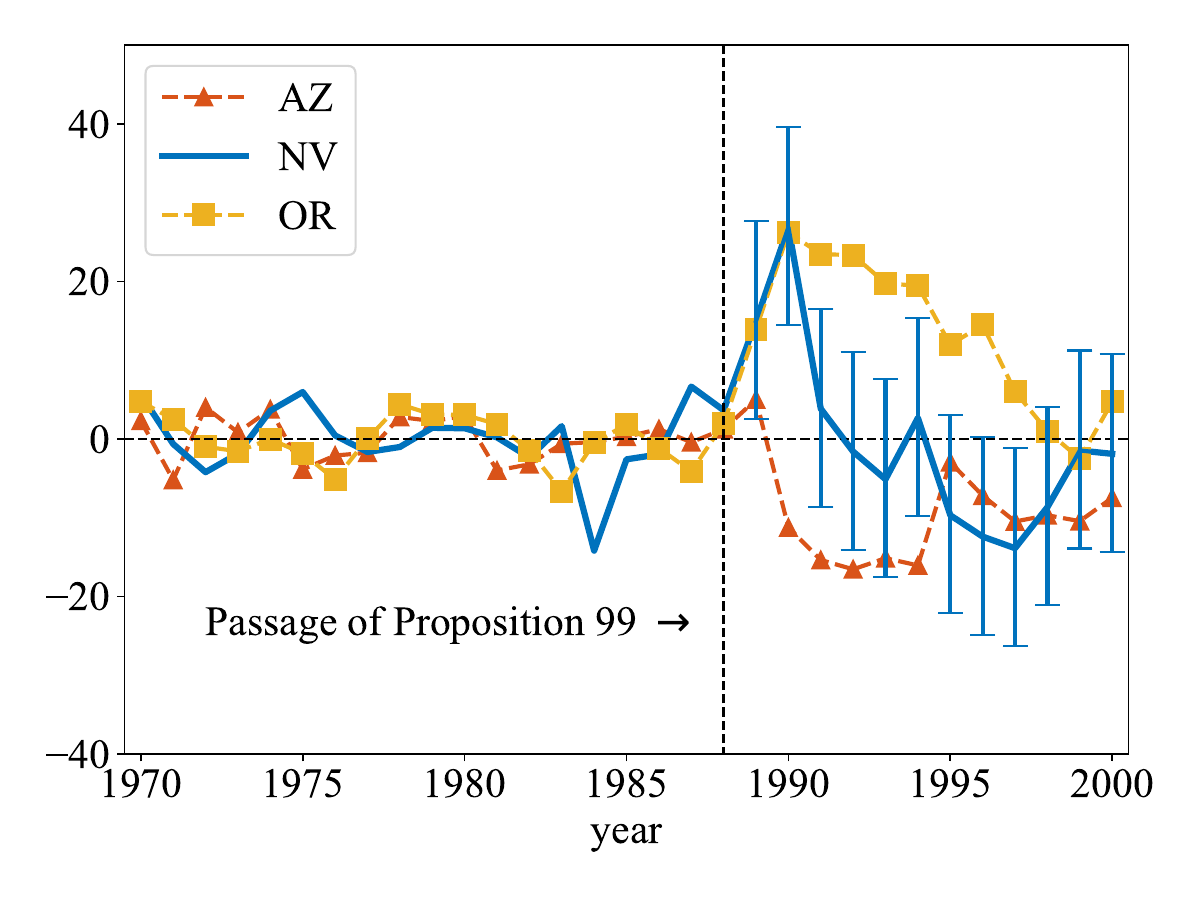}
	\caption{spillover effect estimates}
	\label{fig_spillover}
\end{subfigure}
	\caption{
		\small
Evaluation of Treatment and Spillover Effects on Cigarette Sales.
Panel (a) illustrates the trends in per-capita cigarette sales in California against two synthetic controls: one derived from the standard synthetic control method (SCM) and the other from our spillover-adjusted method (SP). The intervention onset is marked by the vertical dashed line at the passage of Proposition 99.
Panel (b) displays the estimated treatment effect on per-capita cigarette sales, showing the difference between actual sales in California and those predicted by both SCM and SP models. The shaded regions highlight the time periods where our test rejects the null hypothesis of no spillover effect at the 5\% significance level. Error bars represent 95\% confidence intervals for the SP estimates post-treatment.
Panel (c) presents the estimated spillover effects on cigarette sales in the neighboring states of Arizona (AZ), Nevada (NV), and Oregon (OR). {\color{revisionblue}Later estimates for Arizona and Oregon may also reflect their own tobacco-control policies and should not be interpreted as spillover effects alone.} For visual clarity, 95\% confidence intervals are included only for Nevada.
}
	\label{fig_prop99}
\end{figure}

To demonstrate our method, we apply it to the classic SCM example from \citeauthor{Abadie2010} (\citeyear{Abadie2010}, hereafter ADH), which looks at the effect of Proposition 99 on California cigarette {\color{revisionblue}sales}.
Furthermore, we compare the proposed method and the pure-donor one in a sensitivity analysis. 

\subsection{Set-up}

Proposition 99 intended to disincentivize smoking, which was primarily achieved by introducing a \$$0.25$ tax on each pack of cigarettes. 
{\color{revisionblue}However, traditional SCM is not guaranteed to produce an unbiased treatment effect estimator when post-treatment donor outcomes are contaminated. In this tobacco control program example, we are concerned about two sources of such contamination. The first is possible ``leakage''. A common problem with cigarette taxes (and other vice taxes like gambling and alcohol) is that measured local sales might fall as people move their purchasing behavior across legal boundaries, particularly in early years. We therefore allow for possible purchase spillovers in Arizona, Nevada, and Oregon, the three states neighboring California. The second source is the adoption of similar tobacco-control policies by other states during the post-treatment period. This type of own-policy contamination is closely related to the staggered-adoption setting studied by \cite{cao_synthetic_2025}. ADH took this concern into account by removing 12 states that experienced legislative changes in the ensuing years.}

We use annual per-capita cigarette {\color{revisionblue}sales} for the 50 states and the District of Columbia from 1970 to 2000.
In 1989 California enacted Proposition 99, so all periods from 1989 onwards are considered post-treatment periods. {\color{revisionblue}We replicate this program evaluation using the method introduced in previous sections and the structured specification in Example \ref{nonparametric spillover}.}
{\color{revisionblue}The synthetic control weights are constructed using only pre-treatment per-capita cigarette sales from 1970--1988 together with an unrestricted intercept, and no additional state-level covariates are used.}
{\color{revisionblue}The set of potentially contaminated states includes AK, AZ, DC, FL, HI, MA, MD, MI, NJ, NV, NY, OR, and WA. Among them, Arizona, Nevada, and Oregon border California and may experience purchase spillovers from Proposition 99. The ten non-border states adopted their own tobacco-control policies during the post-treatment period. These policies are separate interventions rather than spillovers from Proposition 99, although the resulting deviations enter our estimator through the same structured post-treatment deviation specification. The deviations are allowed to vary across states and time periods. For robustness, we consider an alternative specification that excludes the ten non-border states with own-policy contamination, as detailed in Section \suppref{sec_app_robustness}{S.5.1} of the Supplement. The results are qualitatively the same and the estimated treatment effects for California are virtually unchanged.}
We also perform hypothesis testing on both treatment effects and existence of spillover effects.
{\color{revisionblue}We hold the pre-treatment estimates and spillover structure fixed while estimating a separate effect for each post-treatment year. These are total event-time effects, which may incorporate earlier exposure, and the annual inference is pointwise. Section \suppref{sec_multi_post_period}{S.2.3} of the Supplement discusses the required stability and no-anticipation conditions.}

\subsection{Main results}

The results are shown in Figure \ref{fig_prop99}. 
{\color{revisionblue}The corresponding numerical results are reported in Section \suppref{sec_app_numbers}{S.5.2} of the Supplement.}
The standard synthetic control method that is similar to \cite{Abadie2010} is indexed by SCM.
Our method is SP. 
{\color{revisionblue}For the spillover structure used here, the smallest eigenvalue of the $14\times14$ matrix $A'\widehat{M}A$ is $0.22$, which is reasonably away from zero.}
{\color{revisionblue}The synthetic control weights are also uniquely determined here.}
Figure \ref{fig_trend} shows the ``synthetic California'' and \ref{fig_gap} elaborates on  this by specifically looking at the estimated treatment effects. 
Figure \ref{fig_spillover} plots the estimated spillover effects for the three neighboring states of California. 
The error bars denote 95\% confidence intervals that are built by inverting the test proposed in Section \ref{spillover effect test}. 
{\color{revisionblue}The confidence intervals reflect uncertainty in our spillover-adjusted estimator, which depends on a weighted combination of prediction errors across units. They are constructed using the pre-treatment distribution of this combination, rather than the individual prediction errors shown in the figure, and can therefore be wider than those pre-treatment fluctuations suggest.}

As Figure \ref{fig_trend} shows, our estimated {\color{revisionblue}sales} in the ``synthetic California'' do not differ substantially from what a standard SCM would predict, especially for later periods.
However, our estimates of the first two post-treatment periods (1989 and 1990) are not significantly different from zero at the 5\% significance level, in contrast with SCM.
This difference may result from the over-estimation of the scale of the treatment effects by SCM in the presence of spillover effects. 
From the tests of spillover effects (shaded area of Figure \ref{fig_gap}), we see that likely there were substantial spillover effects.
Here, spillover effects may include both cross-border purchasing effects of Proposition 99 and contamination from other states' own tobacco-control policies.
One potential cause of spillovers may be that consumers in California shifted their purchasing to the nearby states, Arizona, Nevada, and Oregon. 
{\color{revisionblue}Similar laws with a tax increase on cigarettes were passed in Arizona and Oregon in 1994 and 1996, respectively.
The subsequent estimates for Arizona and Oregon should therefore be interpreted as a combination of possible spillover effects of Proposition 99 and direct or anticipatory effects of their own policies, which our analysis cannot separately identify.}
Nevada, however, did not pass any such laws in this period, so the spillover effect estimates for Nevada are more reliable.
From Figure \ref{fig_spillover}, we observe that Nevada experienced significant spillover effects in the first two periods after the passage of Proposition 99 and mostly insignificant effects afterward, with the exception of 1997. 
This is consistent with our conjecture and may provide more evidence on how the effects of the policy have propagated. 

The non-significant effects of policy right after the implementation could be explained by the addictive behavior of cigarette consumption. 
The persistence of cigarette consumption has been extensively studied and well-understood by both the rational addiction and the medical literature \citep{Baltagi2001,Baumeister2017,Becker1994,Benowitz1992,Christelis2011,Labeaga1999,Miura2019,Vleeming2002}.
Compared with SCM, our results are more consistent with an addiction story, that tobacco consumption is addictive and unlikely to drop immediately after the policy, but rather slowly transition to a lower equilibrium.

\begin{figure}[h!]
	\centering
	\begin{subfigure}{.5\textwidth}
		\centering
		\includegraphics[width=.7\linewidth,trim={4cm 6.5cm 4cm 6.5cm}]{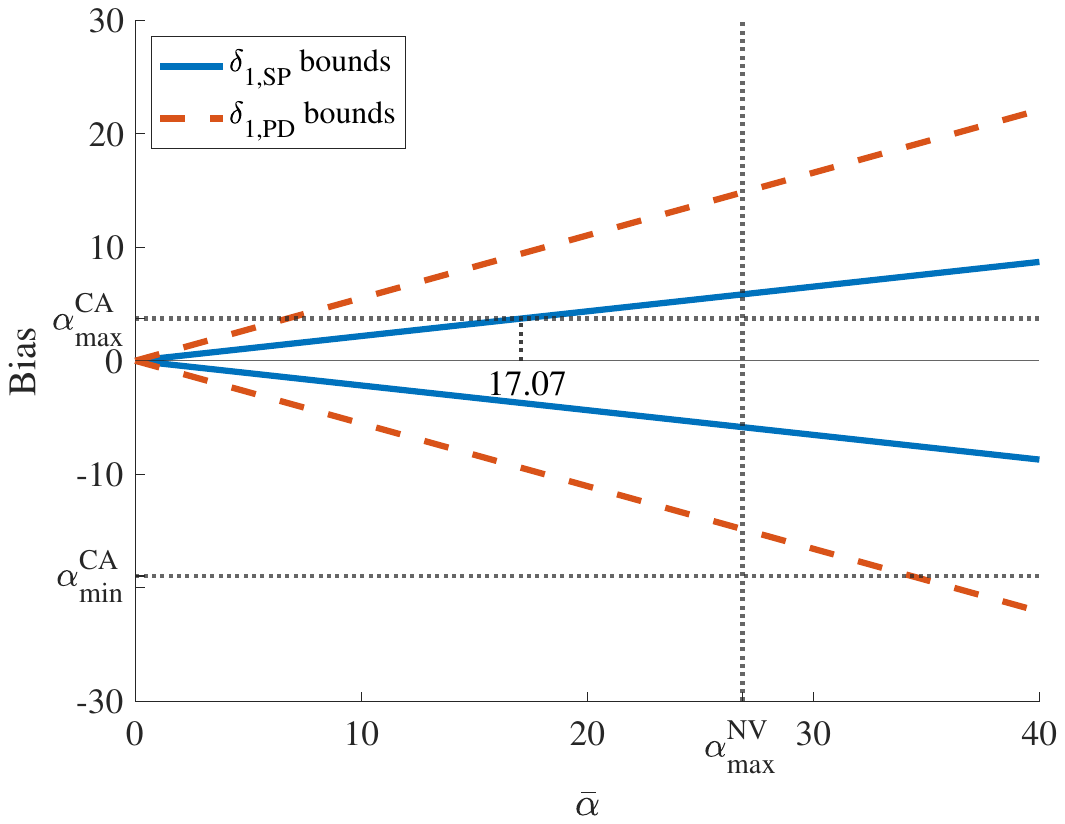}
		\caption{$p=1$}
	\end{subfigure}%
	\begin{subfigure}{.5\textwidth}
		\centering
		\includegraphics[width=.7\linewidth,trim={4cm 6.5cm 4cm 6.5cm}]{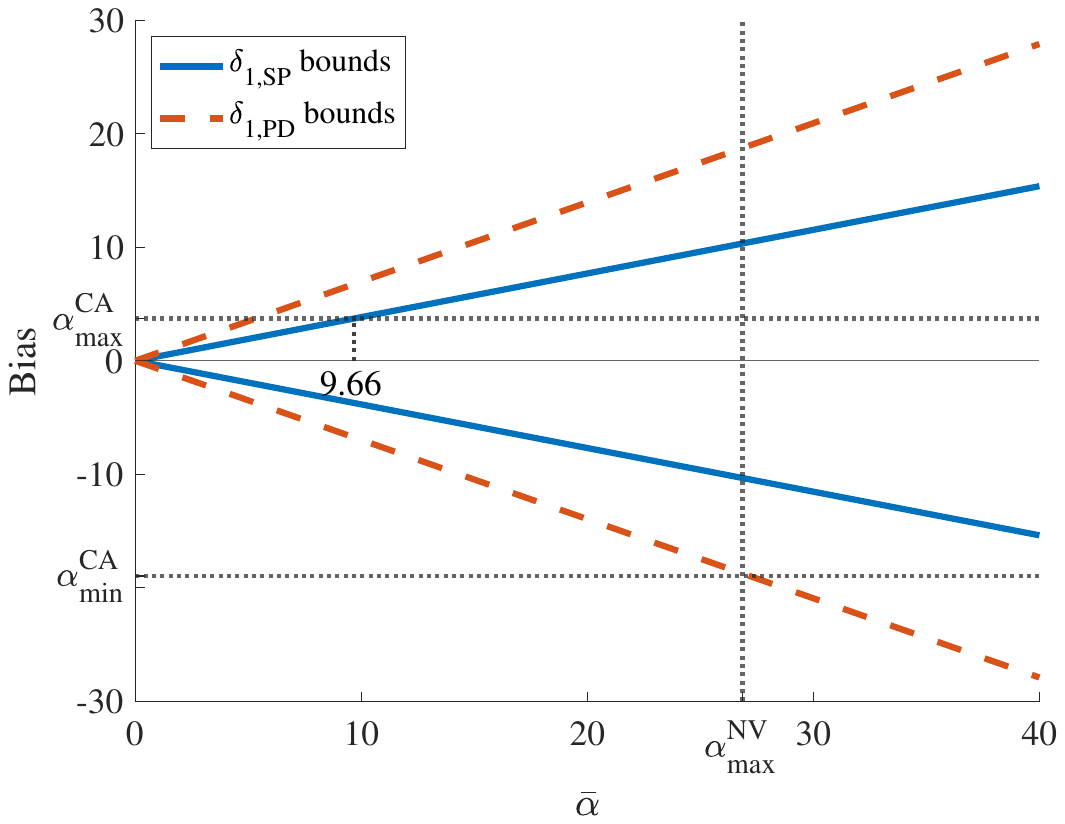}
		\caption{$p=2$}
	\end{subfigure}
	\caption{
		\small
		Sensitivity Analysis of Spillover Structure Misspecification. 
	 This figure illustrates the relationship between the magnitude of missed spillover effects ($\bar{\alpha}$) and the potential bias in estimators. The parameter $p$ denotes the number of units where spillovers have been missed. The biases $\delta_{1,SP}$ and $\delta_{1,PD}$ represent the (asymptotic) biases of the proposed method and the usual synthetic control method using only pure donors, respectively. The bounds of the identified bias sets for both methods are shown. Benchmarks within the figure include the largest and smallest treatment effect estimates, $\alpha_{\max}^{CA}$ and $\alpha_{\min}^{CA}$, alongside the largest estimated spillover effect, $\alpha_{\max}^{NV}$. Notably, the intersection point of $\alpha_{\max}^{CA}$ with the upper bound of the $\delta_{1,SP}$ bias set is highlighted. Panels (a) and (b) correspond to misspecification scenarios with $p=1$ and $p=2$ missed spillover units, respectively.
	}
	\label{fig_sensitivity_analysis}
\end{figure}

\subsection{Sensitivity analysis}
\label{section_sensitivity_analysis_misspecification}

Structure misspecification could be common in comparative case studies, especially with datasets of moderate sizes.
As a result, it is crucial to evaluate the robustness of these methods under structure misspecification. To this end, we introduce a sensitivity analysis to assess the impact when the spillover structure is incorrectly specified.

Suppose the effect vector is $\alpha = (\alpha_1, \dots, \alpha_N)'$. We proceed with estimation assuming units $i = 2, \dots, k$ are exposed to spillovers, though in reality additional units in $i = k+1, \dots, N$ may also be contaminated. Let $p$ be the number of units mistakenly assumed not to be exposed to spillovers, and let $\bar{\alpha}$ denote the maximum possible spillover effects in absolute value. We want to assess the asymptotic bias for both methods, $\delta_{1,SP} = \sum_{i=k+1}^N c_i \alpha_i$ and $\delta_{1,PD} = -\sum_{i=k+1}^N b_{1,i}^{PD} \alpha_i$ as characterized in Section \ref{section_misSpeBiasCharacterization}.
Since the identities of the neglected spillover units are unknown, the bias for each method is characterized by the identified set
$\Delta_{1,SP}(\bar\alpha) = \{ \delta_{1,SP} : \alpha \in \mathcal{A}_p(\bar{\alpha}) \}$ and $\Delta_{1,PD}(\bar\alpha) = \{ \delta_{1,PD} : \alpha \in \mathcal{A}_p(\bar{\alpha}) \}$,
where
$\mathcal{A}_p(\bar\alpha) = \{ \alpha : \exists S \subset \{k+1, \dots, N\} \text{ s.t. } |S| = p, |\alpha_i| \leq \bar\alpha \text{ for } i \in S, \alpha_i = 0 \text{ for } i \in \{k+1, \dots, N\}\setminus S \}.$
The sensitivity analysis is conducted by plotting these identified sets against $\bar{\alpha}$ to compare the potential biases under varying levels of misspecification.



Figure \ref{fig_sensitivity_analysis} presents the comparison. The figure contains two panels that delineate the upper and lower bounds of the bias under spillover structure misspecification scenarios with $p=1$ and $p=2$. Here, $p$ is the number of incorrectly omitted units that are actually subject to spillover effects. The solid lines in both panels represent the bounds for the bias of our proposed method ($\delta_{1,SP}$), while the dashed lines represent that of the pure-donor method ($\delta_{1,PD}$). Key reference points, $\alpha_{\max}^{CA}$ and $\alpha_{\min}^{CA}$, represent the largest and smallest estimated treatment effects observed post-treatment, which are 3.72 (2 years post-treatment) and -18.96 (11 years post-treatment), respectively. The largest estimated spillover effect in absolute value, $\alpha_{\max}^{NV}=26.87$, is recorded two years post-treatment in Nevada, which serves as a benchmark measure of the magnitude of spillover effects. The intersection of $\alpha_{\max}^{CA}$ with the upper bound of $\delta_{1,SP}$ is also shown (17.07 and 9.66 for the $p=1$ and the $p=2$ case, respectively).

We have two key findings. 
First, we find that the pure-donor method has a much larger identified bias. 
In this application, its effective donor weights are more concentrated, so contamination of the most influential donors produces larger worst-case biases.
Second, the intersection of $\alpha_{\max}$ and the upper bound of $\delta_{1,SP}$ can be interpreted as the smallest possible spillover effect needed to invalidate the non-zero treatment effect estimate. 
For example, when $p=1$, they intersect at 17.07. This means that in the extreme case where a spillover of 17.07 occurs at the missed control with the largest weight, our estimate should have been zero, but is not so due to the misspecification of the spillover structure. 
Of course, in the application we are more interested in the negative effect. 
We see that $\alpha_{\min}^{CA}$ intersects the lower bound of $\delta_{1,SP}$ at a point much larger than $\alpha_{\max}^{NV}$, suggesting the robustness of our results even allowing for some level of spillover structure misspecification.

\section*{Acknowledgments}
We thank Max Farrell and Christian Hansen for their invaluable guidance. We are grateful to Tetsuya Kaji, Giovanni Mellace, Alexander Torgovitsky, Ruey Tsay, Yinchu Zhu, and other seminar participants at {\color{revisionblue}the} 2018 Midwest Econometrics Group Conference, the 2019 China Meeting of the Econometric Society, and the 2019 North America Summer Meeting of the Econometric Society.
We thank Ziyao Wang for his excellent research assistantship.

	\section*{Data availability statement}
	The data that support the findings of this study are openly available on the Centers for Disease Control and Prevention website: \url{https://data.cdc.gov/Policy/The-Tax-Burden-on-Tobacco-1970-2019/7nwe-3aj9/about_data}.

\startsupplement
	
	{\color{revisionblue}
	\section{Empirical Examples of Spillover Structures}
	\label{section_spillover_examples}
	The framework is conditional on a researcher-specified spillover structure $A$.
We do not claim that specifying $A$ is equally difficult in every application, nor do we provide a universal algorithm for discovering it.
Its credibility is application-specific and must be defended using substantive knowledge before the post-treatment outcomes are examined.
Depending on the setting, that knowledge may come from geography, institutional boundaries, economic or epidemiological networks, the mechanism through which the treatment operates, or auxiliary evidence about the reach of exposure.

Table \ref{tab:applied_spillover_structures} provides a descriptive inventory of seventeen studies identified in our review of empirical work that specifies potentially exposed units.
Some impose an exposure structure and use our estimator to estimate both treatment and spillover effects, whereas others identify potentially contaminated units only to remove them from a donor pool or combine them with the treated unit.
The table is therefore evidence about the kinds of substantive judgments used in applied work, rather than evidence that a credible $A$ is available in every application.
Across the studies, common inputs include adjacency and distance, named economic or institutional links, infection rates, and auxiliary distance-band regressions.
Several studies also impose common effects within substantively defined groups or allow effects to vary across distance bands.


\nocite{araujo_avaliacao_2024,auerbach_regression_2024,
bluszcz_economic_2022,DiStefano2020,duncan_tempting_2025,
elenev_staggered_2024,goli_making_2024,Grossi2020,
krajewski_accounting_2025,
lim_estimating_2021,liu_three_2021,montero_impacts_2024,
peri_economic_2024,rizzo_pay_2024,tello_effects_2024,
tello_pacheco_spillovers_2023,wiltshire_estimating_2022}

\begingroup
\captionsetup{labelfont={color=revisionblue},textfont={color=revisionblue}}
\begin{singlespace}
\small
\renewcommand{\arraystretch}{1.12}
\setlength{\LTpre}{0.8em}
\setlength{\LTpost}{0.5em}
\begin{ThreePartTable}
\begin{TableNotes}[flushleft]
\footnotesize\upshape
\setlength{\labelsep}{0pt}
\item[] Notes:
Descriptive, non-exhaustive inventory based on the empirical implementations in our review of sixty works citing this paper as of August 2026. It includes studies that estimate spillovers or use potential exposure to adjust donor pools or treated aggregates.
\end{TableNotes}
\begin{longtable}{@{}>{\raggedright\arraybackslash}p{1.25in}>{\raggedright\arraybackslash}p{2.02in}>{\raggedright\arraybackslash}p{2.75in}@{}}
\caption{Examples of potential-spillover structures used in empirical work}
\label{tab:applied_spillover_structures}\\
\hline
\hline
Paper & Setting & Structure of potential spillovers \\
\hline
\endfirsthead
\multicolumn{3}{l}{\small Table \thetable\ continued}\\
\hline
Paper & Setting & Structure of potential spillovers \\
\hline
\endhead
\hline
\multicolumn{3}{r}{\small Continued on the next page}\\
\endfoot
\hline
\hline
\insertTableNotes\\
\endlastfoot
Araujo and Linhares (2024) & Effect of the Community Protection Cell community-policing program on neighbourhood theft and robbery rates & Untreated neighbourhoods are assigned to 0--2, 2--4, 4--6, and 6--8 kilometre rings according to their distance from the nearest treated neighbourhood. Each ring has a common spillover coefficient, and neighbourhoods beyond 8 kilometres form the comparison group. \\
Auerbach et al.\ (2024) & Effect of cell-phone coverage on electoral fraud in the 2009 Afghan presidential election & Potential neighbours are polling stations within a circular Euclidean radius. The analysis considers radii of 2, 4, 6, and 8 kilometres and a 6.1-kilometre radius chosen by cross-validation, with common local coefficients on mean neighbour outcomes and the treated-neighbour share. \\
Bluszcz and Valente (2022) & Effect of the Donbass war on Ukraine's GDP per capita and Donetsk's and Luhansk's GRP per capita & At the national level, Belarus, Moldova, Hungary, Poland, Romania, and the Slovak Republic are designated as potentially exposed. The regional sets comprise Dnipropetrovsk, Kharkiv, and Zaporizhzhya for Donetsk and Kharkiv for Luhansk. \\
Di Stefano and Mellace (2024) & Effect of German reunification on real GDP per capita in West Germany and Austria & Austria is the sole potentially exposed donor in the main specification. \\
Duncan et al.\ (2025) & Effect of the Federal Reserve's adoption of flexible average inflation targeting on U.S. inflation and inflation expectations & Canada is the sole potentially exposed donor, selected because of U.S.--Canada economic integration and geographic proximity. \\
Elenev et al.\ (2025) & Effect of county stay-at-home orders on smartphone-based mobility and social interaction & Counties are organized into contiguous triplets. A neighbour shares a border with the treated county and adopts an order at least four days later; a hinterland county borders the neighbour but not the treated county or another early adopter. A common spillover coefficient is assigned to the neighbour group. \\
Goli et al.\ (2024) & Effect of the Massachusetts menthol-cigarette ban on menthol, nonmenthol, and total cigarette sales & Non-Massachusetts stores in ZIP codes within 30 miles of the state border are designated as potentially exposed. The radius is selected using distance-band estimates, and this ring is aggregated with Massachusetts in the synthetic-control analysis. \\
Grossi et al.\ (2025) & Effect of a new Florence tramway line on street-level retail density & Talenti Street is directly treated, and the other five streets in its Legnaia neighbourhood are potentially exposed. The neighbourhood is defined using Italian Ministry of Finance Real Estate Observatory areas, and street-specific spillovers are allowed. \\
Krajewski and Hudgens (2025) & Effect of initial comuna-level COVID-19 lockdowns in Chile on the instantaneous reproduction number & Neighbours are comunas that share a border, and a comuna is exposed if at least one neighbour is locked down. Units are partitioned by own-lockdown and neighbour-lockdown status; each treated comuna's direct-effect donor pool excludes its own neighbours. \\
Lim et al.\ (2021) & Effect of the 2020 Sabah state election on district-level COVID-19 case counts in Malaysia & Each Sabah district has its own direct-effect parameter. Exposure outside Sabah equals a common time-specific coefficient scaled by district connectivity, measured alternatively by airport distance, October 2012 passenger volume, or total 2010--2017 passenger volume. \\
Liu (2021) & Effect of the armed conflict beginning in late 2013 on Ukraine's real GDP & The baseline structure assigns separate parameters to twelve countries selected through contiguity, near-contiguity, former-Soviet ties, or major trading links. An expanded structure adds France, Italy, the Netherlands, Sweden, and Switzerland. \\
Montero de Espinosa Garcia (2024) & Effect of Bolivia's government--soybean-sector partnership on multidimensional poverty among Indigenous households in Santa Cruz & In Santa Cruz, matching uses only Indigenous households, excluding potentially exposed non-Indigenous households. A robustness specification also excludes households in the four bordering departments of Beni, Cochabamba, Chuquisaca, and Tarija. \\
Peri et al.\ (2024) & Effect of post-Hurricane Maria Puerto Rican migration to Orlando on employment, earnings, and establishment counts & The treated Orlando commuting zone aggregates Lake, Orange, Osceola, Seminole, and Sumter counties. Donor zones with at least 10 hurricane-related FEMA applications per 100,000 residents are excluded, including all zones bordering Orlando. \\
Rizzo and Secomandi (2024) & Effect of Ferrara's pay-as-you-throw threshold tariff on recycling shares and total waste per capita & Potential exposure includes all municipalities within 50 kilometres of Ferrara. \\
Tello Pacheco (2024) & Effect of COVID-19 and contemporaneous government policies on monetary poverty in Peruvian provinces & 59 high-infection provinces are treated; 31 low-infection provinces are controls. For each of nine behavioural indicators, candidate structures select the one, three, or five most extreme controls, or all controls on the relevant side of its mean, as distinct spillover sources. \\
Tello Pacheco (2023) & Effect of COVID-19 and contemporaneous government policies on employment and real income in Peruvian provinces & Same treated-control partition, nine behavioural indicators, and candidate structures as Tello Pacheco (2024), with selected controls entering as distinct spillover sources. \\
Wiltshire (2022) & Effect of Walmart Supercenter entry on county employment, earnings, labour-force participation, and labour-market concentration & For each treated county, donors in the same commuting zone are potentially exposed and excluded. \\
\end{longtable}
\end{ThreePartTable}
\end{singlespace}
\endgroup

	}

	\section{Extensions}
	\label{section_extension}

	\subsection{Large-$(N,T)$ asymptotics}
	\label{section_largeN}
	{\color{revisionblue}
The main results keep $N$ fixed as $T$ grows. If $N$ also grows, two issues need further attention. The matrix used to recover the effects must not become nearly singular, and the first-stage estimates must remain sufficiently accurate as the number of coefficients increases. We describe sufficient conditions for the estimation approximation in Theorem \mainref{unbiasedness}{1}, keeping the number of effect parameters $k$ fixed.

To state these conditions, write $N_T=N(T)$ and index the model by $T$. Let $A_T\in\mathbb R^{N_T\times k}$, $\alpha_T=A_T\gamma_T$, $a_T\in\mathbb R^{N_T}$, and $B_T\in\mathbb R^{N_T\times N_T}$. Write the untreated outcomes and disturbances as $Y_{t,T}(0)$ and $u_{t,T}$, and the feasible weight sets as $\mathcal W_T^{(i)}$ for $i=1,\ldots,N_T$. Define $M_T=(I_{N_T}-B_T)'(I_{N_T}-B_T)$, and use the same $T$ subscript for $G$ in Theorem \mainref{unbiasedness}{1} and for all estimators. Throughout this section, $\|\cdot\|_2$ denotes the Euclidean norm of a vector and $\|\cdot\|_{\mathrm{op}}$ the induced operator norm of a matrix.

First consider the matrix that must be inverted. Let $X_T=(I_{N_T}-B_T)A_T$. The matrix $A_T'M_TA_T=X_T'X_T$ is invertible if and only if $\operatorname{rank}(X_T)=k$. However, full rank at each $T$ does not prevent this matrix from becoming nearly singular as $N_T$ grows. To compare exposure columns on the same scale, define $S_{A,T}=\operatorname{diag}(\|A_{T,\cdot1}\|_2,\ldots,\|A_{T,\cdot k}\|_2)$, where all columns of $A_T$ are nonzero. A sufficient condition is
\[
\inf_T\lambda_{\min}\!\left(S_{A,T}^{-1}A_T'M_TA_TS_{A,T}^{-1}\right)>0.
\]
After this normalization, the smallest eigenvalue must stay bounded away from zero. Thus, subtracting each unit's synthetic control must leave enough information to distinguish the effects represented by the columns of $A_T$.


The choice of first-stage weights matters for this condition. As the donor pool grows, regularization may set some weights to zero and remove connections between affected units and informative pure donors. What matters is whether these connections are needed to distinguish the effects, not simply how many weights are zero.
As discussed in Section \mainref{sec_IB_rank_Nminus1}{3.4.2} of the main text, one possible remedy is to omit disconnected groups unrelated to the effects of interest, together with any effect parameters specific to those groups. This can restore full rank if the identification problem is confined to the omitted groups. The rank and stability conditions must then be checked again for the reduced system.

Next consider the accuracy of the first-stage estimates. Errors that are small for each unit can still accumulate as $N_T$ grows. Suppose that the second moments of the coordinates of $u_{t,T}$ are uniformly bounded. Together with the eigenvalue condition above, a sufficient strengthening of Assumption \mainref{unbiased assumption}{1}(b)--(c) is
\[
\begin{aligned}
\text{(b)}\qquad &\|\widehat a_T-a_T\|_2=o_p(1),
\qquad \sqrt{N_T}\|\widehat B_T-B_T\|_{\mathrm{op}}=o_p(1),\\
\text{(c)}\qquad &\|(\widehat B_T-B_T)Y_{T+1,T}(0)\|_2=o_p(1).
\end{aligned}
\]
To see why these rates suffice, first note that part (b) and the eigenvalue condition imply that $A_T'\widehat M_TA_T$ is invertible with probability approaching one. On this event, the exact remainder is
\[
\widehat\alpha_T-\alpha_T-G_Tu_{T+1,T}=(\widehat G_T-G_T)u_{T+1,T}-\widehat G_T\big[(\widehat a_T-a_T)+(\widehat B_T-B_T)Y_{T+1,T}(0)\big].
\]
The eigenvalue condition gives $\|\widehat G_T\|_{\mathrm{op}}=O_p(1)$. Together with the stated rate for $\widehat B_T-B_T$, it also gives $\|\widehat G_T-G_T\|_{\mathrm{op}}=o_p(N_T^{-1/2})$. Since $\|u_{T+1,T}\|_2=O_p(\sqrt{N_T})$, the first term on the right is $o_p(1)$. Parts (b) and (c) make the second term $o_p(1)$ as well.

The separate prediction-error condition in part (c) is important. Untreated outcome levels can become more variable over time even when $u_{t,T}$ is stationary. In that case, small errors in the weights can produce substantial prediction errors, so the coefficient rate in part (b) alone need not suffice.

These conditions control the remainder for the entire $N_T$-dimensional effect vector. If only a single treatment effect or a fixed number of contrasts is of interest, weaker conditions may suffice. Only the corresponding components or linear combinations of the remainder then need to be $o_p(1)$. For the treated-unit effect, for example, it is enough that the first component of the displayed remainder is $o_p(1)$.

The remaining step is to establish these rates for a particular first-stage estimator. Regularization may be needed when $N_T$ is comparable to $T$. For related approaches to synthetic-control estimation in larger panels, see \citet{Abadie2021}, \citet{arkhangelsky_synthetic_2021}, and \citet{ben-michael_synthetic_2022}. Extending the inference results would also require verifying the inferential assumptions as the number of units grows.
}

\subsection{Multiple treated units}

Our method readily extends to cases where multiple units are treated. In our setting, the treatment and spillover effects can be estimated in the same way, since the spillover effects can be interpreted as the indirect treatment effect. 
With a correctly specified structure matrix $A$, we can perform estimation and inference just as in previous sections. For example, suppose $N=4$, unit 1 and unit 2 are treated, unit 3 is affected by spillover effects, and unit 4 is neither treated nor exposed to spillover effects. Then we can specify $A=[I_3, 0_{3\times 1}]'$,
and the resulting estimator $\widehat{\gamma}=(\widehat{\gamma}_1,\widehat{\gamma}_2,\widehat{\gamma}_3)'$ by \mainequationref{gamma estimation}{6} is such that $\widehat{\gamma}_1$ and $\widehat{\gamma}_2$ are the treatment effect estimator for unit 1 and unit 2, respectively, and $\widehat{\gamma}_3$ is the spillover effect estimator for unit 3. Tests can be performed accordingly.
If the researcher wants to test for the hypothesis that there are no spillover effects, the null is then $H_0:C\alpha=d$, where $C=(0,0,1,0)$ and $d=0$. 

\subsection{Multiple post-treatment time periods}
\label{sec_multi_post_period}

The availability of multiple post-treatment periods facilitates a more flexible set of estimands, such as dynamic treatment effects, as illustrated in the empirical application section. More importantly, under stability and separation conditions, repeated post-treatment observations can help discriminate among a finite set of prespecified spillover structures \citep{Manresa2013-cr,de_paula_identifying_2023}.
In this section, we introduce estimation and inference strategies when multiple post-treatment periods are available and state conditions under which such discrimination is consistent.

\paragraph{Period-by-period estimation.}
Suppose we have observations $\{y_{i,t}\}$ for $i=1,\dots,N$ and $t=1,\dots,T+T_1$. Treatment is received at $t=T+1$. The model becomes
\[
Y_{t}=\begin{cases}
	Y_{t}(0), & \text{if } t \le T,\\
	Y_{t}(0) + \alpha_t, & \text{otherwise}.
\end{cases}
\]
For each $t=T+1,\dots,T+T_1$, we need to specify the spillover structure matrix $A_t$. Then, an estimator of $\alpha_t$ is
\[
\widehat{\alpha}_t = A_t(A_t'\widehat{M}A_t)^{-1}A_t'(I-\widehat{B})'((I-\widehat{B})Y_{t}-\widehat{a}).
\]
For each $t=T+1,\dots, T+T_1$, we can perform separate tests as introduced in previous sections.

{\color{revisionblue}In the empirical application, we estimate $\widehat{a}$ and $\widehat{B}$ once from the 1970--1988 pre-treatment sample and apply the same estimator in each post-treatment year. We use the same spillover structure $A_t=A$ in every post-treatment period, while allowing $\alpha_t=A\gamma_t$ to vary freely over time. Thus, $\alpha_{T+s}$ is the total effect at event time $s$, rather than a new period-$s$ increment. Its path can reflect gradual adjustment and the reduced-form consequences of earlier exposure, although the method does not separately identify contemporaneous, carryover, or lagged mechanisms. Identification requires no anticipation and that both the pre-treatment relationship used to approximate untreated counterfactual outcomes and the spillover structure $A$ remain stable. Because these requirements become more demanding at longer horizons, later estimates should be interpreted more cautiously than near-term estimates. Inference is pointwise by year.}

\paragraph{Joint testing across periods.}
To answer questions such as whether there is a spillover effect at all, we can extend Andrews' instability test discussed above. Consider the null hypothesis $H_0: C_t\alpha_t=d_t$ for $t=T+1,\dots,T+T_1$. Let $\widehat{P}_t$ be constructed as in Section \mainref{spillover effect test}{4.2} for $t=1,\dots, T$. For $t=T+1,\dots,T+T_1$, let $\widehat{P}_{t}=(C_t\widehat{\alpha}_t-d_t)'{\color{revisionblue}\Xi_T}(C_t\widehat{\alpha}_t-d_t)$. For $t=1,\dots, T+1$, let $P^{(t)}=\sum_{s=0}^{T_1-1}\widehat{P}_{t+s}$, each of which contains information from period $t$ through $t+T_1-1$. The test statistic is then $P^{(T+1)}$. When $1\le T_1\le T$, we use its pre-treatment counterparts $\{P^{(t)} \}_{t=1}^{T-T_1+1}$ to form the null distribution.

\paragraph{Comparing spillover structures.}
Suppose the spillover structure is stable, so that $A_t=A$ for every post-treatment period, and let $\mathcal A$ be a finite set of structures specified before examining the post-treatment outcomes.
Following Section \mainref{section_kappaA_statistic}{5.1.2}, define for each $A\in\mathcal A$
\[
\kappa_A = \left\| \frac{1}{\sqrt{T_1}} \sum_{s=1}^{T_1} [(I - \widehat{B}) (Y_{T+s} - \widehat{\alpha}_{T+s}) - \widehat{a}] \right\|,
\]
where $\widehat{\alpha}_{T+s}$ depends on $A$.
For notation, let
\[
\widehat{A} \in {\arg\min}_{A \in \mathcal{A}} \kappa_A,
\]
where any ties are resolved by a deterministic rule specified in advance.
This minimizer is a useful device for describing discrimination among the candidates, but it is not by itself a generally valid automatic selection rule and does not justify treating $\widehat A$ as fixed in subsequent inference.
The proposition below states when the criterion separates a candidate from the data-generating effect path.

Given $A$, define $\Gamma_A = (I-B)A(A'(I-B)'(I-B)A)^{-1}A'(I-B)'$, the projection onto the span of columns of $(I-B)A$. Define the sample version accordingly by $\widehat \Gamma_A = (I-\widehat{B})A(A'(I-\widehat{B})'(I-\widehat{B})A)^{-1}A'(I-\widehat{B})'$. Define $\bar{\alpha}_{T_1} = T_1^{-1}\sum_{s=1}^{T_1}\alpha_{T+s}$ as the average effects over time.

\begin{assum}
	\label{assum multiple periods}
	\emph{(a)} $\{ u_t\}_{t\ge 1}$ is stationary, has mean zero, and satisfies $\sum_{s=0}^\infty \|E[u_t u_{t+s}']\|_F<\infty$.
	
	\emph{(b)} $\widehat \Gamma_A-\Gamma_A=o_p(1)$, $\sqrt{T_1}[(I-\widehat \Gamma_A)(I-\widehat{B})-(I-\Gamma_A)(I-B)]\bar{\alpha}_{T_1}=O_p(1)$;
	
	\emph{(c)} $(I-\widehat \Gamma_A)(\widehat{B}-B) T_1^{-1/2}\sum_{s=1}^{T_1}Y_{T+s}(0)=O_p(1)$, $\sqrt{T_1}(I-\widehat \Gamma_A)(a-\widehat{a})=O_p(1)$;
	
	\emph{(d)} $A'MA$ is non-singular.
\end{assum}
This is a multiperiod counterpart of Assumption \mainref{unbiased assumption}{1}. Parts (b) and (c) restrict the rate of $\widehat{B}$ and $\widehat{a}$. Under weak conditions, they imply $T$ and $T_1$ are of similar magnitude.

\begin{prop}
	\label{prop_multiple_period_structure_selection}
	Under Assumption \ref{assum multiple periods}, $\kappa_A=O_p(1)$ if and only if $\|\sqrt{T_1}(I-\Gamma_A)(I-B)\bar{\alpha}_{T_1}\|=O(1)$.
\end{prop}

If $A$ accommodates the effect path, then $\sqrt{T_1}(I-\Gamma_A)(I-B)\bar{\alpha}_{T_1}=0$ and $\kappa_A=O_p(1)$.
By contrast, suppose a candidate is sufficiently separated in the sense that
\[
\left\|\sqrt{T_1}(I-\Gamma_A)(I-B)\bar{\alpha}_{T_1}\right\|\longrightarrow\infty.
\]
Its criterion then diverges while the criterion for an accommodating candidate remains bounded in probability.
Consequently, for a finite candidate set, comparisons of $\kappa_A$ consistently eliminate every sufficiently separated candidate.
If exactly one candidate has a bounded criterion and every other candidate is sufficiently separated, $\widehat A$ selects that candidate with probability approaching one.

This conclusion requires both uniqueness and separation.
It does not generally identify a minimal structure among nested candidates.
For example, under the limited-range specification in Example \mainref{nonparametric spillover}{1}, an oversized structure that contains all truly exposed units may accommodate the same effect path as the smallest correct structure, so both criteria can remain $O_p(1)$.
Distinguishing such candidates requires an additional complexity restriction or penalty, which we do not develop here.
Likewise, averaging effects over time can prevent separation when heterogeneous effects cancel.
Finally, the fixed-$A$ inferential results in this paper do not automatically become valid post-selection inference after searching over $\mathcal A$.
Absent a uniquely separated candidate and the additional conditions needed for oracle equivalence, researchers should report the fixed-$A$ results separately across a small set of prespecified structures or use a separate selection sample or a selection-adjusted inferential procedure.

{\color{revisionblue}
\subsection{Connection to Powell (2022)}
\label{sec_powell_connection}

\citet{powell_synthetic_2022} develops a fixed-$N$, large-$T$ synthetic control estimator that can accommodate selection into treatment on time-varying unobservables. Identification comes from repeatedly observing treatment across units and periods, which generates sufficient treatment variation after each unit is differenced from its synthetic control. Our main setting focuses on one post-treatment period, so Powell's approach does not apply directly, although it could be used in the many-period extension considered below.

To see this, first note that Powell's estimator and ours can be written in a similar form. Let $\mathcal B$ denote the set of $N\times N$ matrices with nonnegative entries, zero diagonal, and rows that sum to one. Consider Powell's homogeneous-effect case with one treatment variable. Let $d_t\in\mathbb R^N$ collect its values across units and let $\alpha\in\mathbb R$ denote the scalar treatment effect. Suppressing Powell's fit-based unit weights, his objective reduces to the form
\[
Q_{\mathrm{Powell}}(B,\alpha)=\sum_t\left\|(I-B)(Y_t-d_t\alpha)\right\|^2
\]
The sum runs over all pre-treatment and post-treatment periods, and the minimization is over $\alpha$ and $B\in\mathcal B$. Here, $\alpha$ is scalar because we have specialized to one treatment variable and a homogeneous effect. Powell's full model permits multiple treatment variables and unit-specific vector coefficients.

Our estimator has a closely related representation. In our setting, $A\in\mathbb R^{N\times k}$ records the known treatment design. Its columns encode direct treatment and spillover exposure. Let $s_t=\mathbbm{1}\{t>T\}$, and let $g\in\mathbb R^k$ collect the direct and spillover-effect coefficients. Then, $s_tAg$ is the counterpart of Powell's $d_t\alpha$. Consider the joint criterion
\[
Q_{\mathrm{SP}}(a,B,g)
=
\sum_{t=1}^{T+1}\left\|(I-B)(Y_t-s_tAg)-a\right\|^2.
\]
Substituting the pre-treatment estimates $(\widehat a,\widehat B)$ into $Q_{\mathrm{SP}}$ and minimizing over $g$ gives exactly the estimator in Equation~\mainequationref{gamma estimation}{6}. Jointly minimizing $Q_{\mathrm{SP}}$ over $a\in\mathbb R^N$, $B\in\mathcal B$, and $g\in\mathbb R^k$ instead defines a simultaneous estimator that, under the paper's regularity conditions, is asymptotically equivalent to the two-step estimator.
This equivalence is because, as $T\to\infty$, the single post-treatment term becomes negligible relative to the $T$ pre-treatment terms and therefore does not affect the limiting estimates of $(a,B)$.

This representation also clarifies the role of Assumption~\mainref{unbiased assumption}{1}(d). Since
\[
A'MA=[(I-B)A]'[(I-B)A],
\]
$A'MA$ is nonsingular exactly when the columns of $(I-B)A$ are linearly independent. This is the analogue, in our fixed-design homogeneous-effect representation, of the treatment-variation part of Assumption A3 of \citet{powell_synthetic_2022}. 

This representation suggests an alternative to the period-by-period extension in Section~\ref{sec_multi_post_period} when the number of post-treatment periods grows. That extension estimates a separate $\gamma_{T+s}$ in every post-treatment period and preserves the dynamic effect path. Suppose instead that $A_t=A$ is stable. A Powell-type extension would jointly estimate a pooled effect and the synthetic control weights from growing pre-treatment and post-treatment samples by solving
\[
(\widehat a_P,\widehat B_P,\widehat\gamma_P)
\in
\underset{a\in\mathbb R^N,\,B\in\mathcal B,\,g\in\mathbb R^k}{\arg\min}
\sum_{t=1}^{T+T_1}
\left\|(I-B)(Y_t-s_tAg)-a\right\|^2.
\]
This criterion imposes one $g$ across all post-treatment periods. Under the pooled rank condition, if effects are constant over time, $\widehat\gamma_P$ estimates their common value. If effects vary over time, it instead has an average-effect interpretation when $A$ and the synthetic control relationship are stable. Because $(a,B,g)$ are estimated simultaneously, the treatment-adjusted post-treatment outcomes also contribute to estimating the synthetic control weights. If an appropriate pooled rank condition and analogous regularity conditions hold in this setting, the repeated post-treatment observations could accommodate selection into treatment on time-varying unobservables.
}

\subsection{Including covariates}

Many empirical researchers are interested in including extra covariates when using synthetic control methods. Our framework can be combined with existing methods such as \cite{Abadie2010} and \cite{Li2019}, and be readily adapted to settings with covariates. 
For example, suppose we have a vector of observable variables $z_{i,t}$ and want to estimate the treatment effects, while being worried about spillover effects. Following \cite{Li2019}, we estimate the least square coefficients for the model
$$y_{i,t}(0)=a_i+\sum_{j\ne i}b_{i,j}y_{j,t}(0)+z_{i,t}'\pi_i+u_{i,t},$$
with the simplex constraints on $b_{i,j}$ and obtain coefficient estimates $(\widehat{a}_i,\widehat{b}_i,\widehat{\pi}_i)$. This is done for each $i$. Let $\widehat{g}_t=(z_{1,t}'\widehat{\pi}_1,\dots,z_{N,t}'\widehat{\pi}_N)'$. Under appropriate regularity conditions, the results of the paper apply when the intercept estimator $\widehat{a}$ is replaced by $\widehat{a}+\widehat{g}_{t}$ at time $t$.
The treatment effects estimator now becomes
$\widehat{\gamma}=(A'\widehat{M}A)^{-1}A'(I-\widehat{B})'((I-\widehat{B})Y_{T+1}-\widehat{a}-\widehat{g}_{T+1}).$

	\subsection{An example without pure donors}
	\label{section_continuous_treatment}

In this section, we explore a case where all control units are affected by the treatment, leaving no pure donors available.
We provide an example of spillover structure $\alpha=A\gamma$ that satisfies this condition. 
Interestingly, this discussion shows a link to the continuous treatment scenario.\footnote{We thank an anonymous reviewer for providing this insight.}

	\begin{eg}
		\label{eg_spatial}
		(Exponential decay)
		Assume that the spillover effect shrinks as the geometric distance goes up. For $i=2,\dots, N$, $\alpha_i=b\exp(-d_i)$ where $d_i$ is the distance between unit $1$ and unit $i$, and $b$ is an unknown parameter of interest. Then, we have $\alpha = A\gamma =(\alpha_1,b\exp(-d_2),\dots, b\exp(-d_N))'$, where 
		$$A=\begin{bmatrix}
			1&0\\0&\exp(-d_2)\\\vdots&\vdots\\0&\exp(-d_N)
		\end{bmatrix}
		\text{ and }
		\gamma=\begin{bmatrix}
			\alpha_1\\b
		\end{bmatrix}. 
		$$
	\end{eg}

	Example \ref{eg_spatial} introduces an interesting case where all control units are affected by spillover effects. 
	This is a challenging scenario, and a sensible estimation strategy requires strong assumptions. 
As in Example \ref{eg_spatial}, one such assumption is that spillover effects are proportional to $\exp(-d_i)$.
	All previously introduced estimation and inference strategies would work, given the assumptions are valid. 
	
	Another perspective on this example is its connection to a continuous treatment model. Namely, $\exp(-d_i)$ is a variable that summarizes the intensity of spillover effects. This links directly to the continuous or moderated treatment literature \citep{callaway_event-studies_2024,erten_trade_2023}. In certain contexts, it may be more useful to focus on intensity as the primary variable of interest. In this scenario, no units are exempt from effects, meaning that the pure counterfactual outcome $y_{i,T+1}(0)$ does not appear for any $i$ and may not be particularly relevant. This situation leads to an alternative parameter of interest, which we discuss below.
	
In our formulation, \(\gamma = (\alpha_1, b)\), where \(\alpha_1\) is the treatment effect, defined by \(y_{1,T+1} - y_{1,T+1}(0)\), and \(b\) is a candidate parameter of interest. Specifically, \(b\) can be interpreted as the marginal contribution to the outcome associated with a one unit increase in the implied intensity metric, \(\exp(-d_i)\).

To illustrate this, consider the proposed estimator for \(\gamma\),
\[
\widehat{\gamma} = (A'\widehat{M}A)^{-1}A'(I-\widehat{B})'((I-\widehat{B})Y_{T+1} - \widehat{a}),
\]
which is the OLS estimator for the linear model (in matrix formulation)
\begin{equation}
	\label{eq_residualized_linear_model}
	(I-\widehat{B})Y_{T+1} - \widehat{a} = (I-\widehat{B})A\gamma + \widehat u_{T+1}.
\end{equation}
Equation \eqref{eq_residualized_linear_model} corresponds to the linear model 
\begin{equation}
	\label{eq_linear_model_exponential_decay}
	Y_{T+1} = a + A\gamma + BY_{T+1}(0) + u_{T+1},
\end{equation}
where \(A\) is the data matrix of some independent variables and the \(i\)-th observation vector of $A$ is \((\mathbbm{1}\{i=1\}, \mathbbm{1}\{i \ne 1\}\exp(-d_i))'\). 
Thus, \(b\) is the coefficient associated with \(\mathbbm{1}\{i \ne 1\}\exp(-d_i)\) and has the standard interpretation of a coefficient in a linear model.
Note that Equation \eqref{eq_residualized_linear_model} can be interpreted as a residualized version of Equation \eqref{eq_linear_model_exponential_decay}, although the residualization is through the linear model learned using pre-treatment data. 

The above analysis implies that \(b\) has the usual good properties of a linear model coefficient, as well as the limitations associated with those, such as linearity, measurement error, etc. This formulation provides one perspective on what we are estimating. Of course, we face greater challenges than those typically encountered in a standard linear regression scenario, due to data limitations.
Moreover, the invertibility of $A'MA$ is now an analog to the common  linear model assumption that the design matrix has full rank in the residualized model \eqref{eq_residualized_linear_model}. 

Statistical inference can be conducted similarly as before. Since \(\widehat{\gamma} - \gamma = (A'MA)^{-1}A'(I-B)'u_{T+1} + o_p(1)\), we can use the pre-treatment residuals to approximate the null distribution of \(\widehat{\gamma}\). For the hypothesis \(H_0: b = b_0\), let the test statistic be \(P = (\widehat{b} - b_0)^2\), where \(\widehat{b}\) is the second entry of \(\widehat{\gamma}\). Its null distribution can be approximated by
\[
P_t = \left[(0,1)(A'\widehat{M}A)^{-1}A'(I-\widehat{B})'\widehat{u}_{t}\right]^2,
\]
for \(t = 1, \dots, T\). Confidence intervals can be constructed by inverting the test.

\begin{table}[h!]
	\centering
	\caption{Summary Statistics of Intensity Measure.}
	\label{table_summary_statistics_intensity}
	\renewcommand{\arraystretch}{1.5}
	\footnotesize
	\begin{threeparttable}
		\begin{tabular}{lcccccccccccc}
			\hline
			\hline
			& $N$ & mean & sd & min & q10 & q25 & q50 & q75 & q90 & max\\
			\hline
			Intensity & 50 & 1.655 & 1.800 & 0.213 & 0.284 & 0.423 & 0.824 & 2.060 & 4.922 & 7.675\\
			\hline
		\end{tabular}
		\begin{tablenotes}
			\small 
			\item Notes: 
			\emph{Intensity} is constructed using the distance between each state and California in terms of the geographic centers of the states. 
			The distance $d_i$ is normalized and the intensity is $\exp(-d_i)$.
		\end{tablenotes}
	\end{threeparttable}
\end{table}

\begin{figure}[h!]
	\centering
	\includegraphics[trim={0 6cm 0 6cm},scale=.5]{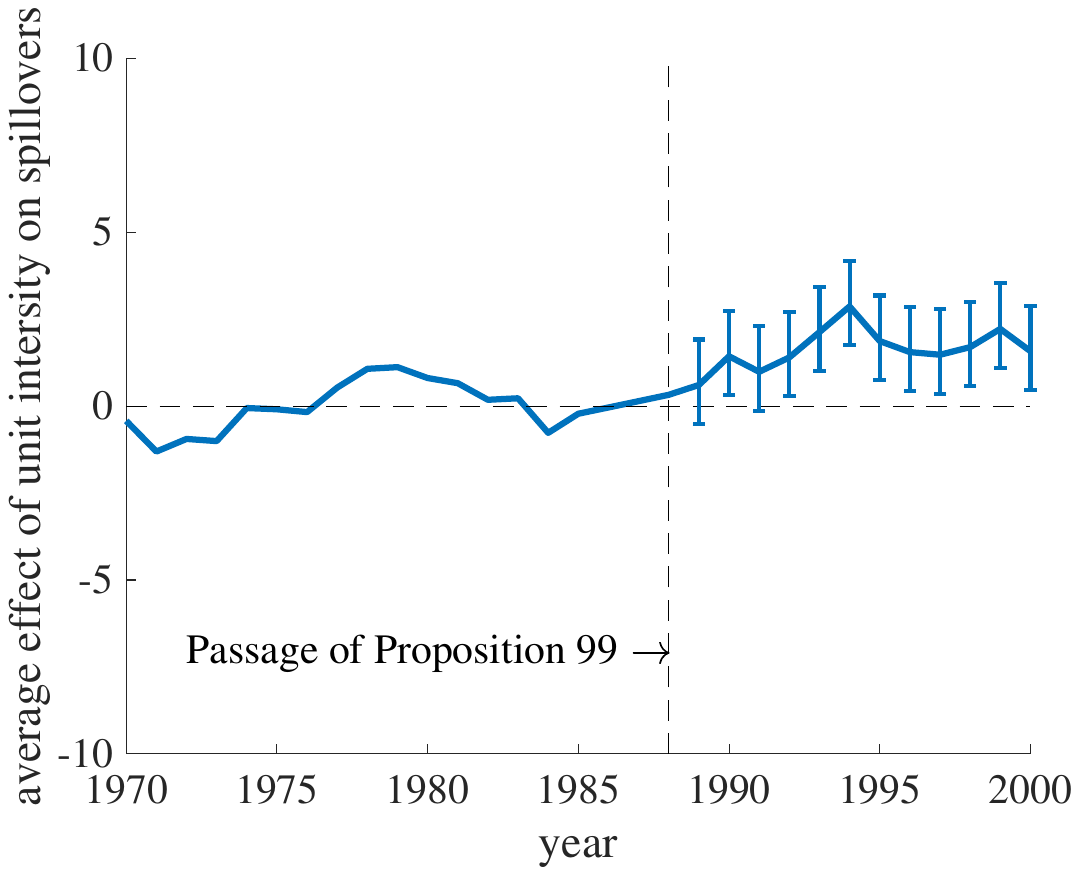}
	\caption{\small $\widehat b$ Estimate over Time with 95\% Confidence Interval. 
	}
	\label{fig_continuous_treatment}
\end{figure}

For illustrative purposes, we apply this method to California's tobacco control policy as in Section \mainref{empirical example}{6}.
We emphasize its illustrative nature due to the strong and debatable assumption that treatment intensity decreases exponentially. The summary statistics of the intensity metric $\exp(-d_i)$ are provided in Table \ref{table_summary_statistics_intensity}. The results are shown in Figure \ref{fig_continuous_treatment}, showing that the estimate $\widehat{b}$ is reasonably stable over time, with most estimates significantly larger than zero. This suggests that closer states suffer more from the spillover effects, leading to {\color{revisionblue}larger increases in cigarette sales}.

	\section{Monte Carlo Simulations}
	\label{section_simulation}
	
{\color{revisionblue}We present the Monte Carlo simulation results in this section. Unless otherwise stated, we use 5,000 simulation replications.}



	\subsection{Estimation with spillover effects}
	\label{sec_estimation_simulation_design}
	
{\color{revisionblue}
In this section we examine the finite sample performance of the estimation procedure proposed in Section \mainref{section_estimation}{3}. 

\paragraph{Monte Carlo design.}
The design places $y_{i,t}(0)$ in a factor model in the spirit of \cite{Li2019}. We report a stationary case and an $\mathcal{I}(1)$ case side by side.
The two cases share the same panel, the same treatment and spillover effects, and the same loading geometry. They differ only in the common factor processes.

There are 51 units, consisting of one treated unit, 13 exposed control units, and 37 pure donors. Each replication generates 50 pretreatment periods and one post-treatment period, and the estimators are evaluated in the post-treatment period. The untreated outcome is
\begin{equation*}
	y_{i,t}(0)=\lambda_i(r)'f_t+\varepsilon_{i,t},
\end{equation*}
where $f_t=(f_{1,t},f_{2,t},f_{3,t})'$ and each $\varepsilon_{i,t}$ is an independent standard normal draw that is independent of the factors and the loadings. In the post-treatment period the treated unit receives a direct effect of 5, each of the 13 exposed control units receives a spillover effect of 3, and the pure donors are unaffected. The estimand is the direct effect 5. Estimation uses the saturated effect design
\begin{equation*}
	A=\begin{bmatrix}I_{14}\\0_{37\times14}\end{bmatrix},
\end{equation*}
so the fact that the 13 true spillover effects happen to share the common value 3 is never imposed on the estimator.

The loading geometry indexes how far the pure donors lie from the treated unit. For unit $i$, let
\begin{equation*}
	d_i=\Bigl(\sqrt{1-\eta^2},\ \eta\cos\bigl(2\pi (i \bmod 3)/3\bigr),\ \eta\sin\bigl(2\pi (i \bmod 3)/3\bigr)\Bigr)',\qquad \eta=0.2,
\end{equation*}
and let $\lambda_i(r)=(-6,0,0)'+s_i(r)d_i$. The treated unit has $s_i(r)=0$. The 13 exposed displacements are equally spaced from $-0.15$ to $0.15$. For the 37 pure donors, $s_i(r)=rq_i$, where the $q_i$ are equally spaced from 3 to 6. Each $d_i$ has unit length, so the pure-donor distances run from $3r$ to $6r$ and $r$ has a direct distance interpretation. We report 13 values of $r$ between 0 and 1. At $r=0$ the pure donors sit at the treated loading and are maximally informative, and larger values of $r$ make them progressively less comparable to the treated unit.

In the stationary case the factors are
\begin{align*}
	f_{1,t}&=0.5f_{1,t-1}+\nu_{1,t},\\
	f_{2,t}&=1+\nu_{2,t}+0.5\nu_{2,t-1},\\
	f_{3,t}&=0.5f_{3,t-1}+\nu_{3,t}+0.5\nu_{3,t-1},
\end{align*}
initialized from their stationary distributions. In the $\mathcal{I}(1)$ case the factors are
\begin{align*}
	f_{1,t}&=f_{1,t-1}+0.5\nu_{1,t},\\
	f_{2,t}&=f_{2,t-1}+0.5\nu_{2,t},\\
	f_{3,t}&=0.5f_{3,t-1}+\nu_{3,t},
\end{align*}
where the first two processes start from independent standard normal draws and the third starts from its stationary distribution. All $\nu_{j,t}$ are independent standard normal draws. 

\paragraph{Estimators.}
We compare five estimators over 5,000 independently generated replications. SP is our proposed method. PD is the conventional synthetic control estimator restricted to the 37 pure donors, with an intercept and simplex weights. iSCM is our adapted inclusive synthetic control benchmark following \citet{DiStefano2020}, which solves the affected-unit subsystem using the same all-unit intercepts and weights. PSCM is the penalized synthetic control benchmark adapted from \citet{Grossi2020}, which uses pure donors without an intercept and selects its penalty over a grid from 0.001 to 1 by leave-one-out post-treatment prediction error.
PD-OLS replaces the simplex restriction in PD with an unrestricted intercept and ordinary least squares coefficients.

}

\begin{figure}[ht]
	\centering
	\captionsetup{font={small,color=revisionblue},labelfont={color=revisionblue}}
	\includegraphics[width=.95\textwidth]{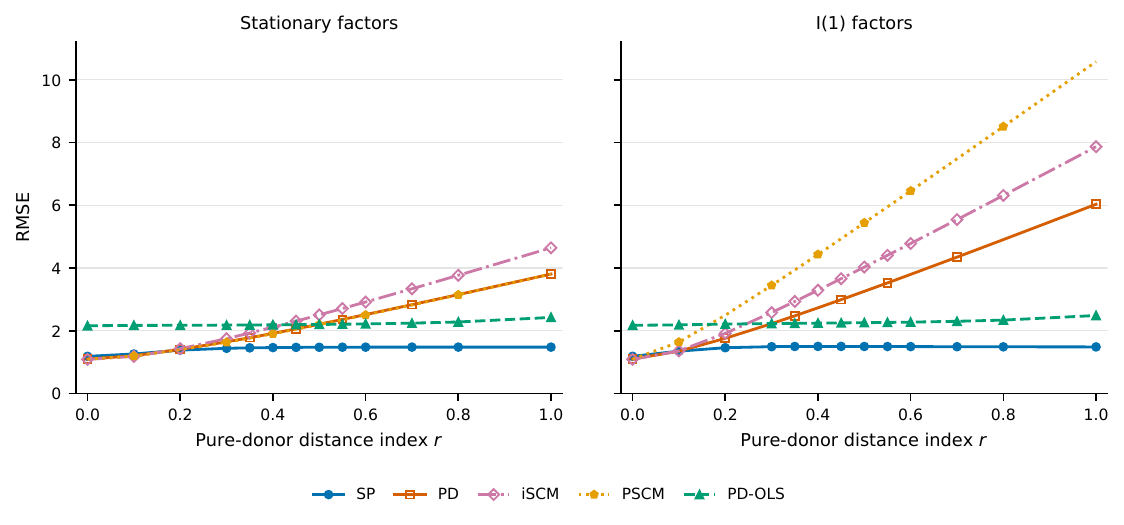}
	\caption{Treatment-effect RMSE under stationary and $\mathcal{I}(1)$ factor designs. 
    }
	\label{fig:estimator_comparison_rmse_n50_t50}
\end{figure}

\begin{table}[p]
\centering
\color{revisionblue}
\captionsetup{font={small,color=revisionblue},labelfont={color=revisionblue}}
\caption{Bias, standard deviation, and RMSE of treatment-effect estimators under stationary and $\mathcal{I}(1)$ factor designs. 
}
\label{tab:estimator_comparison_bias_sd_rmse_n50_t50}
\renewcommand{\arraystretch}{1.05}
\setlength{\tabcolsep}{4pt}
\small
\begin{threeparttable}
\begin{tabular}{@{}l*{6}{>{$}c<{$}}@{}}
\hline
\hline
& \multicolumn{6}{c}{Pure-donor distance index ($r$)} \\
\cline{2-7}
Method & 0 & 0.2 & 0.4 & 0.6 & 0.8 & 1 \\
\hline
\multicolumn{7}{l}{\textit{Panel A. Stationary factors}} \\
SP & 0.006 & 0.012 & 0.009 & 0.006 & 0.004 & 0.003 \\
 & (1.184) & (1.382) & (1.464) & (1.477) & (1.479) & (1.478) \\
 & [1.184] & [\mathbf{1.382}] & [\mathbf{1.464}] & [\mathbf{1.477}] & [\mathbf{1.479}] & [\mathbf{1.478}] \\[1pt]
PD & 0.006 & 0.022 & 0.032 & 0.041 & 0.053 & 0.066 \\
 & (1.100) & (1.407) & (1.916) & (2.515) & (3.152) & (3.807) \\
 & [1.099] & [1.408] & [1.916] & [2.515] & [3.152] & [3.807] \\[1pt]
iSCM & 0.005 & 0.022 & 0.041 & 0.060 & 0.079 & 0.096 \\
 & (1.087) & (1.427) & (2.114) & (2.915) & (3.767) & (4.642) \\
 & [\mathbf{1.087}] & [1.427] & [2.114] & [2.916] & [3.768] & [4.642] \\[1pt]
PSCM & 0.005 & 0.021 & 0.030 & 0.034 & 0.034 & 0.034 \\
 & (1.094) & (1.396) & (1.903) & (2.503) & (3.145) & (3.807) \\
 & [1.094] & [1.396] & [1.903] & [2.503] & [3.145] & [3.807] \\[1pt]
PD-OLS & 0.027 & 0.002 & -0.007 & -0.010 & -0.013 & -0.020 \\
 & (2.158) & (2.173) & (2.189) & (2.218) & (2.279) & (2.424) \\
 & [2.158] & [2.173] & [2.189] & [2.218] & [2.279] & [2.424] \\
\hline
\multicolumn{7}{l}{\textit{Panel B. $\mathcal{I}(1)$ factors}} \\
SP & 0.006 & 0.007 & 0.007 & 0.008 & 0.007 & 0.008 \\
 & (1.185) & (1.459) & (1.501) & (1.495) & (1.489) & (1.485) \\
 & [1.185] & [\mathbf{1.459}] & [\mathbf{1.501}] & [\mathbf{1.495}] & [\mathbf{1.488}] & [\mathbf{1.485}] \\[1pt]
PD & 0.006 & 0.017 & 0.031 & 0.048 & 0.065 & 0.085 \\
 & (1.100) & (1.758) & (2.731) & (3.796) & (4.904) & (6.030) \\
 & [1.099] & [1.758] & [2.731] & [3.796] & [4.904] & [6.030] \\[1pt]
iSCM & 0.005 & 0.025 & 0.051 & 0.075 & 0.099 & 0.123 \\
 & (1.087) & (1.917) & (3.292) & (4.781) & (6.311) & (7.865) \\
 & [\mathbf{1.087}] & [1.917] & [3.293] & [4.781] & [6.312] & [7.865] \\[1pt]
PSCM & 0.005 & 0.051 & 0.089 & 0.129 & 0.170 & 0.214 \\
 & (1.094) & (2.494) & (4.432) & (6.455) & (8.506) & (10.574) \\
 & [1.094] & [2.494] & [4.432] & [6.456] & [8.507] & [10.575] \\[1pt]
PD-OLS & 0.021 & 0.019 & 0.020 & 0.020 & 0.018 & 0.013 \\
 & (2.173) & (2.212) & (2.243) & (2.275) & (2.339) & (2.486) \\
 & [2.173] & [2.211] & [2.243] & [2.275] & [2.339] & [2.486] \\
\hline
\end{tabular}
\begin{tablenotes}
\footnotesize
\item Notes: For each method and $r$, bias appears first, the standard deviation in parentheses appears second, and RMSE in brackets appears third. All statistics use 5,000 Monte Carlo replications. Bias is the mean estimate minus the true treatment effect of 5. The standard deviation is the square root of the sample variance across replications, using denominator $5{,}000-1$. RMSE is the square root of the mean squared estimation error. The lowest RMSE within each factor design and $r$ column is in bold.
The five estimators are defined in Section~\ref{sec_estimation_simulation_design}.
\end{tablenotes}
\end{threeparttable}
\end{table}

{\color{revisionblue}
\paragraph{Results.}
Figure \ref{fig:estimator_comparison_rmse_n50_t50} plots RMSE against $r$ for the two factor designs, and Table \ref{tab:estimator_comparison_bias_sd_rmse_n50_t50} reports bias, standard deviation, and RMSE at six values of $r$. The left panel of Figure \ref{fig:estimator_comparison_rmse_n50_t50} is exactly the stationary result reported in Figure \mainref{fig:estimator_comparison_rmse}{2} of the main text, repeated here so that the two factor designs can be read side by side.

All five estimators are close to unbiased in both designs. The largest absolute bias anywhere in the table is 0.214, and the bias of SP never exceeds 0.012. Since the estimand is 5, these biases are negligible relative to the standard deviations, so the comparison between methods is a comparison of dispersion.

The behavior in $r$ is what separates the methods. The RMSE of SP is nearly flat in both designs. The three benchmarks that rely on pure donors deteriorate steadily as the pure donors become less comparable, and the $\mathcal{I}(1)$ design magnifies this deterioration. PD-OLS is also relatively stable in $r$, but it is uniformly less precise than SP.

At $r=0$ the pure donors are ideal comparisons, and there SP is slightly less precise than the benchmarks, with an RMSE of 1.184 against 1.087 for iSCM. This gap is the price of estimating 14 effect parameters when a pure-donor comparison would already suffice. The price is small, and it is recovered immediately. For every $r$ at or above 0.2, SP has the lowest RMSE in both factor designs, and its advantage widens as $r$ grows. The practical message is that adjusting for spillovers costs little when a clean donor pool exists and gains a great deal when one does not.
}

	{\color{revisionblue}
	\subsection{Monte Carlo comparison of inferential methods}
	\label{sec_simulation_treatment}
	}
	{
\color{revisionblue}

This section studies the finite-sample performance of our procedure and the
alternatives. We report complete results for the stationary and
\(\mathcal{I}(1)\) factor designs, including nominal rejection probabilities
and size-adjusted power.

\paragraph{Monte Carlo design.}
We use the same panel, loading geometry, spillover effects, and shock processes
as the estimation simulation in Section
\ref{sec_estimation_simulation_design}. The stationary and
\(\mathcal{I}(1)\) cases use exactly the same respective factor processes as
in that simulation. We restrict attention to \(r\in\{0,0.5,1\}\), which spans
the range from pure donors that match the treated loading exactly to pure
donors that are far from it. Each exposed unit again receives a spillover
effect of 3. The direct effect assigned to the treated unit varies over
\(\{0,0.5,1,\dots,5\}\), so that the value 0 measures size and the remaining
values trace out power. We use 5,000 Monte Carlo replications and a nominal
level of five percent throughout.

\paragraph{Procedures.}
We compare the six procedures defined in Section
\mainref{section_inference_theory_comparison}{5.3.1}, namely SP-A, PD-P, PD-A,
iSCM-P, PSCM-O, and PSCM-C. 
For both Andrews
procedures, the post-treatment statistic uses the fit from all pretreatment
periods, while each pretreatment reference statistic is formed after omitting
that period. PD-A re-estimates \(\widehat{\beta}_1^{(t)}\), and SP-A refits the
entire first-stage system and recomputes \(G(\widehat{\theta}^{(t)})\). Thus,
both use a leave-one-pretreatment-period-out reference construction, which we
choose to improve finite-sample performance.
In both PSCM-O and
PSCM-C, we use 999 resampling draws.

Raw rejection frequencies are not comparable across procedures whose sizes
differ. For a fair comparison, we therefore also report size-adjusted power.
For each procedure, factor design, and value of \(r\), we replace the nominal
critical value by the Monte Carlo 95th percentile of that procedure's statistic
under the null. The resulting test has size 0.05 by construction. Following
\citet{Davidson1998}, this provides a common basis for comparing power across
procedures with different size distortions. The adjustment is a diagnostic
rather than a usable procedure because it calibrates the critical value on a
known null that a practitioner does not have \citep{Horowitz2000}.

\begin{table}[H]
\centering
\color{revisionblue}
\captionsetup{font={small,color=revisionblue},labelfont={color=revisionblue}}
\caption{Empirical size under a zero direct effect and nominal and size-adjusted power against a direct effect of five.}
\label{tab:inference_comparison}
\renewcommand{\arraystretch}{1.15}
\small
\begin{threeparttable}
\begin{tabular}{lccccccccccc}
\hline
\hline
& \multicolumn{3}{c}{Size under the null}&& \multicolumn{3}{c}{Nominal power}&& \multicolumn{3}{c}{Size-adjusted power} \\
\cline{2-4}\cline{6-8}\cline{10-12}
Procedure & $r=0$ & $r=0.5$ & $r=1$ && $r=0$ & $r=0.5$ & $r=1$ && $r=0$ & $r=0.5$ & $r=1$ \\
\hline
\multicolumn{12}{l}{\emph{Stationary factors}} \\
SP-A              & 0.062 & 0.067 & 0.066 && 0.989 & 0.921 & 0.917 && 0.991 & \textbf{0.923} & \textbf{0.920} \\
PD-P              & 0.052 & 0.029 & 0.024 && 0.988 & 0.515 & 0.203 && 0.994 & 0.606 & 0.254 \\
PD-A              & 0.063 & 0.068 & 0.069 && 0.996 & 0.655 & 0.298 && \textbf{0.996} & 0.627 & 0.261 \\
iSCM-P            & 0.066 & 0.327 & 0.584 && 0.992 & 0.844 & 0.736 && 0.995 & 0.501 & 0.189 \\
PSCM-O            & 0.593 & 0.565 & 0.666 && 1.000 & 0.952 & 0.841 && 0.959 & 0.557 & 0.254 \\
PSCM-C            & 0.059 & 0.258 & 0.475 && 0.991 & 0.862 & 0.737 && 0.988 & 0.613 & 0.277 \\
\hline
\multicolumn{12}{l}{\emph{$\mathcal{I}(1)$ factors}} \\
SP-A              & 0.064 & 0.071 & 0.066 && 0.989 & 0.917 & 0.920 && 0.990 & \textbf{0.918} & \textbf{0.920} \\
PD-P              & 0.052 & 0.061 & 0.061 && 0.988 & 0.474 & 0.243 && 0.994 & 0.449 & 0.217 \\
PD-A              & 0.063 & 0.149 & 0.173 && 0.996 & 0.604 & 0.380 && \textbf{0.996} & 0.325 & 0.121 \\
iSCM-P            & 0.065 & 0.513 & 0.738 && 0.992 & 0.734 & 0.735 && 0.996 & 0.226 & 0.091 \\
PSCM-O            & 0.592 & 0.781 & 0.855 && 1.000 & 0.835 & 0.862 && 0.959 & 0.215 & 0.078 \\
PSCM-C            & 0.059 & 0.595 & 0.771 && 0.991 & 0.730 & 0.788 && 0.988 & 0.158 & 0.075 \\
\hline
\end{tabular}
\begin{tablenotes}
\small
\item Notes: Entries are rejection frequencies over 5,000 Monte Carlo replications under the inference design of Section \ref{sec_simulation_treatment}, which uses the loading geometry and the stationary and $\mathcal{I}(1)$ factor processes of Section \ref{sec_estimation_simulation_design}. PSCM-O and PSCM-C use 999 resampling draws. The left block reports rejection when the direct effect is zero, against a nominal level of 0.05. The middle block reports rejection when the direct effect is five using each procedure's nominal critical value. The right block replaces each procedure's critical value by the Monte Carlo 95th percentile of its own null statistic, so that every procedure has size 0.05 by construction, and reports rejection when the direct effect is five. The highest size-adjusted power in each column is in bold. The six procedures are defined in Section \mainref{section_inference_theory_comparison}{5.3.1}.
\end{tablenotes}
\end{threeparttable}
\end{table}

\paragraph{Results.}
Table \ref{tab:inference_comparison} reports size, nominal power, and
size-adjusted power against a direct effect of five. Figures
\ref{fig:inference_comparison_power_stationary_six_panel} and
\ref{fig:inference_comparison_power_i1_six_panel} trace nominal rejection
probabilities in their top rows and size-adjusted power in their bottom rows.

SP-A is the only procedure that holds size and retains power across the whole
range in both factor designs. Its size ranges from 0.062 to 0.071. Its
size-adjusted power against a direct effect of five falls from 0.991 to 0.920
under stationary factors and from 0.990 to 0.920 under \(\mathcal{I}(1)\)
factors.

PD-P also holds size in both designs, but its size-adjusted power falls as the
pure donors move away from the treated unit. At \(r=1\), it reaches only 0.254
under stationary factors and 0.217 under \(\mathcal{I}(1)\) factors. PD-A has
similar size and power under stationary factors, but under \(\mathcal{I}(1)\)
factors its null rejection probability rises to 0.173 and its size-adjusted
power falls to 0.121 at \(r=1\). The loss of pure-donor support therefore
appears primarily as lost power for PD-P and as both lost power and size
distortion for PD-A in the nonstationary design.

iSCM-P becomes increasingly oversized as \(r\) rises. At \(r=1\), its null
rejection probability is 0.584 under stationary factors and 0.738 under
\(\mathcal{I}(1)\) factors, while its size-adjusted power is only 0.189 and
0.091. Its second stage uses only the affected-unit equations, so its induced
pure-donor weights stay inside the convex hull and the resulting prediction
error grows as the pure donors become less comparable to the treated unit.

PSCM-O over-rejects even at \(r=0\), where its null rejection probability is
0.593 under stationary factors and 0.592 under \(\mathcal{I}(1)\) factors.
PSCM-C reduces this probability to 0.059 in both designs, which supports
the explanation that the omitted treated shock causes the distortion when the
pure donors match the treated loading. The correction does not restore size
as support weakens. At \(r=1\), PSCM-C rejects under the null with probability
0.475 under stationary factors and 0.771 under \(\mathcal{I}(1)\) factors.
After size adjustment, the higher power of the two PSCM procedures is 0.277 in
the stationary case and 0.078 in the \(\mathcal{I}(1)\) case at \(r=1\).

The bottom rows of the figures show why high nominal rejection by an oversized
procedure should not be interpreted as high power. At \(r=1\), SP-A has
size-adjusted power of 0.920 in both factor designs, whereas no alternative
exceeds 0.277 under stationary factors or 0.217 under \(\mathcal{I}(1)\) factors.
The \(\mathcal{I}(1)\) results likewise show SP-A's substantial power advantage when
pure-donor support is weak, while the alternatives' size-adjusted power declines further.

\begin{figure}[H]
	\centering
	\captionsetup{font={small,color=revisionblue},labelfont={color=revisionblue}}
	\includegraphics[width=\textwidth]{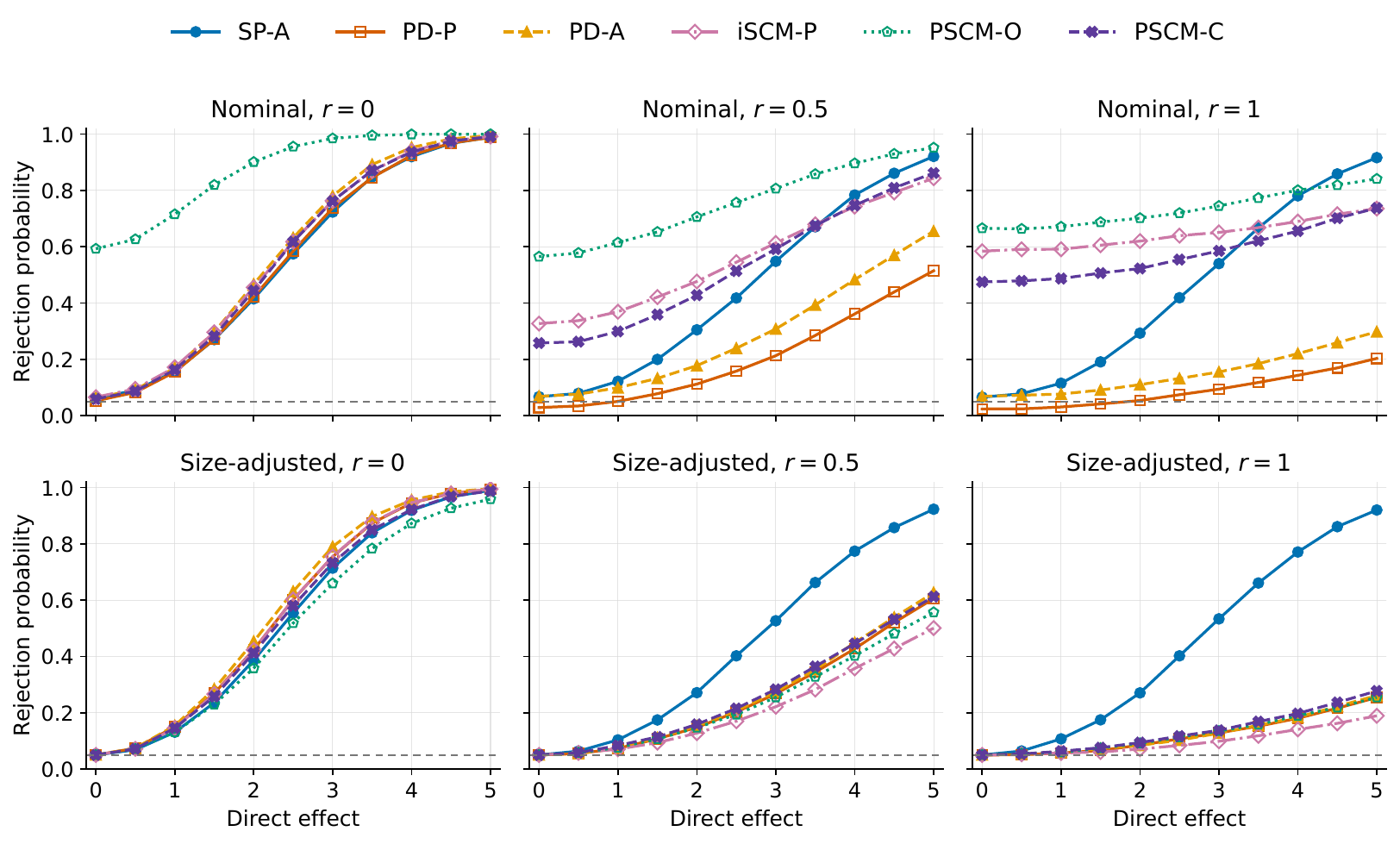}
	\caption{Nominal rejection probabilities (top row) and size-adjusted power
	(bottom row) under the stationary factor design. The top row reproduces
	Figure \mainref{fig:inference_comparison_power}{3}. Columns set the
	pure-donor distance index to \(r=0\), \(0.5\), and \(1\). Size adjustment
	replaces each procedure's critical value by its Monte Carlo 95th percentile
	under the null. The dashed line marks 0.05. Curves use 5,000 Monte Carlo
	replications and 999 resampling draws.}
	\label{fig:inference_comparison_power_stationary_six_panel}
\end{figure}

\begin{figure}[H]
	\centering
	\captionsetup{font={small,color=revisionblue},labelfont={color=revisionblue}}
	\includegraphics[width=\textwidth]{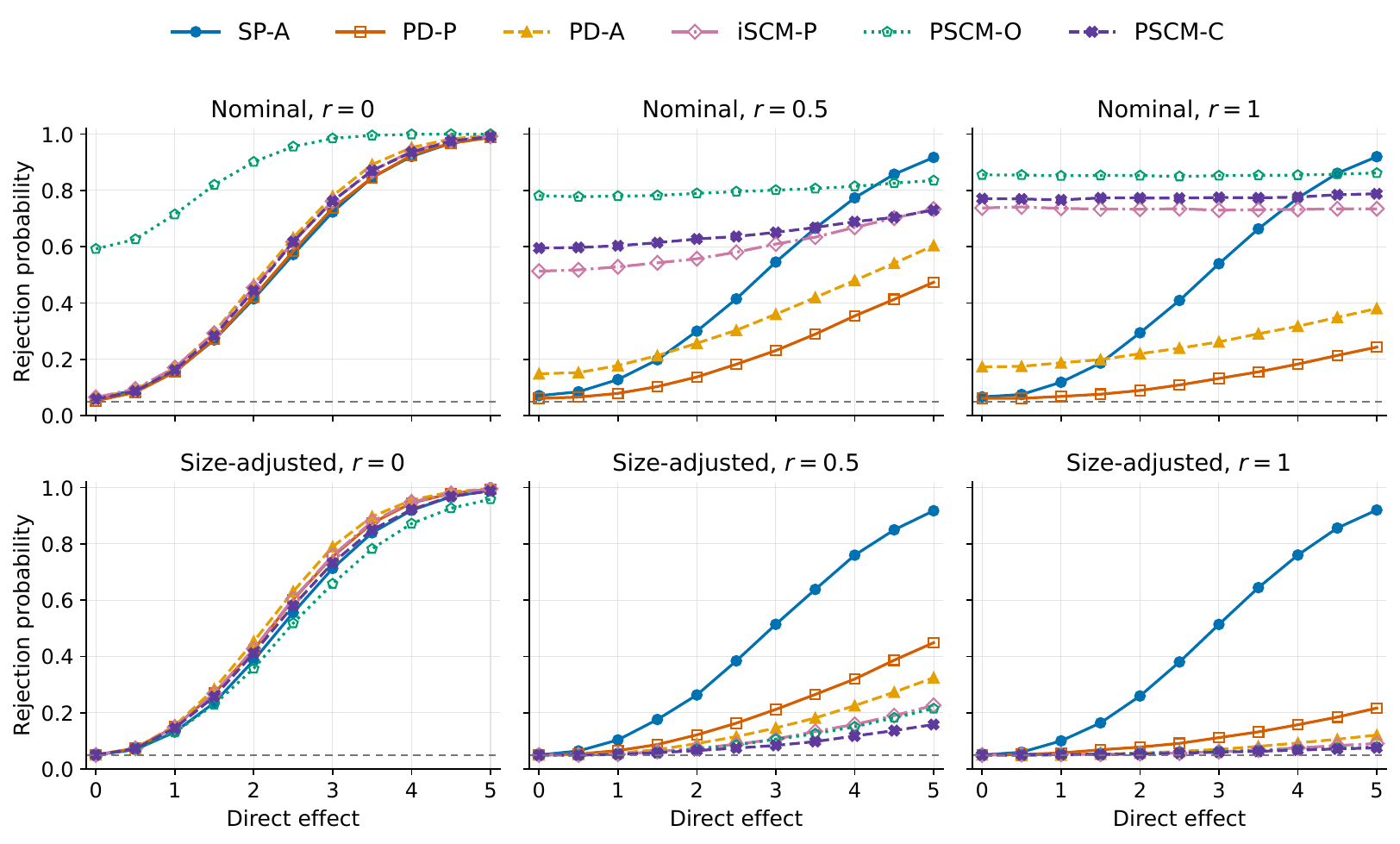}
	\caption{Nominal rejection probabilities (top row) and size-adjusted power
	(bottom row) under the \(\mathcal{I}(1)\) factor design. Columns set the
	pure-donor distance index to \(r=0\), \(0.5\), and \(1\). Size adjustment
	replaces each procedure's critical value by its Monte Carlo 95th percentile
	under the null. The dashed line marks 0.05. Curves use 5,000 Monte Carlo
	replications and 999 resampling draws.}
	\label{fig:inference_comparison_power_i1_six_panel}
\end{figure}

}

	{\color{revisionblue}
	\subsection{Power of the $\kappa_A$ specification test}
	\label{sec_simulation_kappaA}
	This section evaluates the finite-sample size and power of the $\kappa_A$ specification test proposed in Section \mainref{section_kappaA_statistic}{5.1.2} against two types of spillover-structure misspecification. The first omits exposed units from the assumed spillover range. The second correctly specifies the exposed set but incorrectly imposes a common spillover effect.

\paragraph{Monte Carlo design.}
Both experiments use the stationary factor design in Section \ref{sec_estimation_simulation_design} with $r=0.5$. The baseline design has one treated unit, 13 designated exposed controls, and 37 designated pure donors, with 50 pretreatment periods and one post-treatment period. The direct effect is set to zero. Because the corresponding column belongs to both assumed structures, its value does not affect $\kappa_A$. The nominal level is five percent, and each experiment uses 1,000 Monte Carlo replications.

For the pretreatment reference distribution, we use a leave-one-period-out construction. For each $t\leq T$, we omit period $t$, re-estimate all 51 first-stage synthetic-control equations using the other 49 periods, and construct the corresponding $\widehat\Gamma_A^{(-t)}$. Here, $\widehat a^{(-t)}$ and $\widehat B^{(-t)}$ are the leave-one-period-out versions of $\widehat a$ and $\widehat B$ as defined by Equation~\mainequationref{optimization}{3}. The held-out reference statistic is
\[
\kappa_{A,t}^{(-t)}=
\left\|(I-\widehat\Gamma_A^{(-t)})
\left[(I-\widehat B^{(-t)})Y_t-\widehat a^{(-t)}\right]\right\|.
\]
We reject when the post-treatment $\kappa_A$ exceeds the empirical 95th percentile of these 50 reference statistics.

The range experiment assumes the limited-range structure in Example \mainref{nonparametric spillover}{1}. The 13 designated exposed controls each receive a spillover effect of 3. To introduce misspecification, the 13 controls nearest to the treated unit within the baseline pure-donor set also receive a common spillover $s\in\{0,0.25,\ldots,3\}$ but remain excluded from $A$. The specification is therefore correct when $s=0$ and increasingly incomplete as $s$ rises.

The functional-form experiment assumes the homogeneous structure in Example \mainref{equally hit}{2}, which gives the 13 exposed controls a common spillover parameter. Let $h\in[0,1]$ denote the level of spillover heterogeneity. The exposed set is correctly specified, but the true effects are evenly spaced from $3(1-h)$ to $3(1+h)$. In the simulation, $h\in\{0,1/12,\ldots,1\}$. We randomly permute these effects across the exposed units in each replication. Their mean remains 3, so a larger $h$ means greater heterogeneity without changing the average spillover effect. The assumed homogeneous structure is correct when $h=0$ and misspecified when $h>0$.

\begin{figure}[H]
	\centering
	\captionsetup{font={small,color=revisionblue},labelfont={color=revisionblue}}
	\begin{subfigure}[t]{.49\textwidth}
		\centering
		\includegraphics[width=\linewidth]{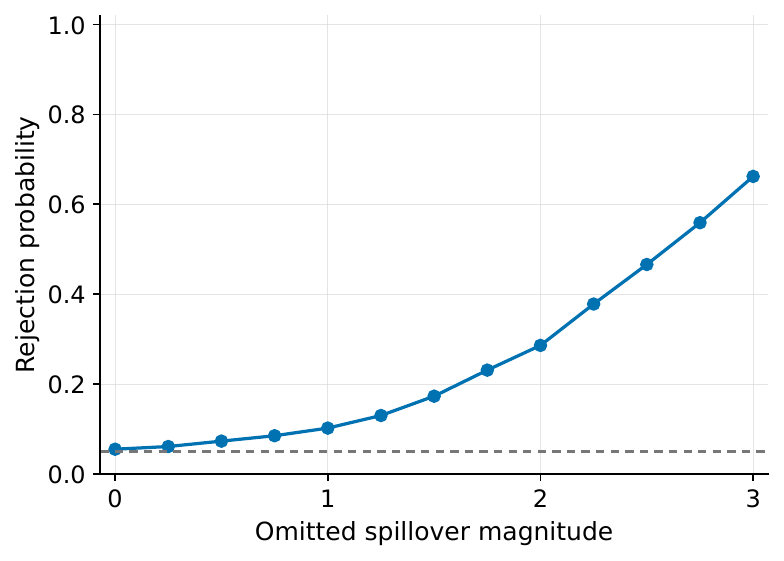}
		\caption{Exposure-range misspecification}
		\label{fig:kappa_power_range}
	\end{subfigure}\hfill
	\begin{subfigure}[t]{.49\textwidth}
		\centering
		\includegraphics[width=\linewidth]{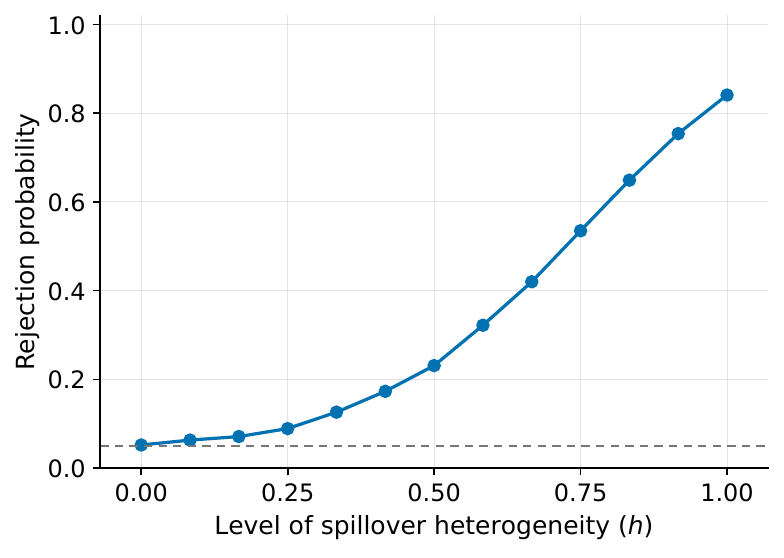}
		\caption{Functional-form misspecification}
		\label{fig:kappa_power_heterogeneity}
	\end{subfigure}
	\caption{Rejection probability of the $\kappa_A$ specification test. Panel (a) varies the common spillover effect on the 13 omitted exposed units. Panel (b) varies the level of spillover heterogeneity $h$ while keeping the exposed set correctly specified. Results are based on 1,000 replications of the stationary design with $r=0.5$ and leave-one-period-out critical values. The dashed lines mark the nominal level of 0.05.}
	\label{fig:kappa_power}
\end{figure}

\paragraph{Results.}
Figure \ref{fig:kappa_power} shows approximately correct size under both correctly specified designs. At zero misspecification, the rejection probabilities are 0.055 under the limited-range structure and 0.052 under the homogeneous structure. Power rises steadily with the departure. It reaches 0.662 when the omitted spillover equals 3 and 0.841 when the heterogeneous effects span 0 to 6. As a check, the more flexible limited-range structure remains correctly specified throughout the heterogeneity experiment and rejects with probability 0.055 at every value of $h$. Thus, in this design, the test has increasing power against both range and functional-form misspecification while a flexible structure that accommodates the heterogeneous effects remains at nominal size.

	}
	\subsection{Test for existence of spillover effects}
	\label{sec_simulation_spillover}
\label{stationary simulation}

We use the stationary factor model $y_{i,t}(0)=\eta_t+\lambda_i'f_t+\varepsilon_{i,t}$, where $f_t=(f_{1,t},f_{2,t},f_{3,t})'$ and
\begin{align*}
\eta_t&=1+0.5\eta_{t-1}+\nu_{0,t},\\
f_{1,t}&=0.5f_{1,t-1}+\nu_{1,t},\\
f_{2,t}&=1+\nu_{2,t}+0.5\nu_{2,t-1},\\
f_{3,t}&=0.5f_{3,t-1}+\nu_{3,t}+0.5\nu_{3,t-1}.
\end{align*}
The innovations $\varepsilon_{i,t}$ and $\nu_{j,t}$ are mutually independent standard normal draws. Each entry of $\lambda_i$ is drawn independently from the uniform distribution on $[0,1]$ and held fixed across replications. At $t=T+1$, the observed outcome is $y_{i,T+1}=y_{i,T+1}(0)+\alpha_i$.

In this section, we examine the power of the proposed test against the null hypothesis that there are no spillover effects. We also look into its behavior when the range of the spillover effect is not correctly specified. In this set of experiments, the level of spillover effects varies from 0 to 2, corresponding to the strength of alternative hypotheses. We set $(N,T)=(20,50)$ and $\alpha_1=5$. There are 9 units that are affected by spillover effects. The model for the range of spillover is as in Example \mainref{nonparametric spillover}{1}.

\begin{figure}[ht]
	\centering
	\includegraphics[trim={0 7cm 0 7cm},scale=.7]{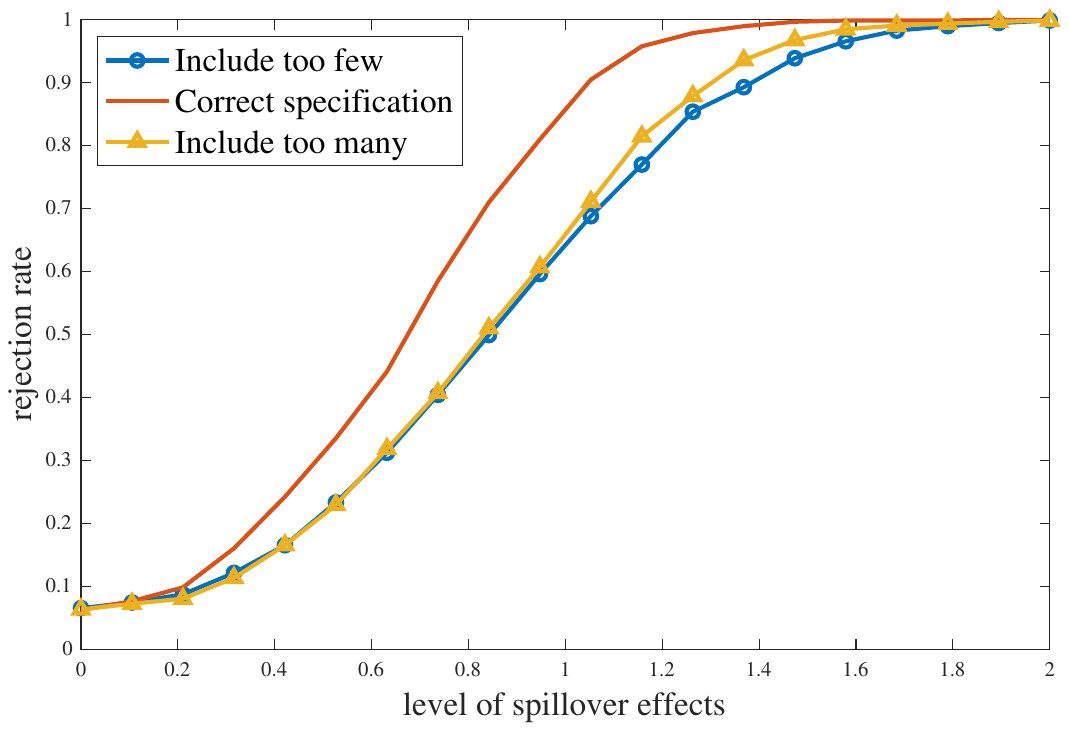}
	\caption{\small Empirical Rejection Rate of Testing for the Existence of Spillover Effects. There are 20 units in total, half of which are affected by the treatment. \emph{Include too few} is assuming only 5 of them are affected by the treatment. \emph{Correct specification} assumes the researcher knows exactly which set of units are affected. \emph{Include too many} assumes 15 units are affected, 5 of which are, in fact, not affected. }
	\label{testing spillover illustration}
\end{figure}

The empirical rejection rates against various levels of spillover effects using our method proposed in Section \mainref{spillover effect test}{4.2} are plotted in Figure \ref{testing spillover illustration}. Here \emph{Include too few} misses half of the units that are actually affected by the treatment (assuming that unit 1 as well as four other units are affected), \emph{Correct specification} assumes we know exactly which units are affected, and \emph{Include too many} assumes 15 units are affected in estimation, five of which are actually not affected by spillover effects.

Among the three cases, \emph{Include too many} is still a correct specification but is supposed to be more conservative, so it has less power than \emph{Correct specification}. 
Note that the range of spillover effects for  \emph{Include too few} is correctly specified only when the level of spillover effects is zero.
Its power curve is similar to that of \emph{Include too many}. 

	\subsection{Spatial correlation and SCM weights}
	\label{section_sim_spatial}
	\label{sec_simulation_spatial}

Spatial settings are common in economics \citep{cao_inference_2025,cao_neighborhood_2025}.
In this section, we conduct simulation studies to illustrate a spatially correlated scenario in which spillover effects increase as the distance between treated and control units decreases. At the same time, the synthetic control method is found to assign greater weights to units that are closer to the treated unit. Although this example is not exhaustive, it highlights the potential risks associated with ignoring spillovers in SCM settings. In particular, SCM may assign greater weights to units that are significantly affected by spillovers due to their relative importance and proximity to the treated unit.

We use the stationary outcome model and factor processes in Section \ref{sec_simulation_spillover}, with $N=10$ and $T=15$, but replace the factor loadings by $\lambda_i=\rho h_i+(1-\rho)\lambda_{i,0}$, where $\rho$ is the degree of spatial correlation, $h_i$ includes location data, and $\lambda_{i,0}$ is free of spatial information. Specifically, $h_i=(h_{i,1}, h_{i,2}, h_{i,3})'$ is generated such that $h_{i,1}$ and $h_{i,2}$ are independently and uniformly distributed over $[0,3]$, and $h_{i,3}$ follows an independent standard normal distribution. Here, $(h_{i,1}, h_{i,2})$ are the geographic coordinates of unit $i$. Control units are sorted by their distance to unit 1, so that unit 2 is the closest one. The non-spatial component $\lambda_{i,0}$ is drawn from an independent uniform distribution on $[0,1]$. All factor loadings are predetermined and remain fixed across replications.

The configuration of the spillover effects also reflects a spatial pattern. Following the framework in Example \ref{eg_spatial}, the spillover effect vector is $\alpha =(\alpha_1, b\exp(-d_2), \ldots, b\exp(-d_N))'$, where $d_i=\sqrt{(h_{1,1}-h_{i,1})^2+(h_{1,2}-h_{i,2})^2}$ represents the distance to the treated unit. The treatment effect $\alpha_1$ is set to 5, and the largest spillover effect $b\exp(-d_2)$ is set to 3.

\begin{figure}[h!]
	\centering
	\includegraphics[trim={0 6cm 0 6cm},scale=.5]{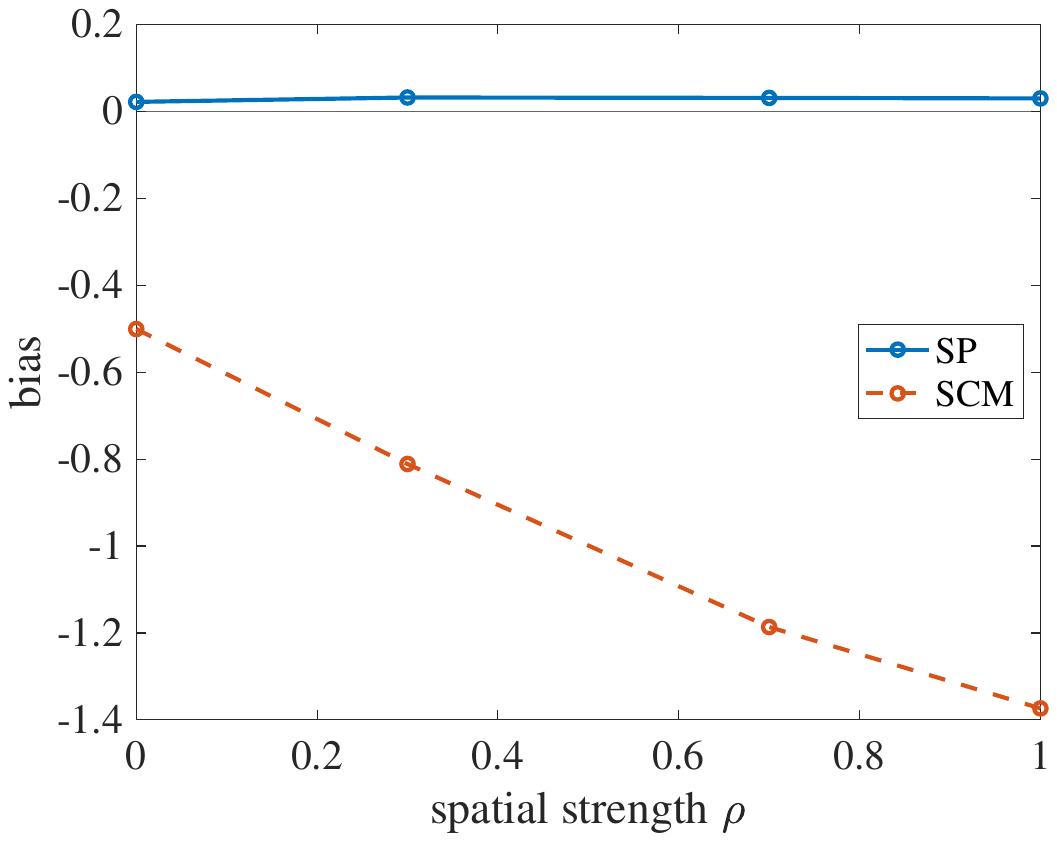}
	\caption{\small Bias of Treatment Effect Estimators in Spatially Correlated Data. This figure shows the simulated bias of two different treatment effect estimators within a spatial model framework. The horizontal axis shows the spatial loading parameter $\rho$, which ranges from 0 to 1. The solid line shows the bias of the SP estimator, the method proposed in this paper, while the dashed line shows the bias of the traditional synthetic control method, which does not account for spillover effects.
}
	\label{fig_spatial}
\end{figure}

The results are shown in Figure \ref{fig_spatial} and Table \ref{table_spatial_corr_and_weights}. 
SCM is the synthetic control method neglecting the possibilities of spillover effects. 
SP is our estimator, where we assume the spillover structure is known as described in Example \ref{eg_spatial}. 
We find that the bias of the usual synthetic control estimator increases in absolute value as spatial correlation increases. In contrast, our method remains relatively unbiased and stable across different levels of spatial correlation. 

\begin{table}[h!]
	\centering
	\caption{Relationship between Spatial Correlation and SCM Weights.}
	\label{table_spatial_corr_and_weights}
	\renewcommand{\arraystretch}{1.5}
	\footnotesize
	\begin{threeparttable}
		\begin{tabular}{lcccccccccccc}
			\hline
			\hline
  unit  $i$ & Distance to treated			&\multicolumn{4}{c}{$Corr[Y_{i,t},Y_{1,t}]$}         &        &    \multicolumn{4}{c}{Ave. SCM weight}    \\
			\cline{3-6} \cline{8-11}
		   &  & $\rho=0$ & $0.3$ & $0.7$ & $1$ && $\rho=0$ & $0.3$ & $0.7$ & $1$  \\
		      \hline
		unit 2  & 0.681    & 0.650      & 0.678       & 0.632       & 0.589      &  & 0.110       & 0.161        & 0.229        & 0.269       \\
		unit 3  & 0.951    & 0.680      & 0.741       & 0.808       & 0.839      &  & 0.126       & 0.244        & 0.401        & 0.465       \\
		unit 4  & 1.407    & 0.659      & 0.667       & 0.673       & 0.677      &  & 0.115       & 0.117        & 0.107        & 0.085       \\
		unit 5  & 1.549    & 0.661      & 0.719       & 0.725       & 0.716      &  & 0.146       & 0.110        & 0.060        & 0.034       \\
		unit 6  & 1.891    & 0.580      & 0.548       & 0.458       & 0.393      &  & 0.083       & 0.103        & 0.096        & 0.088       \\
		unit 7  & 2.144    & 0.587      & 0.636       & 0.693       & 0.721      &  & 0.083       & 0.050        & 0.029        & 0.017       \\
		unit 8  & 2.338    & 0.681      & 0.681       & 0.498       & 0.326      &  & 0.112       & 0.113        & 0.055        & 0.032       \\
		unit 9  & 2.675    & 0.632      & 0.540       & 0.399       & 0.318      &  & 0.113       & 0.054        & 0.016        & 0.007       \\
		unit 10 & 2.949    & 0.682      & 0.658       & 0.508       & 0.415      &  & 0.112       & 0.047        & 0.008        & 0.002      \\
			\hline
		\end{tabular}
		\begin{tablenotes}
			\small 
			\item Notes: 
			The table shows the relationship between spatial correlation and SCM weights. ``Distance to treated'' is the Euclidean distance from unit 1 to each respective unit. $Corr[Y_{i,t}, Y_{1,t}]$ is the correlation of each unit with unit 1. ``Ave. SCM weight'' is the average SCM weight assigned to each unit across replications, showing how SCM weights change in response to the spatial correlation.
		\end{tablenotes}
	\end{threeparttable}
\end{table}

A detailed examination of the Table \ref{table_spatial_corr_and_weights} provides insight into the underlying mechanism. At a spatial strength ($\rho$) of zero, the correlations between the treated unit and the controls are relatively equally-distributed, resulting in an (approximately) uniformly-distributed set of SCM weights. However, as $\rho$ increases, closer units tend to have a stronger correlation with unit 1 than more distant units, leading to more disproportionate weights to closer units. In our simulation, the unit that receives the highest average SCM weight (unit 3) also has the strongest correlation with the treated unit, except when $\rho=0$. This finding is consistent with the premise that the SCM may put more weights on units that are more affected by spillovers.

	\section{Proofs}
	\label{section_proof}

\renewenvironment{proof}{{\bfseries Proof of Proposition \mainref{prop_IminusB_rank_network}{1}.}}{\qed}
\noindent\begin{proof}
	Suppose a non-zero vector $w\in \mathbb R^{N}$ satisfies $(I-B)w=0$. 
	Let $i\in\arg\max_j |w_j|$. Suppose for contradiction that there exists $j$ where $e(i,j)=1$ and $|w_i|\ne |w_j|$. 
	Then, by the $i$-th row of $(I-B)w=0$, 
	$$|w_i|=\left|\sum_{j\ne i} b_{i,j} w_j\right|\le \sum_{e(i,j)=1}b_{i,j}|w_j|< \sum_{e(i,j)=1}b_{i,j}|w_i|=|w_i|,$$
	leading to contradiction, so $w_j$ is either $w_i$ or $-w_i$ for $e(i,j)=1$. 
	However, if $w_j=-w_i$ for any $j$ with $e(i,j)=1$, $w_i=\sum_{e(i,j)=1}b_{i,j}w_j$ will not hold. 
	Therefore, $w_j=w_i$ for any $j$ with $e(i,j)=1$. 
	
	Since we have a connected network, induction implies $w_j=w_i$ for any $j$. 
	This means the null space of $I-B$ has only one dimension, concluding our proof. 
	
\end{proof}

\renewenvironment{proof}{{\bfseries Proof of Theorem \mainref{unbiasedness}{1}.}}{\qed}
\noindent\begin{proof}
	Using formula of $\widehat{\gamma}$ in Equation \mainequationref{gamma estimation}{6}, we have
	\begin{align*}
		\widehat{\gamma} &=  (A'\widehat{M}A)^{-1}A'(I-\widehat{B})'((I-\widehat{B})Y_{T+1}(0)+(I-\widehat{B})\alpha-\widehat{a})\notag\\
		&=  (A'\widehat{M}A)^{-1}A'(I-\widehat{B})'(u_{T+1}+(B-\widehat{B})Y_{T+1}(0)+(a-\widehat{a})+(I-\widehat{B})A\gamma)\notag\\
		&= (A'\widehat{M}A)^{-1}A'(I-\widehat{B})'u_{T+1}+o_p(1)+o_p(1)+\gamma.
	\end{align*}
	The first equality is by $Y_{T+1}=Y_{T+1}(0)+\alpha$. The second equation is because $Y_{T+1}(0)=a+BY_{T+1}(0)+u_{T+1}$. The third equation is by (b) and (c) in Assumption \mainref{unbiased assumption}{1}. Therefore,
	\begin{align*}
		\widehat{\alpha}-(\alpha+Gu_{T+1})&=A(A'\widehat{M}A)^{-1}A'(I-\widehat{B})'u_{T+1}+A\gamma +o_p(1)-\alpha-Gu_{T+1}\notag\\
		&= (A(A'\widehat{M}A)^{-1}A'(I-\widehat{B})'-G)u_{T+1}+o_p(1)\notag\\
		&= o_p(1)O_p(1)+o_p(1)\notag\\
		&= o_p(1).
	\end{align*}
	The third equality is by (b) in Assumption \mainref{unbiased assumption}{1} and stationarity of $\{u_t \}_{t\ge 1}$.
	
\end{proof}

\renewenvironment{proof}{{\bfseries Proof of Lemma \mainref{estimation_lemma}{1}.}}{\qed}
\noindent \begin{proof}
	(i)	First assume that $A'MA$ is non-singular and Condition ST holds. The proof follows \cite{Ferman2016}, except that we do not assume that there is a set of weights that reconstruct the factor loadings and belong to the simplex. 
	Also, we proceed assuming $\eta_t=0$ for proof of part (a) and (b). 
	This is without loss of generality because none of $\widehat{a}$, $\widehat{B}$, and $u_t$ depends on $\{\eta_t\}$. 
	This also implies $\Omega_y=Cov[Y_t(0)]$. 
		
	We first show part (b). It suffices to show $|\widehat{a}_i-a_i|=o_p(1)$ and $\Vert \widehat{b}_i-b_i\Vert=o_p(1)$ for each $i$, i.e., $a_i$ and $b_i$ are well-defined. We show it for the $i=1$ case, and other cases follow the same strategy. Let $\bar{y}_j=T^{-1}\sum_{t=1}^Ty_{j,t}$. Write down an (equivalent) optimization problem
	\begin{equation*}
		\widehat{v}=\underset{v\in V}{\arg\min}\frac{1}{T}\sum_{t=1}^T\left( (y_{1,t}-\bar{y}_1)-\sum_{j=2}^N(y_{j,t}-\bar{y}_{j})v_j\right) ^2,
	\end{equation*}
	where $V=\{v=(v_2,\dots,v_N)\in \mathbb{R}_+^{N-1}:\sum_{j=2}^Nv_j=1 \}$. The objective is strictly convex (with probability approaching one), so the solution is unique. Note that it implies $\widehat{b}_1$ is numerically equivalent to $(0,\widehat{v}')'$, otherwise the minimization problem in forming $\widehat{a}_1$ and $\widehat{b}_1$ may have a lower objective evaluated at $(\bar{y}_1-\sum_{j=2}^N\bar{y}_j\widehat{v}_j,0,\widehat{v}')'$. Now we let  $\widehat{Q}(v)$ denote the objective function such that 
	\begin{equation*}
		\widehat{Q}(v)=\frac{1}{T}\sum_{t=1}^T\left( (y_{1,t}-\bar{y}_1)-\sum_{j=2}^N(y_{j,t}-\bar{y}_{j})v_j\right) ^2,
	\end{equation*}
	and its population analog be $Q(v)=(-1,v')\Omega_y(-1,v')'$. 
	Let $v_0$ be a minimizer of $Q(v)$ in $V$. We verify the conditions for consistency \citep[see][Theorem 2.1]{Newey1994}: (i) Since $\Omega_y$ is positive definite, $Q(v)$ is strictly convex. Also, $V$ is convex. Therefore, $Q(v)$ is uniquely minimized at $v_0$. (ii) $V$ is compact since it is an $(N-1)$-dimensional simplex. (iii) $Q(v)$ is continuous since it has a quadratic form. (iv) To see uniform convergence, note
	\begin{align*}
		\sup_{v\in V}|\widehat{Q}(v)-Q(v)| &=\sup_{v\in V}\left| \begin{bmatrix}
			-1\\v
		\end{bmatrix}'\left(\frac{1}{T}\sum_{t=1}^T(Y_t-\bar{Y})(Y_t-\bar{Y})'-\Omega_y \right)   \begin{bmatrix}
			-1\\v
		\end{bmatrix}\right| \notag\\
		&\le \sup_{v\in V}\left\Vert \begin{bmatrix}
			-1\\v
		\end{bmatrix}\right\Vert^2\left\Vert\frac{1}{T}\sum_{t=1}^T(Y_t-\bar{Y})(Y_t-\bar{Y})'-\Omega_y \right\Vert_F \notag\\
		&\le N\cdot o_p(1)\notag\\
		&= o_p(1),
	\end{align*}
	where $\Vert\cdot\Vert_F$ is the Frobenius norm. The second inequality is by ergodicity for the second moments. Therefore, $\widehat{v}\rightarrow_p v_0$. This implies $\Vert \widehat{b}_1-b_1\Vert=o_p(1)$. By ergodicity,
	\begin{equation*}
		\widehat{a}_1=\bar{y}_{1}-[\bar{y}_2\ \bar{y}_3\ \dots\ \bar{y}_N]\widehat{v}\rightarrow_p E[y_{1,t}(0)-Y_t(0)'b_1]=a_1. 
	\end{equation*}
	This shows part (b) and $E[u_{1,t}]=0$ by definition of $u_{i,t}$. We also have that $\{u_t \}_{t\ge 1}$ is stationary since it is a linear combination of stationary and ergodic processes. This shows part (a) in Assumption \mainref{unbiased assumption}{1}.
	
	Part (c) follows from part (b), the stationarity of $\{(\lambda_t,\varepsilon_t)\}_{t\ge 1}$, and $\Vert (\widehat B-B)\eta_{T+1}\Vert=o_p(1)$. Part (d) is assumed. 
	Thus, Assumption \mainref{unbiased assumption}{1} holds under the invertibility of $A'MA$ and Condition ST.
	
	\
	
	\noindent (ii)	Now, we assume that $A'MA$ is non-singular and Condition CO holds. 
	
	We first show part (c). We will show $\Vert Y_{T+1}(0)'(\widehat{b}_1-b_1)\Vert =o_p(1)$ and other $i$'s follows the same strategy. Since the synthetic control estimator can be written as a projection of the OLS estimator onto a closed convex set, we will first derive the asymptotic properties of the OLS estimator, and then use the properties of projections to obtain the desired results. For examples of this strategy, see \cite{Li2019} and \cite{Yu2019}. For some positive definite matrix $D\in \mathbb{R}^{N\times N}$, let $\mathbb{R}^{N}$ be a Hilbert space with the inner product $\langle\cdot,\cdot\rangle_D$ such that for $\theta_1,\theta_2\in \mathbb{R}^{N}$, 
	$		\langle \theta_1,\theta_2\rangle_D=\theta_1'D\theta_2.$
	The norm $\Vert \cdot\Vert_D$ is defined accordingly, i.e. $\Vert\theta\Vert_D=\sqrt{\theta'D\theta}$, for $\theta\in \mathbb{R}^{N}$. For a closed convex set $\Lambda\subset\mathbb{R}^{N}$, define a projection $\Pi_D$ such that for each $\theta\in \mathbb{R}^{N}$, $\Pi_D\theta = \arg\min_{\theta'\in \Lambda}\Vert \theta-\theta'\Vert_D$. \cite{Zarantonello1971} shows that for each $\theta,\theta'\in \mathbb{R}^{N}$, 
	\begin{equation}
		\label{projection inequality}
		\Vert \Pi_D\theta-\Pi_D\theta'\Vert_D\le \Vert \theta-\theta'\Vert_D. 
	\end{equation}
	
	With some abuse of notation, let $x_t=Y_t-T^{-1}\sum_{s=1}^TY_s$. Then, $\widehat{b}_1$ is the synthetic control weight estimators of regressing $(y_{1,t}-T^{-1}\sum_{s=1}^Ty_{1,s})$ on $x_t$, subject to $\{0 \}\times \Delta_{N-1}$ with $\Delta_{N-1}$ being an $(N-1)$-dimensional simplex. Let $\tilde{b}_1$ be the OLS estimator of regressing $(y_{1,t}-T^{-1}\sum_{s=1}^Ty_{1,s})$ on $x_t$. Let $\Sigma_T=T^{-1}\sum_{t=1}^Tx_tx_t'$.
	
	Appendix A.2 in \cite{Li2019} establishes that $\widehat{b}_1=\Pi_{\Sigma_T}\tilde{b}_1$. Thus, we have
	\begin{eqnarray}
		\label{square root convergence}
		\Vert\widehat{b}_1-b_1\Vert&=& \Vert\Sigma_T^{-1/2}\Sigma_T^{1/2}(\widehat{b}_1-b_1)\Vert\notag\\
		&\le&\Vert \Sigma_T^{-1/2}\Vert_F\cdot\Vert\Sigma_T^{1/2}(\widehat{b}_1-b_1)\Vert\notag\\
		&=&\Vert \Sigma_T^{-1/2}\Vert_F\cdot\Vert\widehat{b}_1-b_1\Vert_{\Sigma_T}\notag\\
		&=&\Vert \Sigma_T^{-1/2}\Vert_F\cdot\Vert \Pi_{\Sigma_T}\tilde{b}_1-\Pi_{\Sigma_T}b_1\Vert_{\Sigma_T}\notag\\
		&\le & \Vert \Sigma_T^{-1/2}\Vert_F\cdot \Vert \tilde{b}_1-b_1\Vert_{\Sigma_T}\notag\\
		&\le&\Vert \Sigma_T^{-1/2}\Vert_F\cdot\Vert \Sigma_T^{1/2}\Vert_F\cdot \Vert \tilde{b}_1-b_1\Vert\notag\\
		&=& O_p(1)o_p(T^{-1/2})\notag\\
		&=&o_p(T^{-1/2}),
	\end{eqnarray}
	where $\Vert \cdot\Vert_F$ is the Frobenius norm of a matrix. 	The third equality is because $b_1\in \{0 \}\times \Delta_{N-1}$. The second inequality is by \eqref{projection inequality}. For the stochastic-order bound, note
	\begin{equation*}
		\Sigma_T=T\left( \frac{1}{T^2}\sum_{t=1}^TY_tY_t'-\left( \frac{1}{T^{3/2}}\sum_{t=1}^TY_t\right)\left( \frac{1}{T^{3/2}}\sum_{t=1}^TY_t\right)'   \right),
	\end{equation*}
	so 
	\begin{equation*}
		\Vert \Sigma_T^{-1/2}\Vert_F\cdot\Vert \Sigma_T^{1/2}\Vert_F=\sqrt{\text{tr}(\Sigma_T^{-1})\text{tr}(\Sigma_T)}=\sqrt{O_p(1)\cdot \frac{1}{T}\cdot T\cdot O_p(1)}=O_p(1),
	\end{equation*}
	where the second equality is standard results for $\mathcal{I}_1$ process \citep[see][part (g) and (i) of Proposition 18.1]{Hamilton1994}. Also, $\Vert \tilde{b}_1-b_1\Vert =o_p(T^{-1/2})$ is by Proposition 19.2 in \cite{Hamilton1994}. This shows \eqref{square root convergence}. Apply part (a) of Proposition 18.1 in \cite{Hamilton1994}, we have 
	\begin{equation*}
		\Vert Y_{T+1}(0)'(\widehat{b}_1-b_1)\Vert =	\Vert (T^{-1/2}Y_{T+1}(0))'(T^{1/2}(\widehat{b}_1-b_1))\Vert = O_p(1)o_p(1)=o_p(1).
	\end{equation*}
	
	Now we show part (b). Again, it suffices to show $|\widehat{a}_i-a_i|=o_p(1)$ and $\Vert \widehat{b}_i-b_i\Vert=o_p(1)$. We consider the $i=1$ case and other cases follow the same strategy. We have showed  $\Vert \widehat{b}_i-b_i\Vert =o_p(1)$  in part (c) of the proof. Proposition 5 in the supplementary material to \cite{Ferman2016} establishes that
	\begin{equation}
		\label{cointegrated factor loadings}
		[\mu_1^1\ \mu_2^1\ \dots \ \mu_N^1](b_1-e_1)=0,
	\end{equation}
	where $e_i$ is the unit vector with one at the $i$-th entry. Thus,
	\begin{eqnarray}
		\label{intercept convergence}
		\widehat{a}_1
		&=&[\bar{y}_1\ \bar{y}_2\ \dots\ \bar{y}_N](e_1-\widehat{b}_1)\notag\\
		&=&[\bar{y}_1\ \bar{y}_2\ \dots\ \bar{y}_N](e_1-b_1)+[\bar{y}_1\ \bar{y}_2\ \dots\ \bar{y}_N](b_1-\widehat{b}_1)\notag\\
		&=&\left\lbrace \frac{1}{T}\sum_{t=1}^T\left(  (\lambda_t^0)'[\mu_1^0\ \dots\ \mu_N^0]+[\varepsilon_{1,t}\ \dots\ \varepsilon_{N,t}]\right) \right\rbrace (e_1-b_1)+\notag\\
		&&\left( \frac{1}{\sqrt{T}}[\bar{y}_1\ \bar{y}_2\ \dots\ \bar{y}_N]\right) \sqrt{T}(b_1-\widehat{b}_1)\notag\\
		&=& E[\lambda_t^0]'[\mu_1^0\ \dots\ \mu_N^0](e_1-b_1)+o_p(1)+O_p(1)o_p(1)\notag\\
		&\rightarrow_p& E[\lambda_t^0]'[\mu_1^0\ \dots\ \mu_N^0](e_1-b_1).\notag\\
		&=& a_1.
	\end{eqnarray}
	The third equality is by \eqref{cointegrated factor loadings}. The fourth equality is by stationarity of $\{(\lambda_t^0,\varepsilon_{t})\}_{t\ge 1}$ and results in part (d) of the proof. This shows part (b) of the Assumption \mainref{unbiased assumption}{1}.
	
	Combining \eqref{cointegrated factor loadings} and \eqref{intercept convergence}, we have part (a) in Assumption \mainref{unbiased assumption}{1}. Part (d) is assumed.
	
\end{proof}

\renewenvironment{proof}{{\bfseries Proof of Theorem \mainref{thm_no_spillover}{2}.}}{\qed}
\noindent\begin{proof}
	We follow the proof of Theorem 2 in \cite{Andrews2006}. Let 
	\begin{equation*}
		\begin{aligned}
			&L_{1,T}(\epsilon)=\left\lbrace \Vert C_T(\widehat{\beta}_1-\beta_1)\Vert\le \epsilon, \max_{t=1,\dots,T}\Vert C_T(\widehat{\beta}_1^{(t)}-\beta_1) \Vert \le \epsilon \right\rbrace, \\
			&L_{2,T}(c)=\left\lbrace \max_{t\le T+1}\Vert C^{-1}_Tx_t\Vert\le c \right\rbrace .
		\end{aligned}
	\end{equation*}
	By Assumption 2(d), there exists a positive sequence $\{\epsilon_T \}_{T\ge 1}$ such that $\epsilon_T\rightarrow 0$ and $\Pr(L_{1,T}(\epsilon_T))\rightarrow 1$. Let $c_T=1/\sqrt{\epsilon_T}$. So we have $c_T\rightarrow \infty$ and $c_T\epsilon_T\rightarrow 0$. By Assumption 2(c), we must have $\Pr(L_{2,T}(c_T))\rightarrow 1$. Let $L_T=L_{1,T}(\epsilon_T)\cap L_{2,T}(c_T)$, then we have $\Pr(L_T)\rightarrow 1$ and $\Pr(L_T^c)\rightarrow 0$. 
	
	Suppose $L_T$ holds. Then, for $\beta=\widehat{\beta}_1$ or $\beta=\widehat{\beta}_{1}^{(t)}$ for some $t=1,\dots,T$, we have 
	\begin{eqnarray*}
		|P_t(\beta)-P_t(\beta_1)|&=&\left| (\beta-\beta_1)'x_tx_t'(\beta-\beta_1)-2x_t'(\beta-\beta_1)u_{1,t}\right|    \notag\\
		&=&\left|  (\beta-\beta_1)'C_T'(C_T')^{-1}x_tx_t'C_T^{-1}C_T(\beta-\beta_1)-2x_t'C_T^{-1}C_T(\beta-\beta_1)u_{1,t}\right|    \notag\\
		&\le & \Vert C_T(\beta-\beta_1)\Vert^2\Vert C_T^{-1}x_t\Vert^2+2\Vert C_T^{-1}x_t\Vert \Vert C_T(\beta-\beta_1)\Vert |u_{1,t}|\notag\\
		&\le &  \epsilon_T^2c_T^2+2\epsilon_Tc_T|u_{1,t}|.
	\end{eqnarray*}
	Define $g_t(\epsilon_T,c_T)=\epsilon_T^2c_T^2+2\epsilon_Tc_T|u_{1,t}|$. Note that $g_t(\epsilon_T,c_T)$ is identically distributed across $t$ for a fixed $T$, by Assumption 2(a). 
	
	We first prove part (a). Let $x$ be some continuous point of distribution function of $P_{T+1}(\beta_1)$. Then,
	\begin{eqnarray*}
		\Pr(P_{T+1}(\widehat{\beta}_1)\le x)&=&\Pr(\{P_{T+1}(\widehat{\beta}_1)\le x \}\cap L_T)+\Pr(\{P_{T+1}(\widehat{\beta}_1)\le x \}\cap L_T^c)\notag\\
		&\le & \Pr(P_{T+1}(\widehat{\beta}_1)\le x +g_t(\epsilon_T,c_T))+\Pr(L_T^c)\notag\\
		&\le &\Pr(P_{T+1}(\beta_1)\le x)+o(1).
	\end{eqnarray*}
	To see the last equality, pick $\epsilon>0$. By continuity, $\exists \delta>0$ such that for each $y\in (x-\delta,x+\delta)$, $|\Pr(P_{T+1}(\beta_1)\le y)-\Pr(P_{T+1}(\beta_1)\le x)|<\epsilon$. Therefore, 
	\begin{eqnarray*}
		&&\Pr(P_{T+1}(\widehat{\beta}_1)\le x +g_t(\epsilon_T,c_T))\\
		&=&\Pr(\{P_{T+1}(\widehat{\beta}_1)\le x +g_t(\epsilon_T,c_T)\}\cap \{ |g_t(\epsilon_T,c_T)|\ge \delta\})\notag\\
		&&+\Pr(\{P_{T+1}(\widehat{\beta}_1)\le x +g_t(\epsilon_T,c_T)\}\cap \{|g_t(\epsilon_T,c_T)|<\delta\})\notag\\
		&\le &\Pr(|g_t(\epsilon_T,c_T)|\ge \delta)+\Pr(P_{T+1}(\widehat{\beta}_1)\le y)\notag\\
		&<&\Pr(P_{T+1}(\beta_1)\le x)+o(1).
	\end{eqnarray*}
	Similarly, 
	$		\Pr(P_{T+1}(\widehat{\beta}_1)\le x)\ge \Pr(P_{T+1}(\beta_1)\le x)+o(1).$
This shows part (a). 
	
	To see part (b), let $k:\mathbb{R}\rightarrow\mathbb{R}$ be a monotonically decreasing and everywhere differentiable function that has bounded derivative and satisfies $k(x)=1$ for $x\le 0$, $k(x)\in [0,1]$ for $x\in (0,1)$, and $k(x)=0$ for $x\ge 1$. For example, let $k(x)=\cos(\pi x)/2+1/2$ for $x\in (0,1)$. Given some $\{\beta^{(t)} \}_{t=1}^T$, a smoothed df is defined by
	\begin{equation*}
		\widehat{F}_T(x,\{\beta^{(t)} \},h_T)=\frac{1}{T}\sum_{t=1}^{T}k\left( \frac{P_t(\beta^{(t)})-x}{h_T} \right), 
	\end{equation*}
	for some sequence of positive constants $\{h_T \}$ such that $h_T\rightarrow 0$ and $c_T\epsilon_T/h_T\rightarrow 0$. For example, we let $h_T=\epsilon_T^{1/4}$ when $c_T=1/\sqrt{\epsilon_T}$. Also, define 
	\begin{equation*}
		\widehat{F}_T(x,\{\beta_1\})=\frac{1}{T}\sum_{t=1}^{T}\mathbbm{1} \{ P_t(\beta_1)\le x \}, 
	\end{equation*}
	i.e., $\widehat{F}_T(x,\{\beta_1\})$ is the empirical cdf of $P_t$ as if the true parameter $\beta_1$ is known.
	
	We write
	\begin{equation*}
		|\widehat{F}_{P,T}(x)-F_P(x)|\le \sum_{i=1}^{4}D_{i,T},
	\end{equation*}	
	for 
	\begin{align*} 
		D_{1,T}&= |\widehat{F}_{P,T}(x)-\widehat{F}_T(x,\{\widehat{\beta}_1^{(t)}\}_{t=1}^T,h_T)|,\notag\\
		D_{2,T}&= |\widehat{F}_T(x,\{\widehat{\beta}_1^{(t)}\}_{t=1}^T,h_T)-\widehat{F}_T(x,\{\beta_1 \},h_T)|,\notag\\
		D_{3,T}&= |\widehat{F}_T(x,\{\beta_1 \},h_T)-\widehat{F}_T(x,\{\beta_1\})|,\text{ and}\notag\\ 
		D_{4,T}&= |\widehat{F}_T(x,\{\beta_1\})-{F}_P(x)|. 
	\end{align*}
	We want to show that all four terms vanish. First note that 
	\begin{equation*}
		D_{1,T}\le \frac{1}{T}\sum_{t=1}^{T}\mathbbm{1}\left\lbrace \frac{P_t(\widehat{\beta}_1^{(t)})-x}{h_T}\in (0,1) \right\rbrace .
	\end{equation*} 
	Thus, for any $\delta>0$, 
	\begin{align}
		\label{bound D1T} 
		\Pr(D_{1,T}>\delta)&\le \Pr(\{D_{1,T}>\delta \}\cap L_T)+\Pr(L_T^c)\notag\\
		&\le \Pr \left( \frac{1}{T}\sum_{t=1}^{T}\mathbbm{1}\left\lbrace P_t(\widehat{\beta}_1^{(t)})-x\in (-g_t(\epsilon_T,c_T),h_T+g_t(\epsilon_T,c_T)) \right\rbrace>\delta \right) +o(1)\notag\\
		&\le \frac{E\mathbbm{1}\left\lbrace P_t(\widehat{\beta}_1^{(t)})-x\in (-g_t(\epsilon_T,c_T),h_T+g_t(\epsilon_T,c_T)) \right\rbrace}{\delta}+o(1),
	\end{align}
	where the last inequality is by Markov's inequality. Recall $\Pr(P_1(\beta_1)\ne x)=1$ and $g_t(\epsilon_T,c_T)\rightarrow 0$ a.s., so $\mathbbm{1}\{P_t(\beta_1)-x\in (-g_t(\epsilon_T,c_T),h_T+g_t(\epsilon_T,c_T))\}\rightarrow 0$ a.s. By the dominated convergence theorem, \eqref{bound D1T} implies $\Pr(D_{1,T}>\delta)\le o(1)$ and thus $D_{1,T}=o_p(1)$. 
	
	For $D_{2,T}$, we have
	\begin{align*}
		D_{2,T}= \left| \frac{1}{T} \sum_{t=1}^{T} k'\left( \frac{\tilde{P}_t-x}{h_T} \right) \frac{P_t(\widehat{\beta}_1^{(t)})-P_t(\beta_1)}{h_T} \right|
		\le \frac{\bar{k}}{T}\sum_{t=1}^{T}\frac{g_t(\epsilon_T,c_T)}{h_T}.
	\end{align*}
	The equality is by the mean value theorem and we have $\tilde{P}_t$ lies between $P_t(\widehat{\beta}_1^{(t)})$ and $P_t(\beta_1)$. In the inequality, $\bar{k}$ is a bound for the derivative of $k$. Also, note
	\begin{equation*}
		E\left[ \frac{g_t(\epsilon_T,c_T)}{h_T}\right] =\frac{\epsilon_T^2c_T^2}{h_T}+2\frac{\epsilon_Tc_T}{h_T}E|u_{1,t}|=o(1).
	\end{equation*}
	Therefore, 
	\begin{align*}
		\Pr(D_{2,T}>\delta)&\le \Pr (\{D_{2,T}>\delta \}\cap L_T)+\Pr (L_T^c)\notag\\
		&\le \Pr \left( \frac{\bar{k}}{T}\sum_{t=1}^{T}\frac{g_t(\epsilon_T,c_T)}{h_T}>\delta \right) +o(1)\notag\\
		&\le \bar{k} \frac{Eg_t(\epsilon_T,c_T)}{\delta h_T}\notag\\
		&\rightarrow 0. 
	\end{align*}
	The third inequality is by Markov's inequality. This shows $D_{2,T}=o_p(1)$. 
	
	$D_{3,T}$ is similar to the $D_{1,T}$ case. Finally, by stationary and ergodicity of $u_{1,t}$, we have $D_{4,T}=o_p(1)$. This shows part (b). 
	
	Now we show part (c). Pick any small $\epsilon$ such that $\widehat{F}_{P,T}(x)\rightarrow_pF_P(x)$ for $x\in (q_{P,1-\tau}-\epsilon,q_{P,1-\tau}+\epsilon)$. Note 
	\begin{eqnarray*}
	&&	\Pr (\widehat{q}_{P,1-\tau}>q_{P,1-\tau}+\epsilon)\\
		&\le &\Pr (\widehat{F}_{P,T}({q}_{P,1-\tau}+\epsilon)<1-\tau)\notag\\
		&= &\Pr (\widehat{F}_{P,T}({q}_{P,1-\tau}+\epsilon)-{F}_P({q}_{P,1-\tau}+\epsilon)<(1-\tau)-{F}_P({q}_{P,1-\tau}+\epsilon))\notag\\
		&\rightarrow &0.
	\end{eqnarray*}
	The inequality is by definition of $\widehat{q}_{P,1-\tau}$. The convergence is because of part (e) of Assumption 2 and part (b) of Theorem 2. Similarly, 
	\begin{eqnarray*}
	&&	\Pr (\widehat{q}_{P,1-\tau}<q_{P,1-\tau}-\epsilon)\\
		&\le &\Pr (\widehat{F}_{P,T}({q}_{P,1-\tau}-\epsilon)\ge 1-\tau)\notag\\
		&=& \Pr (\widehat{F}_{P,T}({q}_{P,1-\tau}-\epsilon)-{F}_P({q}_{P,1-\tau}-\epsilon)\ge (1-\tau)-{F}_P({q}_{P,1-\tau}-\epsilon))\notag\\
		&\rightarrow& 0.
	\end{eqnarray*}
	Again, the inequality is by definition of $\widehat{q}_{P,1-\tau}$, and the convergence is because of part (e) of Assumption 2 and part (b) of Theorem 2. 
	
	Finally, we show part (d). Under null, $P_\infty$ and $P_1(\beta_1)$ have the same distribution, so $q_{P,1-\tau}$ is $(1-\tau)$-quantile of $P_\infty$. Therefore, 
$$		\Pr(P>\widehat{q}_{P,1-\tau})=1-\Pr (P\le \widehat{q}_{P,1-\tau})=1-\Pr (P+( {q}_{P,1-\tau}- \widehat{q}_{P,1-\tau})\le {q}_{P,1-\tau})\rightarrow \tau,$$
	where the convergence is by combining part (a) and (c). This concludes our proof.
	
\end{proof}

\renewenvironment{proof}{{\bfseries Proof of Lemma \mainref{lemma_inference_no_spillover}{2}.}}{\qed}
\noindent\begin{proof}
	Since Assumption 3 implies Assumption 2, we only need to show Lemma 3. 
	
\end{proof}

\renewenvironment{proof}{{\bfseries Proof of Theorem \mainref{thm_spillover}{3}.}}{\qed}
\noindent\begin{proof}
	We use similar strategy as we do in the proof of Theorem \mainref{thm_no_spillover}{2}. Let $H=G'C'{\color{revisionblue}\Xi}CG$, $H_T(\theta)=G(\theta)'C'{\color{revisionblue}\Xi_T}CG(\theta)$, and
	\begin{equation*}
		\begin{aligned}
			&L_{1,T}(\epsilon)=\left\lbrace \Vert (\widehat{\theta}-\theta_0)D_T\Vert_F\le \epsilon, \max_{t=1,\dots,T}\Vert (\widehat{\theta}^{(t)}-\theta_0)D_T \Vert_F \le \epsilon \right\rbrace, \\
			&L_{2,T}(c)=\left\lbrace \max_{t\le T+1}\Vert D^{-1}_Tx_t\Vert\le c \right\rbrace,\\
			&L_{3,T}(\eta)=\left\lbrace \max\left\{\Vert H_T(\widehat{\theta})-H\Vert_F,\max_{t=1,\dots,T}\Vert H_T(\widehat{\theta}^{(t)})-H\Vert_F\right\}<\eta \right\rbrace.
		\end{aligned}
	\end{equation*}
	
	By Assumption \mainref{spillover assumption}{3}(d), there exists a positive sequence $\{\epsilon_T \}_{T\ge 1}$ such that $\epsilon_T\rightarrow 0$ and $\Pr(L_{1,T}(\epsilon_T))\rightarrow 1$. Let $c_T=1/\sqrt{\epsilon_T}$. So we have $c_T\rightarrow \infty$ and $c_T\epsilon_T\rightarrow 0$. By Assumption \mainref{spillover assumption}{3}(c), we must have $\Pr(L_{2,T}(c_T))\rightarrow 1$. By Assumption \mainref{spillover assumption}{3}(a), (d), and (f), continuity of $G(\theta)$ at $\theta_0$ implies that there exists a positive sequence $\{\eta_T \}_{T\ge 1}$ such that $\eta_T\rightarrow 0$ and $\Pr(L_{3,T}(\eta_T))\rightarrow 1$. Let $L_T=L_{1,T}(\epsilon_T)\cap L_{2,T}(c_T)\cap L_{3,T}(\eta_T)$, then we have $\Pr(L_T)\rightarrow 1$ and $\Pr(L_T^c)\rightarrow 0$.
	
	Suppose $L_T$ holds. Then, for some $\theta=\widehat{\theta}$ or $\theta=\widehat{\theta}^{(t)}$ and for some $t=1,\dots, T$, we have 
	\begin{equation}
		\label{P_hat diff}
		|\widehat{P}_t(\theta)-P_t(\theta_0)|\le |\widehat{P}_t(\theta)-P_t(\theta)|+|{P}_t(\theta)-P_t(\theta_0)|.
	\end{equation}
	Note that 
	\begin{align}
		\label{P_hat diff_1}
		|\widehat{P}_t(\theta)-P_t(\theta)|
		&= \left| (Y_t-\theta x_t)'\{H_T(\theta)-H\}(Y_t-\theta x_t) \right| \notag\\
		&\le  \Vert Y_t-\theta x_t\Vert^2 \Vert H_T(\theta)-H\Vert_F\notag\\
		&\le \Vert u_t+(\theta_0-\theta)x_t\Vert^2 \cdot \eta_T\notag\\
		&\le (\Vert u_t\Vert+\Vert (\theta_0-\theta)D_TD_T^{-1}x_t\Vert)^2\eta_T\notag\\
		&\le (\Vert u_t\Vert +\Vert (\theta_0-\theta)D_T\Vert_F \Vert D_T^{-1}x_t\Vert)^2\eta_T\notag\\
		&\le (\Vert u_t\Vert +\epsilon_Tc_T)^2\eta_T
	\end{align} 
	and
	\begin{align}
		\label{P_hat diff_2}
		|{P}_t(\theta)-P_t(\theta_0)|
		=&\ |(Y_t-\theta x_t)'G'C'{\color{revisionblue}\Xi}CG(Y_t-\theta x_t)-(Y_t-\theta_0 x_t)'G'C'{\color{revisionblue}\Xi}CG(Y_t-\theta_0 x_t)|\notag\\
		\le&\  |(Y_t-\theta x_t)'G'C'{\color{revisionblue}\Xi}CG(Y_t-\theta x_t)-(Y_t-\theta x_t)'G'C'{\color{revisionblue}\Xi}CG(Y_t-\theta_0 x_t)|\notag\\
		&\ +	|(Y_t-\theta x_t)'G'C'{\color{revisionblue}\Xi}CG(Y_t-\theta_0 x_t)-(Y_t-\theta_0 x_t)'G'C'{\color{revisionblue}\Xi}CG(Y_t-\theta_0 x_t)|\notag\\
		=&\ |(u_t+(\theta_0-\theta) x_t)'G'C'{\color{revisionblue}\Xi}CG(\theta_0-\theta) x_t|+|((\theta_0-\theta)x_t)'G'C'{\color{revisionblue}\Xi}CGu_t|\notag\\
		\le &\  \Vert u_t+(\theta_0-\theta)D_TD_T^{-1} x_t\Vert \Vert G'C'{\color{revisionblue}\Xi}CG\Vert_F \Vert (\theta_0-\theta)D_TD_T^{-1} x_t\Vert \notag\\
		&\ +\Vert (\theta_0-\theta)D_TD_T^{-1}x_t\Vert \Vert G'C'{\color{revisionblue}\Xi}CG\Vert_F \Vert u_t\Vert \notag\\
		\le &\ (\Vert u_t\Vert +\epsilon_Tc_T) \Vert G'C'{\color{revisionblue}\Xi}CG\Vert_F \epsilon_Tc_T+\epsilon_Tc_T \Vert G'C'{\color{revisionblue}\Xi}CG\Vert_F\Vert u_t\Vert\notag\\
		= &\  (2\Vert u_t\Vert +\epsilon_Tc_T) \Vert G'C'{\color{revisionblue}\Xi}CG\Vert_F \epsilon_Tc_T.
	\end{align}
	Combining \eqref{P_hat diff}, \eqref{P_hat diff_1}, and \eqref{P_hat diff_2}, we have 
$		|\widehat{P}_t(\theta)-P_t(\theta_0)|\le g_t(\epsilon_T,c_T,\eta_T),$
	where
$		g_t(\epsilon_T,c_T,\eta_T)= (\Vert u_t\Vert +\epsilon_Tc_T)^2\eta_T+(2\Vert u_t\Vert +\epsilon_Tc_T) \Vert G'C'{\color{revisionblue}\Xi}CG\Vert_F \epsilon_Tc_T.$
	By Assumption 1(a), $g_t(\epsilon_T,c_T,\eta_T)$ is identically distributed across $t$ for a fixed $T$. 
	
	To show part (a), note that under null,
	\begin{align*}
		P&=(C\widehat{\alpha}-d)'{\color{revisionblue}\Xi_T}(C\widehat{\alpha}-d)\notag\\
		&=(C(\alpha+Gu_{T+1}+o_p(1))-d)'({\color{revisionblue}\Xi}+o_p(1))(C(\alpha+Gu_{T+1}+o_p(1))-d)\notag\\
		&= (CGu_{T+1}+o_p(1))'({\color{revisionblue}\Xi}+o_p(1))(CGu_{T+1}+o_p(1))\notag\\
		&= u_{T+1}'G'C'{\color{revisionblue}\Xi}CGu_{T+1}+o_p(1).
	\end{align*}
	The second equality is by Theorem 1. Since $P_\infty=u_1'G'C'{\color{revisionblue}\Xi}CGu_1$, we have $P\rightarrow_d P_\infty$ by stationary of $\{u_t \}_{t\ge 1}$.
	
	Part (b)-(d) can be shown using the same strategy as in the proof of Theorem 2, with $g_t(\epsilon_T,c_T,\eta_T)$ in place of $g_t(\epsilon_T,c_T)$, and $\theta$ in place of $\beta$, so is omitted here. 
	
\end{proof}

\renewenvironment{proof}{{\bfseries Proof of Lemma \mainref{lemma inference spillover}{3}.}}{\qed}
\noindent\begin{proof}
	(i) Assume Condition ST holds. 
	
	By Lemma 1, part (a) of Assumption \mainref{spillover assumption}{3} holds.
	
	Part (b) follows because $u_t$ is a linear combination of the stationary components in Condition ST, which have finite $(2+\delta)$ moments.
	
	For part (c), pick some $\tau$ such that $1/(2+\delta)<\tau<1/2$, where $\delta$ is defined in Condition ST. Let 
	\begin{equation}
		\label{D_T}
		D_T=\begin{bmatrix}
			1 & 0 \\ 0 & T^\tau I_N
		\end{bmatrix}.
	\end{equation}
	Then, we have
	\begin{equation}
		\label{D_Tx_t_expansion}
		\max_{t\le T+1}\Vert D_T^{-1}x_t\Vert = \max_{t\le T+1}\left\Vert\begin{bmatrix}
			1\\T^{-\tau}Y_t
		\end{bmatrix} \right\Vert 
		=\sqrt{1+\left( \max_{t\le T+1}\Vert T^{-\tau}Y_t\Vert\right)^2 }.
	\end{equation}
	Also, for any $\epsilon>0$, note 
	\begin{align}
		\label{TY_t_vanish}
		\Pr\left( \max_{t\le T+1}\Vert T^{-\tau}Y_t\Vert >\epsilon \right) 
		&=\Pr \left( \bigcup_{t\le T+1}\Vert Y_t\Vert >T^\tau \epsilon \right) \notag\\
		&\le\left(  \sum_{t=1}^T\Pr (\Vert Y_t\Vert >T^\tau \epsilon)\right) +\Pr (\Vert Y_{T+1}(0)+\alpha\Vert >T^\tau \epsilon )\notag\\
		&\le\frac{TE[\Vert Y_t\Vert ^{2+\delta}]}{T^{\tau (2+\delta)}\epsilon^{2+\delta}}+o(1)\notag\\
		&=o(1).
	\end{align}
	The second inequality is due to Markov inequality and stationarity of $\{Y_t(0)\}_{t\ge 1}$. The last equality is because $\tau>1/(2+\delta)$. Combining \eqref{D_Tx_t_expansion} and \eqref{TY_t_vanish}, we obtain part (c).
	
	For part (d), we use $D_T$ defined in \eqref{D_T}. Following the same reasoning as in \eqref{square root convergence}, for each $i=1,\dots,N$, we have 
	\begin{align}
		\label{square_root_converge_ST}
		\Vert \widehat{b}_i-b_i\Vert& \le \Vert \Sigma_T^{-1/2}\Vert_F\cdot \Vert \Sigma_T^{1/2}\Vert_F\cdot\Vert \tilde{b}_i-b_i\Vert  \notag\\
		&=O_p(1) O_p(T^{-1/2})\notag\\
		&=O_p(T^{-1/2}).
	\end{align}
	The first equality is because $\{Y_t(0) \}_{t\ge 1}$ is ergodic for the second moment, and $\tilde{b}_i$ is the OLS estimator for $b_i$. Thus, 
	\begin{equation*}
		\Vert D_T(\widehat{\beta}_i-\beta_i)\Vert \le |\widehat{a}_i-a_i|+T^\tau\Vert\widehat{b}_i-b_i\Vert=o_p(1)+O_p(T^{\tau-1/2})=o_p(1).
	\end{equation*}
	The first equality follows from Assumption \mainref{unbiased assumption}{1}(b) and \eqref{square_root_converge_ST}, and the last equality uses $\tau<1/2$. Therefore,
	$	\Vert (\widehat{\theta}-\theta_0)D_T\Vert_F=\sqrt{\sum_{i=1}^N\Vert D_T(\widehat{\beta}_i-\beta_i) \Vert^2 }=o_p(1). $
	Also, since $\widehat{\theta}^{(t)}=\widehat{\theta}$ for each $t$,
	$	\max_{t=1,\dots,T}\Vert (\widehat{\theta}^{(t)}-\theta_0)D_T\Vert_F=	\Vert (\widehat{\theta}-\theta_0)D_T\Vert_F=o_p(1). $
	Moreover, Assumption 1(b) and the full-sample construction imply
	$\max_{t=1,\dots,T}\Vert \widehat{B}^{(t)}-B\Vert_F=\Vert \widehat{B}-B\Vert_F=o_p(1)$.
	This shows part (d).
	
	Part (e) is assumed. 
	
	Part (f) is trivial if ${\color{revisionblue}\Xi_T}=I$. Assume now ${\color{revisionblue}\Xi_T}=(C\widehat{G}(T^{-1}\sum_{t=1}^T\widehat{u}_t\widehat{u}_t')\widehat{G}'C')^{-1}$. Then,
	\begin{eqnarray*}
		&&\frac{1}{T}\sum_{t=1}^T\widehat{u}_t\widehat{u}_t'\\
		&=&(I-\widehat{B})\left( \frac{1}{T}\sum_{t=1}^TY_tY_t'\right) (I-\widehat{B})'-(I-\widehat{B})\left( \frac{1}{T}\sum_{t=1}^TY_t\right)\widehat{a}'\\
		&&-\widehat{a}\left( \frac{1}{T}\sum_{t=1}^TY_t'\right)(I-\widehat{B})'+\widehat{a}\widehat{a}'\notag\\
		&\rightarrow &E[u_tu_t'],
	\end{eqnarray*}
	by ergodicity and Assumption 1(b). Therefore, ${\color{revisionblue}\Xi_T}\rightarrow_p {\color{revisionblue}\Xi}=(CGE[u_tu_t']G'C')^{-1}$.
	
	This concludes part (i) of Lemma 3. 
	
	\
	
	\noindent (ii) Assume Condition CO holds. 
	
	By Lemma 1, Assumption 1 holds. This shows Part (a). 
	
	By \eqref{cointegrated factor loadings}, $u_t$ is a linear combination of $\lambda_t^0$ and $\varepsilon_t$, so $\{u_t \}_{t\ge 1}$ is ergodic and has finite second moment under Condition CO's finite fourth-moment requirement. This shows Part (b).
	
	Now we show Part (c). Let  
	\begin{equation*}
		D_T=\begin{bmatrix}
			1 & 0 \\ 0 & \sqrt{T}\cdot I_N
		\end{bmatrix}.
	\end{equation*}
	Then, we have 
	\begin{align*}
		\max_{t\le T+1} \Vert D_T^{-1}x_t\Vert 
		&=\sqrt{1+\left( \max_{t\le T+1}\Vert T^{-1/2}Y_t\Vert  \right)^2 }\notag\\
		&\le \sqrt{1+\sum_{i=1}^N\left( \max_{t\le T+1}|T^{-1/2}y_{i,t}|  \right)^2 }\notag\\
		&\le \sqrt{1+\sum_{i=1}^N\left( T^{-1/2}|\alpha_i|+\max_{t\le T+1}|T^{-1/2}y_{i,t}(0)|  \right)^2 }\notag\\
		&= \sqrt{1+\sum_{i=1}^N\left( o(1)+O_p(1)  \right)^2 }\notag\\
		&=O_p(1)
	\end{align*}
	The second equality is because 
	\begin{equation*}
		\max_{1\le t\le T+1}|T^{-1/2}y_{i,t}(0)|\Rightarrow \sup_{r\in[0,1]}|\nu_i(r)|,
	\end{equation*} 
	by the continuous mapping theorem and $\sqrt{(T+1)/T}\to1$.
	
	To show Part (d), we combine \eqref{square root convergence} and \eqref{intercept convergence}, and have 
	\begin{equation*}
		\Vert D_T(\widehat{\beta}_i-\beta_i)\Vert =\left\Vert \begin{bmatrix}\widehat{a}_i-a_i\\\sqrt{T}(\widehat{b}_i-b_i)
		\end{bmatrix} \right\Vert=o_p(1). 
	\end{equation*}
	Therefore, 
	$	\Vert (\widehat{\theta}-\theta_0)D_T\Vert_F=\sqrt{\sum_{i=1}^N\Vert D_T(\widehat{\beta}_i-\beta_i) \Vert^2 }=o_p(1). $
	The second condition in Part (d) is also satisfied since $\widehat{\theta}^{(t)}=\widehat{\theta}$ for each $t$.
	Moreover, Assumption 1(b) and the full-sample construction imply
	$\max_{t=1,\dots,T}\Vert \widehat{B}^{(t)}-B\Vert_F=\Vert \widehat{B}-B\Vert_F=o_p(1)$.
	
	Part (e) is assumed.
	
	{\color{revisionblue}
Part (f) is trivial for ${\color{revisionblue}\Xi_T}=I$.
	Now assume $\Xi_T=(C\widehat{G}(T^{-1}\sum_{t=1}^T\widehat{u}_t\widehat{u}_t')\widehat{G}'C')^{-1}$.
	The expansion used in part (i) is unavailable here, because $T^{-1}\sum_{t=1}^TY_tY_t'$ and $T^{-1}\sum_{t=1}^TY_t$ both diverge when $\{\lambda_t^1\}$ is $\mathcal{I}(1)$.
	We instead treat the estimation error as a perturbation of $u_t$.
	Write $\delta_B=\widehat{B}-B$ and $\delta_a=\widehat{a}-a$, so that for $t\le T$,
	\begin{equation*}
		\widehat{u}_t=u_t+\Delta_t,\qquad \Delta_t=-\delta_BY_t-\delta_a.
	\end{equation*}
	Under Condition CO, $u_t$ is a linear combination of $\lambda_t^0$ and $\varepsilon_t$, hence stationary and ergodic with finite second moment, so
	$T^{-1}\sum_{t=1}^Tu_tu_t'\rightarrow_pE[u_tu_t']$, $T^{-1}\sum_{t=1}^T\Vert u_t\Vert^2=O_p(1)$, and $T^{-1}\sum_{t=1}^Tu_t=o_p(1)$.
	Condition CO's weak convergence $T^{-1/2}y_{i,[rT]}(0)\Rightarrow\nu_i(r)$ and the continuous mapping theorem give
	$T^{-2}\sum_{t=1}^T\Vert Y_t\Vert^2\Rightarrow\sum_{i=1}^N\int_0^1\nu_i(r)^2dr$, so that
	\begin{equation}
		\label{eq_co_Y_second_moment}
		T^{-1}\sum_{t=1}^T\Vert Y_t\Vert^2=O_p(T),
		\qquad
		T^{-1}\sum_{t=1}^T\Vert Y_t\Vert=O_p(T^{1/2}),
	\end{equation}
	the second bound following from the first by Cauchy--Schwarz.
	The display establishing Part (d) above gives $\Vert\delta_a\Vert=o_p(1)$ and $\Vert\delta_B\Vert=o_p(T^{-1/2})$.
	
	Expanding,
	\begin{equation*}
		\frac{1}{T}\sum_{t=1}^T\widehat{u}_t\widehat{u}_t'
		=\frac{1}{T}\sum_{t=1}^Tu_tu_t'
		+\frac{1}{T}\sum_{t=1}^T\left( u_t\Delta_t'+\Delta_tu_t'\right)
		+\frac{1}{T}\sum_{t=1}^T\Delta_t\Delta_t'.
	\end{equation*}
	For the quadratic term, \eqref{eq_co_Y_second_moment} gives
	\begin{equation*}
		\left\Vert \frac{1}{T}\sum_{t=1}^T\delta_BY_tY_t'\delta_B'\right\Vert
		\le \Vert\delta_B\Vert^2\cdot\frac{1}{T}\sum_{t=1}^T\Vert Y_t\Vert^2
		=o_p(T^{-1})O_p(T)=o_p(1),
	\end{equation*}
	while the remaining pieces are
	$\delta_B(T^{-1}\sum_{t=1}^TY_t)\delta_a'=o_p(T^{-1/2})O_p(T^{1/2})o_p(1)=o_p(1)$
	and $\delta_a\delta_a'=o_p(1)$.
	For the cross term,
	$T^{-1}\sum_{t=1}^Tu_t\Delta_t'=-(T^{-1}\sum_{t=1}^Tu_tY_t')\delta_B'-(T^{-1}\sum_{t=1}^Tu_t)\delta_a'$,
	and by Cauchy--Schwarz together with \eqref{eq_co_Y_second_moment},
	\begin{equation*}
		\left\Vert \frac{1}{T}\sum_{t=1}^Tu_tY_t'\right\Vert
		\le\left( \frac{1}{T}\sum_{t=1}^T\Vert u_t\Vert^2\right)^{1/2}
		\left( \frac{1}{T}\sum_{t=1}^T\Vert Y_t\Vert^2\right)^{1/2}
		=O_p(1)O_p(T^{1/2}),
	\end{equation*}
	so that piece is $O_p(T^{1/2})o_p(T^{-1/2})=o_p(1)$, and the second piece is $o_p(1)$.
	Therefore $T^{-1}\sum_{t=1}^T\widehat{u}_t\widehat{u}_t'\rightarrow_pE[u_tu_t']$.
	Since $\Vert\delta_B\Vert=o_p(1)$ implies $\widehat{M}\rightarrow_pM$, non-singularity of $A'MA$ and the continuous mapping theorem give $\widehat{G}\rightarrow_pG$, and thus
	$\Xi_T\rightarrow_p\Xi=(CGE[u_tu_t']G'C')^{-1}$.
	
	}
	
\end{proof}

\renewenvironment{proof}{{\bfseries Proof of Proposition \mainref{prop_test_for_A}{2}.}}{\qed}
\noindent\begin{proof}
	The main statement follows from the same argument as in the proof of Theorem \mainref{thm_spillover}{3}, so is omitted.
	
	For the correct specification case, note that
	\begin{align*}
		\kappa_A = & \Vert (I-\widehat \Gamma_A)u_{T+1}+(I-\Gamma_A)(I-B)\alpha+[(I-\widehat \Gamma_A)(I-\widehat{B})-(I-\Gamma_A)(I-B)]\alpha \\
		& +(I-\widehat \Gamma_A)(B-\widehat{B}) Y_{T+1}(0)+(I-\widehat \Gamma_A)(a-\widehat{a})\Vert \\
		=& \Vert (I-\widehat \Gamma_A)u_{T+1}+(I-\Gamma_A)(I-B)\alpha+o_p(1)\Vert. 
	\end{align*}
	When $A$ is correctly specified, $(I-\Gamma_A)(I-B)\alpha=0$, which concludes the proof. 
	
\end{proof}

%

\renewenvironment{proof}{{\bfseries Proof of Proposition \ref{prop_multiple_period_structure_selection}.}}{\qed}
\noindent\begin{proof}
	Let $\widehat{G} = A(A'\widehat{M}A)^{-1}A'(I-\widehat{B})'$. Note that
	\[
	\alpha_{T+s}-\widehat{\alpha}_{T+s}=(I-\widehat{G}(I-\widehat{B}))\alpha_{T+s}-\widehat{G}(I-\widehat{B})Y_{T+s}(0)+\widehat{G}\widehat{a},
	\]
	so we have
	\begin{align*}
		&(I-\widehat{B})(Y_{T+s}-\widehat{\alpha}_{T+s})-\widehat{a}\\
		= & (I-\widehat{B})Y_{T+s}(0)+(I-\widehat{B})(\alpha_{T+s}-\widehat{\alpha}_{T+s})-\widehat{a} \\
		= & (I-\widehat \Gamma_A)u_{T+s}+(I-\Gamma_A)(I-B)\alpha_{T+s}+[(I-\widehat \Gamma_A)(I-\widehat{B})-(I-\Gamma_A)(I-B)]\alpha_{T+s} \\
		& +(I-\widehat \Gamma_A)(B-\widehat{B}) Y_{T+s}(0)+(I-\widehat \Gamma_A)(a-\widehat{a}).
	\end{align*}
	Thus, the vector inside the norm defining $\kappa_A$ is
	\begin{align*}
		&(I-\widehat \Gamma_A)\frac{1}{\sqrt{T_1}}\sum_{s=1}^{T_1} u_{T+s}+\sqrt{T_1}(I-\Gamma_A)(I-B)\bar{\alpha}_{T_1}\\
		&\quad +\sqrt{T_1}[(I-\widehat \Gamma_A)(I-\widehat{B})-(I-\Gamma_A)(I-B)]\bar{\alpha}_{T_1} \\
		&\quad +(I-\widehat \Gamma_A)(B-\widehat{B}) \frac{1}{\sqrt{T_1}}\sum_{s=1}^{T_1}Y_{T+s}(0)+\sqrt{T_1}(I-\widehat \Gamma_A)(a-\widehat{a})\\
		&=  (I-\widehat \Gamma_A)\frac{1}{\sqrt{T_1}}\sum_{s=1}^{T_1} u_{T+s}+\sqrt{T_1}(I-\Gamma_A)(I-B)\bar{\alpha}_{T_1}+O_p(1).
	\end{align*}
	Since $I-\widehat{\Gamma}_A\rightarrow_p I-\Gamma_A$ and $u_t$ is stationary, $(I-\widehat{\Gamma}_A)T_1^{-1/2}\sum_{s=1}^{T_1}u_{T+s}=O_p(1)$ by Proposition 7.5 of \cite{Hamilton1994}. It follows that $\kappa_A=O_p(1)$ if and only if $\|\sqrt{T_1}(I-\Gamma_A)(I-B)\bar\alpha_{T_1}\|=O(1)$.
	
\end{proof}

	{\color{revisionblue}
	\section{Additional Details on the Empirical Application}
	\label{section_app_details}
	}
	
{\color{revisionblue}
\captionsetup{labelfont={color=revisionblue},textfont={color=revisionblue}}

\subsection{Robustness to excluding states with their own policy changes}
\label{sec_app_robustness}

The exposure structure in Section \mainref{empirical example}{6} treats two different
sources of contamination in the same way. Cross-border purchasing in Arizona, Nevada and
Oregon is plausibly a spillover from Proposition 99. The remaining states in the exposure
set experienced their own tobacco-control legislation, and a deviation there need not be an
indirect effect of the California policy. The estimator does not require the two to be
distinguished, because identification uses only the restriction $\alpha=A\gamma$ and is
silent about why a unit departs from its untreated path. The interpretation of the
individual components of $\gamma$ does differ, however, and we therefore report a
specification that separates them.

We drop the ten policy-contaminated states that do not border California, namely AK, DC,
FL, HI, MA, MD, MI, NJ, NY and WA, and retain the three neighbouring states in the exposure
structure. Arizona and Oregon are both policy-contaminated and neighbours, so they remain in
the analysis. This leaves 41 units and an exposure structure of rank
4. Own-policy contamination is handled by exclusion, as in \citet{Abadie2010},
while genuine spillovers continue to be modelled.

Figure \ref{fig_prop99_rob} reproduces the three panels of Figure
The estimated effect on California is close to the main specification throughout. The
average effect over the post-treatment period is $-10.06$ packs
against $-9.44$ in the main specification, and the two paths
differ by about one pack in every year. Conditioning improves, with the smallest eigenvalue
of $A'\widehat{M}A$ rising from $0.22$ to $0.54$.
The joint test for the existence of spillover effects behaves differently, rejecting in
1989, 1990, 1991, 1992, 1993, 1994 here against 1990, 1991, 1992, 1994, 1995, 1996, 1997, 1998, 1999, 2000 in the main specification. The later rejections in the main
specification therefore appear to reflect the other states' own policy changes rather than
spillovers from Proposition 99, which is consistent with the timing of those changes.

\mainref{fig_prop99}{4} under this specification.

\begin{figure}[htbp]
	\centering
	\color{revisionblue}
	\begin{subfigure}{.5\textwidth}
		\centering
		\includegraphics[width=\linewidth]{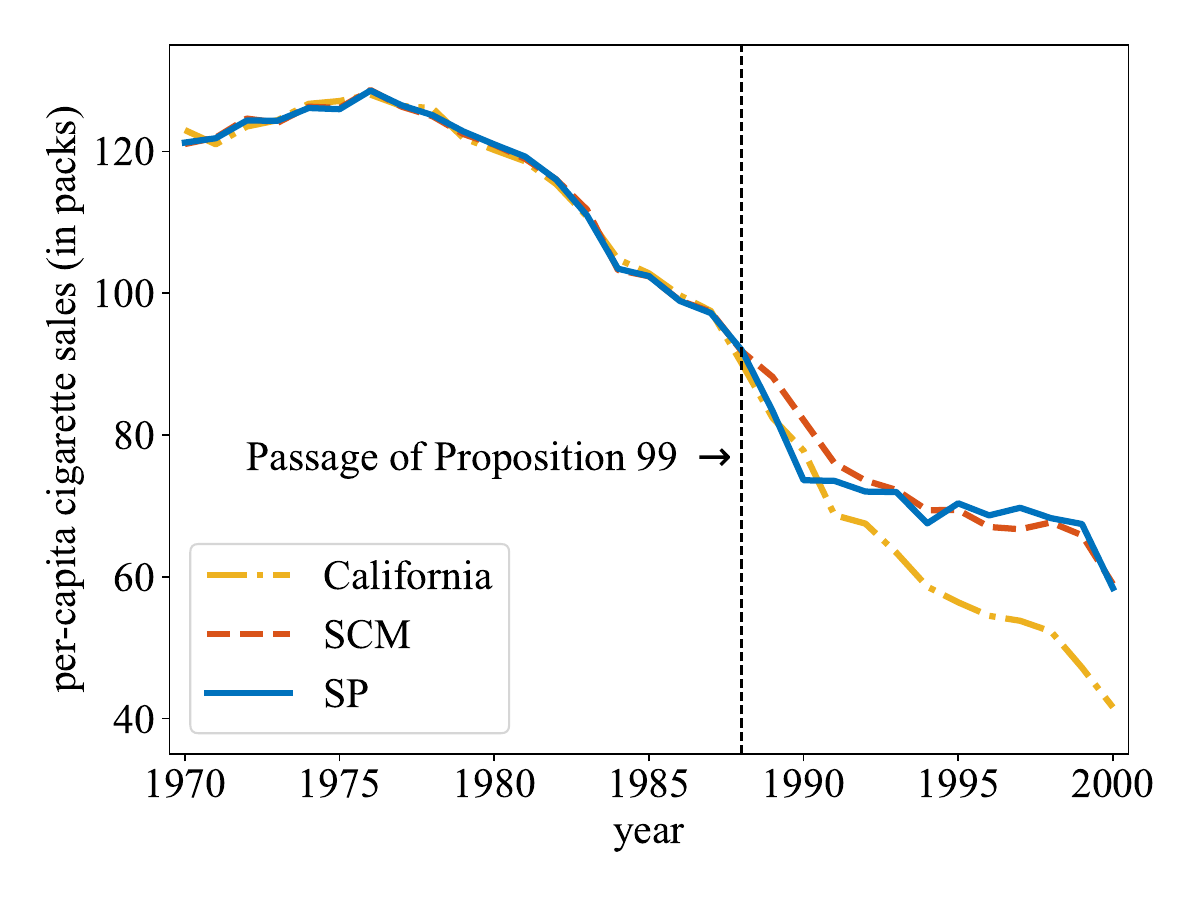}
		\caption{trends in per-capita cigarette sales}
		\label{fig_trend_rob}
	\end{subfigure}%
	\begin{subfigure}{.5\textwidth}
		\centering
		\includegraphics[width=\linewidth]{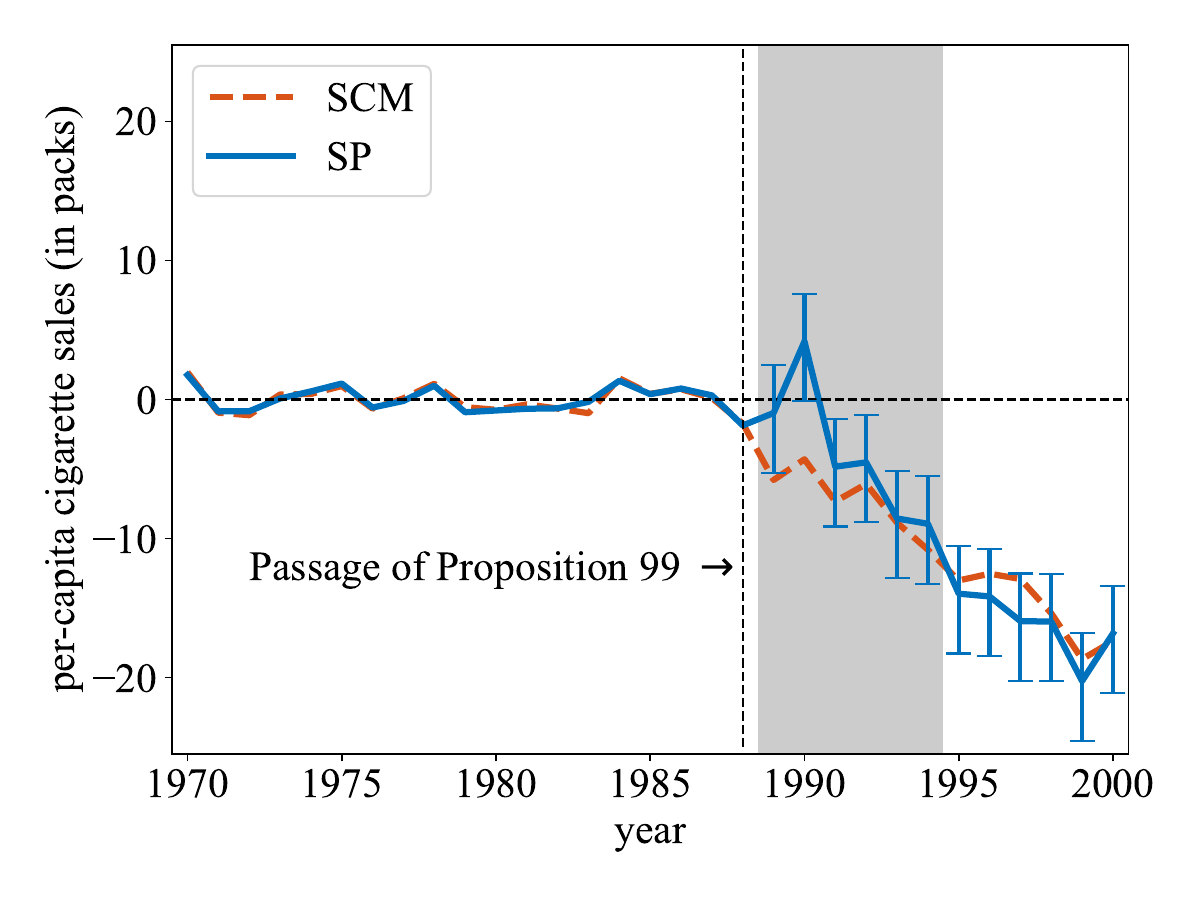}
		\caption{treatment effect estimates}
		\label{fig_gap_rob}
	\end{subfigure}
	\begin{subfigure}{.5\textwidth}
		\centering
		\includegraphics[width=\linewidth]{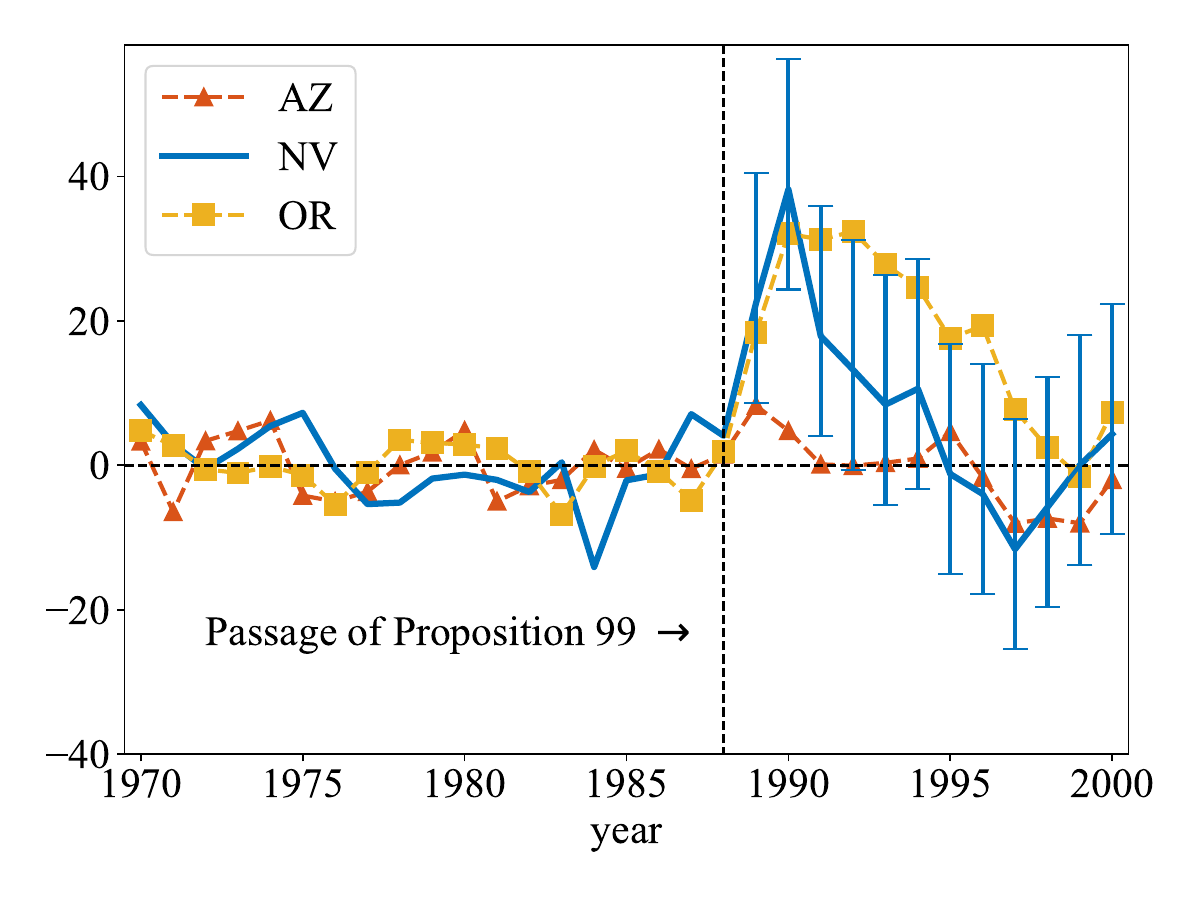}
		\caption{spillover effect estimates}
		\label{fig_spillover_rob}
	\end{subfigure}
	\caption{\small Treatment and spillover effects of excluding the ten policy-contaminated
	states that do not border California. The panels correspond to those of
	Figure \mainref{fig_prop99}{4}. The SCM line is the Abadie, Diamond and Hainmueller
	donor pool and is unchanged across specifications, so the only difference is the SP
	line. Shaded regions mark periods where the test of no spillover effect rejects at
	the 5 percent level.}
	\label{fig_prop99_rob}
\end{figure}

Two qualifications are worth stating. Dropping ten states discards donor information that
the main specification retains, which is the cost the method is designed to avoid. And
Arizona and Oregon raised their own cigarette taxes in 1994 and 1997, respectively, so their coefficients
remain a combination of leakage and own-policy response in the later years. Nevada never
enacted a comparable change, which is why its estimates carry the clearest spillover
interpretation.

\subsection{Numerical results}
\label{sec_app_numbers}

This subsection reports the numerical results underlying Figure
\mainref{fig_prop99}{4} in the main text and Figure \ref{fig_prop99_rob} in the supplement.
Tables \ref{tab:app_main_effects} and
\ref{tab:app_rob_effects} report the annual treatment-effect estimates for California and
spillover-effect estimates for Arizona, Nevada and Oregon under the main and robustness
specifications, respectively, together with 95\% confidence intervals and $p$-values.
Confidence intervals are obtained by inverting the test of Section
\mainref{spillover effect test}{4.2}, and $p$-values are computed from the
pre-treatment distribution of the corresponding statistic. 
Table \ref{tab:app_weights} reports the synthetic control weights for California, together
with the intercept and pre-treatment RMSPE, under both specifications.

\begin{table}[htbp]
\centering
\footnotesize
\color{revisionblue}
\caption{Treatment and spillover effects, main specification}
\label{tab:app_main_effects}
\setlength{\tabcolsep}{4pt}
\newcommand{\appresult}[4]{\begin{tabular}[t]{@{}c@{}}#1\\$[#2,\ #3]$\\$(#4)$\end{tabular}}
\begin{tabular*}{0.85\textwidth}{@{\extracolsep{\fill}}ccccc@{}}
\hline\hline
Year & CA & NV & AZ & OR \\
\hline
1989 & \appresult{0.08}{-3.87}{3.27}{1.000} & \appresult{14.97}{2.52}{27.65}{0.000} & \appresult{4.99}{0.00}{11.71}{0.158} & \appresult{13.90}{5.34}{22.86}{0.000} \\
1990 & \appresult{3.72}{-0.24}{6.90}{0.053} & \appresult{26.87}{14.42}{39.55}{0.000} & \appresult{-11.24}{-16.23}{-4.52}{0.000} & \appresult{26.22}{17.66}{35.18}{0.000} \\
1991 & \appresult{-3.76}{-7.72}{-0.57}{0.053} & \appresult{3.83}{-8.61}{16.52}{0.526} & \appresult{-15.37}{-20.35}{-8.64}{0.000} & \appresult{23.46}{14.89}{32.42}{0.000} \\
1992 & \appresult{-3.42}{-7.38}{-0.24}{0.053} & \appresult{-1.61}{-14.05}{11.08}{0.789} & \appresult{-16.55}{-21.54}{-9.83}{0.000} & \appresult{23.33}{14.77}{32.29}{0.000} \\
1993 & \appresult{-7.61}{-11.57}{-4.43}{0.000} & \appresult{-5.12}{-17.56}{7.57}{0.474} & \appresult{-15.14}{-20.12}{-8.42}{0.000} & \appresult{19.76}{11.19}{28.72}{0.000} \\
1994 & \appresult{-10.91}{-14.87}{-7.73}{0.000} & \appresult{2.67}{-9.77}{15.36}{0.579} & \appresult{-16.05}{-21.04}{-9.33}{0.000} & \appresult{19.43}{10.86}{28.39}{0.000} \\
1995 & \appresult{-12.83}{-16.79}{-9.65}{0.000} & \appresult{-9.68}{-22.13}{3.00}{0.211} & \appresult{-2.96}{-7.95}{3.76}{0.421} & \appresult{11.96}{3.39}{20.92}{0.000} \\
1996 & \appresult{-13.08}{-17.04}{-9.89}{0.000} & \appresult{-12.40}{-24.84}{0.29}{0.158} & \appresult{-7.20}{-12.18}{-0.47}{0.053} & \appresult{14.47}{5.90}{23.43}{0.000} \\
1997 & \appresult{-14.91}{-18.87}{-11.72}{0.000} & \appresult{-13.87}{-26.31}{-1.18}{0.000} & \appresult{-10.48}{-15.46}{-3.76}{0.000} & \appresult{6.00}{-2.56}{14.97}{0.211} \\
1998 & \appresult{-16.08}{-20.04}{-12.89}{0.000} & \appresult{-8.66}{-21.10}{4.03}{0.211} & \appresult{-9.69}{-14.68}{-2.97}{0.000} & \appresult{0.99}{-7.57}{9.95}{0.789} \\
1999 & \appresult{-18.96}{-22.92}{-15.77}{0.000} & \appresult{-1.46}{-13.91}{11.22}{0.842} & \appresult{-10.43}{-15.42}{-3.71}{0.000} & \appresult{-2.52}{-11.09}{6.44}{0.526} \\
2000 & \appresult{-15.49}{-19.45}{-12.30}{0.000} & \appresult{-1.89}{-14.34}{10.79}{0.737} & \appresult{-7.47}{-12.46}{-0.75}{0.053} & \appresult{4.71}{-3.86}{13.67}{0.263} \\
\hline\hline
\end{tabular*}
\begin{minipage}{0.85\textwidth}\vspace{0.4em}\footnotesize\textup{Notes:} Each state-year cell reports the point estimate on the first line, the 95\% confidence interval in square brackets on the second line, and the $p$-value in parentheses on the third line. CA reports treatment effects, while NV, AZ and OR report spillover effects. The outcome is per-capita cigarette sales in packs. Confidence intervals are obtained by inverting the test in Section \mainref{spillover effect test}{4.2}, and $p$-values are computed from the pre-treatment distribution of the corresponding statistic.
\end{minipage}
\end{table}

\begin{table}[htbp]
\centering
\footnotesize
\color{revisionblue}
\caption{Treatment and spillover effects, robustness specification with 41 units}
\label{tab:app_rob_effects}
\setlength{\tabcolsep}{4pt}
\newcommand{\appresult}[4]{\begin{tabular}[t]{@{}c@{}}#1\\$[#2,\ #3]$\\$(#4)$\end{tabular}}
\begin{tabular*}{0.85\textwidth}{@{\extracolsep{\fill}}ccccc@{}}
\hline\hline
Year & CA & NV & AZ & OR \\
\hline
1989 & \appresult{-0.97}{-5.27}{2.46}{0.579} & \appresult{22.46}{8.59}{40.46}{0.000} & \appresult{8.27}{1.17}{14.46}{0.000} & \appresult{18.36}{6.98}{30.43}{0.000} \\
1990 & \appresult{4.18}{-0.12}{7.60}{0.053} & \appresult{38.20}{24.34}{56.21}{0.000} & \appresult{4.78}{-2.31}{10.97}{0.368} & \appresult{32.06}{20.68}{44.13}{0.000} \\
1991 & \appresult{-4.83}{-9.13}{-1.41}{0.000} & \appresult{17.88}{4.02}{35.89}{0.053} & \appresult{0.13}{-6.97}{6.32}{1.000} & \appresult{31.26}{19.88}{43.33}{0.000} \\
1992 & \appresult{-4.52}{-8.82}{-1.10}{0.053} & \appresult{13.21}{-0.65}{31.21}{0.158} & \appresult{-0.06}{-7.15}{6.14}{1.000} & \appresult{32.38}{21.00}{44.45}{0.000} \\
1993 & \appresult{-8.55}{-12.85}{-5.13}{0.000} & \appresult{8.41}{-5.46}{26.41}{0.368} & \appresult{0.35}{-6.75}{6.54}{1.000} & \appresult{27.77}{16.39}{39.84}{0.000} \\
1994 & \appresult{-8.94}{-13.24}{-5.51}{0.000} & \appresult{10.59}{-3.27}{28.60}{0.263} & \appresult{0.94}{-6.16}{7.13}{0.842} & \appresult{24.66}{13.28}{36.73}{0.000} \\
1995 & \appresult{-13.96}{-18.26}{-10.54}{0.000} & \appresult{-1.21}{-15.08}{16.79}{0.789} & \appresult{4.63}{-2.47}{10.82}{0.368} & \appresult{17.58}{6.20}{29.65}{0.000} \\
1996 & \appresult{-14.16}{-18.46}{-10.74}{0.000} & \appresult{-3.97}{-17.84}{14.03}{0.632} & \appresult{-1.69}{-8.79}{4.50}{0.789} & \appresult{19.30}{7.92}{31.37}{0.000} \\
1997 & \appresult{-15.94}{-20.24}{-12.51}{0.000} & \appresult{-11.63}{-25.49}{6.37}{0.158} & \appresult{-8.05}{-15.15}{-1.86}{0.000} & \appresult{7.67}{-3.71}{19.74}{0.211} \\
1998 & \appresult{-15.97}{-20.27}{-12.55}{0.000} & \appresult{-5.82}{-19.68}{12.18}{0.474} & \appresult{-7.37}{-14.46}{-1.17}{0.000} & \appresult{2.44}{-8.94}{14.51}{0.579} \\
1999 & \appresult{-20.24}{-24.55}{-16.82}{0.000} & \appresult{0.01}{-13.85}{18.02}{1.000} & \appresult{-8.02}{-15.12}{-1.83}{0.000} & \appresult{-1.56}{-12.94}{10.50}{0.579} \\
2000 & \appresult{-16.82}{-21.13}{-13.40}{0.000} & \appresult{4.32}{-9.55}{22.32}{0.579} & \appresult{-2.01}{-9.10}{4.19}{0.737} & \appresult{7.37}{-4.01}{19.44}{0.211} \\
\hline\hline
\end{tabular*}
\begin{minipage}{0.85\textwidth}\vspace{0.4em}\footnotesize\textup{Notes:} Each state-year cell reports the point estimate on the first line, the 95\% confidence interval in square brackets on the second line, and the $p$-value in parentheses on the third line. CA reports treatment effects, while NV, AZ and OR report spillover effects. The outcome is per-capita cigarette sales in packs. Confidence intervals are obtained by inverting the test in Section \mainref{spillover effect test}{4.2}, and $p$-values are computed from the pre-treatment distribution of the corresponding statistic.
\end{minipage}
\end{table}

\begin{table}[htbp]
\centering
\footnotesize
\color{revisionblue}
\caption{Synthetic control weights for California}
\label{tab:app_weights}
\begin{threeparttable}
\begin{tabular}{@{}lrlr@{}}
\hline\hline
\multicolumn{2}{c}{Main specification} & \multicolumn{2}{c}{Robustness, 41 units} \\
\cline{1-2}\cline{3-4}
State & Weight & State & Weight \\
\hline
OR & 0.276 & CT & 0.241 \\
MA & 0.206 & NV & 0.164 \\
AZ & 0.148 & NE & 0.155 \\
AK & 0.101 & OR & 0.121 \\
NV & 0.069 & CO & 0.120 \\
CT & 0.061 & IL & 0.079 \\
MN & 0.036 & NC & 0.036 \\
HI & 0.035 & NH & 0.033 \\
KS & 0.033 & KS & 0.031 \\
NH & 0.031 & MT & 0.021 \\
DC & 0.005 &  &  \\
\hline
Intercept & -16.16 & Intercept & -20.03 \\
Pre-period RMSPE & 0.589 & Pre-period RMSPE & 0.918 \\
\hline\hline
\end{tabular}
\begin{tablenotes}[flushleft]
\footnotesize\upshape
\setlength{\labelsep}{0pt}
\item[] Notes: Donors receiving a weight below $0.001$ are omitted.
\end{tablenotes}
\end{threeparttable}
\end{table}

\subsection[Sensitivity Analysis of Including Spillover units in Pure Donor Method]{Sensitivity Analysis of Including Spillover units in Pure Donor Method}
\label{sec_app_nevada_sensitivity}

Table \ref{tab:pd_nevada_comparison} compares the pure-donor method (PD), using the
37 donors classified as unaffected in the main specification, with the same method
after adding Nevada to the donor pool. Both specifications estimate an unrestricted
intercept and nonnegative weights summing to one using outcomes from 1970--1988,
then apply the fitted relationships to 1989--2000. Including Nevada shifts the
estimated treatment effects downward in the first six post-treatment years, making
the policy appear to reduce California cigarette sales by more than the pure-donor
comparison suggests. This pattern is consistent with the positive spillover effects
estimated for Nevada in the early years. Spillover-driven sales in Nevada can raise
the synthetic counterfactual and exaggerate the apparent reduction in California
sales. The differences change sign in later years, so this distortion is not uniform
over time.

\begin{table}[H]
\centering
\footnotesize
\color{revisionblue}
\caption{Treatment-effect estimates with and without Nevada in the donor pool}
\label{tab:pd_nevada_comparison}
\setlength{\tabcolsep}{2pt}
\renewcommand{\arraystretch}{1.5}
\begin{tabular*}{\textwidth}{@{\extracolsep{\fill}}l*{12}{r}@{}}
\hline\hline
Method & 1989 & 1990 & 1991 & 1992 & 1993 & 1994 & 1995 & 1996 & 1997 & 1998 & 1999 & 2000 \\
\hline
PD & -4.32 & 0.83 & -5.26 & -3.88 & -6.24 & -7.55 & -13.41 & -13.63 & -15.54 & -16.60 & -17.92 & -17.39 \\
PD plus Nevada & -5.78 & -4.30 & -7.33 & -6.07 & -8.85 & -10.80 & -13.01 & -12.53 & -12.91 & -15.37 & -18.67 & -17.38 \\
Difference & -1.47 & -5.12 & -2.07 & -2.19 & -2.61 & -3.25 & 0.39 & 1.10 & 2.63 & 1.22 & -0.74 & 0.01 \\
\hline\hline
\end{tabular*}
\begin{minipage}{\textwidth}\vspace{0.4em}\footnotesize\textup{Notes.}
Entries are estimated treatment effects on California's per-capita cigarette sales,
in packs. PD excludes all 13 potentially contaminated states. PD plus Nevada adds
only Nevada, leaving the other 12 excluded, and reproduces the SCM benchmark in
Figure \mainref{fig_prop99}{5}. The difference is PD plus Nevada minus PD, computed
before rounding. Negative values indicate a more negative estimate when Nevada is
included. Both the weights and the intercept are re-estimated, so the difference
includes changes in the fitted relationship as well as possible donor contamination.
\end{minipage}
\end{table}

}

\bibliographystyle{apalike}
\bibliography{ref}

\end{document}